\documentclass[10pt,a4paper]{article}
\usepackage[utf8]{inputenc} 
\usepackage{CJKutf8}
\usepackage{hyperref}       
\usepackage{url}            
\usepackage{booktabs}       
\usepackage{amsfonts}       
\usepackage{nicefrac}       
\usepackage{microtype}      
\usepackage{lipsum}
\usepackage{graphicx}
\usepackage{mathtools}
\usepackage{amsthm}
\usepackage{bm}
\graphicspath{{./images/}}
\newtheorem{theorem}{theorem}
\newtheorem{lemma}{lemma}
\title{The principle of minimum virtual work and its application in bridge engineering}
\author{
  Xiang Lukai\\
  China Railway Eryuan Engineering Group Co., Ltd\\ 
  \texttt{247073858@qq.com}\\
}
\begin{document}
\maketitle
\begin{CJK}{UTF8}{gbsn}
%
\bgroup
\hrule
\egroup
%

\begin{abstract}
\noindent \textit{In mechanics, common energy principles are based on fixed boundary conditions. However, in bridge engineering structures, it is usually necessary to adjust the boundary conditions to make the structure's internal force reasonable and save materials. However, there is currently little theoretical research in this area. To solve this problem, this paper proposes the principle of minimum virtual work for movable boundaries in mechanics through theoretical derivation such as variation method and tensor analysis. It reveals that the exact solution of the mechanical system minimizes the total virtual work of the system among all possible displacements, and the conclusion that the principle of minimum potential energy is a special case of this principle is obtained. At the same time, proposed virtual work boundaries and control conditions, which added to the fundamental equations of mechanics. The general formula of multidimensional variation method for movable boundaries is also proposed, which can be used to easily derive the basic control equations of the mechanical system. The incremental method is used to prove the theory of minimum value in multidimensional space, which extends the Pontryagin's minimum value principle. Multiple bridge examples were listed to demonstrate the extensive practical value of the theory presented in this article. The theory proposed in this article enriches the energy principle and variation method, establishes fundamental equations of mechanics for the structural optimization of movable boundary, and provides a path for active control of mechanical structures, which has important theoretical and engineering practical significance.
}\\


\end{abstract}%

\bgroup
\hrule
\egroup


\section {Overview}
In mechanics, common energy principles such as the principle of minimum potential energy and the principle of minimum complementary energy have fixed boundary conditions, such as fixed integration domains, force and displacement boundaries. However, in engineering structures, it is often necessary to actively control boundary conditions to ensure that the structure is subjected to reasonable forces and has good economic efficiency. Especially for bridge engineering, this demand is quite urgent, such as how to configure prestressed tendon to make the internal force of prestressed concrete beams reasonable; Determine the reasonable span of continuous beams and continuous rigid frames \cite{wayajun2016, daigonglian2015} to minimize their internal energy and save the most materials; How to adjust the cable tension of a cable-stayed bridge to achieve the most reasonable force \cite{daijie2019, zhouyungang2017,xiaorucheng1998}, the minimum bending moment of the bridge tower and main beam; How to adjust the suspension force of an arch bridge to achieve optimal internal force distribution \cite{liuzhao2009, fujinlong2014}; how to adjust the arch axis of an arch bridge to minimize its bending; how to adjust the node position of the truss to make its stress more reasonable.

The energy principle is the foundation of structural analysis. The principle of virtual displacement was proposed by John Bernoulli in 1717, and the reciprocal theorem of displacement was established by Maxwell in the United Kingdom in 1864. In 1872, Betti in Italy extended the reciprocal theorem of displacement to the reciprocal theorem of work;Castigliano in Italy proposed the Castigliano's first and second theorem in 1879, and Ergesser in Germany proposed the complementary energy method in 1889.In 1950, Reissener proposed the generalized variation principle for the two kind of variables in the theory of elasticity \cite{qianfeng2022}, and in 1954, Hu Haichang proposed the generalized variation principle for the three kind of variables \cite{huhaichang1954}.In 1983, Long Yuqiu proposed the partitioned mixed generalized variation principle \cite{longyuqiu1983}.The above structural energy principles can be analyzed using the variation method with fixed boundaries.

The research on variation method has always been researched by many scholars \cite{penghaijun2011,zhanghongwu2006, reza2007, longyuqiu1987}. Besides the fixed boundary variation method, there is also the movable boundary variation method.In 1981, Niu Xiangjun \cite{niuxiangjun1981} used the variation method of movable boundary and based on the stationary condition of zero first-order variation, established a discrete variation form for solids, eliminating the errors introduced at the element boundaries during finite element discretization.In 1985, Zhong Wanxie \cite{zhongwanxie1997} proposed the parametric variation principle, which divides the variables of the functional into state variables that participate in the variation process and control variables (also known as parametric variables) that do not participate in the variation process. After nearly 40 years of continuous development, the parametric variation principle has been successfully applied in various fields, including elastic-plastic analysis, contact problems, lubrication mechanics, geotechnical mechanics, and other engineering fields. In 2024, Wu Chengwei \cite{wuchengwei2024} reviewed this method.In 2006, Lao Dazhong \cite{laodazhong2011} proposed and proved the variation problem of complete functional for partial derivatives of any number of independent variables, any number of multivariate functions, and any order multivariate functions with fixed boundaries.In 2007, Reza Memarbashi proposed a solution to the variation problem of moving boundary based on decomposition method.

In mechanics, the principle of energy can be analyzed using variation methods, but variation methods are only applicable when the control domain is an open set. When the control variable is constrained, the variation method is no longer applicable. Soviet scholar L S. Pontryagin's minimum principle, proposed by Pontryagin in 1958, significantly extended the variation method, which can be applied to cases where the control domain is a closed set, laying the foundation for modern optimal control. However, there is still little research on the application of this method in the field of mechanics.

This article attempts to derive the energy principle applicable to mechanical structures with movable boundaries using mathematical tools such as tensor analysis \cite{huangkezhi2020} and functional analysis \cite{sunjiong2018, liuzhongkan1986}, by using the variation method and drawing on Pontryagin's idea of minima. The principle will be applied to bridge engineering in order to establish a relatively complete theoretical basis for optimal design of bridge structures.

\section {Definition, Theorem and Explanation} \label{sec:dingyi}
\numberwithin{equation}{subsection}
This article presents some new concepts and theories, which are first defined for ease of reading.
\begin{theorem} [virtual work]
Virtual work: refers to the product of the system's load and displacement, which is called virtual work, not the actual work done by the load. For example, for a linear elastic system, the actual work is half of the virtual work.
\end{theorem}
\begin{theorem} [principle minimum virtual work]
The principle of minimum virtual work: For a movable boundary mechanical system, the exact solution of the mechanical system minimizes the total virtual work of the system among all possible displacements. In addition to satisfying the equilibrium equations, constitutive equations，geometric relationships, force and displacement boundary conditions of conventional mechanical systems, the movable boundary system also needs to meet control conditions and virtual work boundary conditions.
\end{theorem}
\begin{theorem} [Control condition]
Control condition: The control boundary conditions that need to be satisfied by the control boundary in a movable boundary mechanics system to minimize the total virtual work of the mechanics system.
\end{theorem}
\begin{theorem} [Virtual work boundary]
Virtual work boundary: For a movable boundary mechanical system, when the total virtual work of the mechanical system is minimized, the sum of the virtual work density of the volumetric force and the strain energy density of the surface force must be zero.
\end{theorem}

\begin{theorem} [optimal control index]
The optimal control index $ op $ : 1- total virtual work $ W $ /fixed load virtual work $ W_0$ without control load. When $ op=0 $ , it indicates no control,
When $ op=1 $ , it indicates optimal control, and when $ op<0 $ , it indicates over control.
\end{theorem}

\begin{theorem} [Definition of functional extremum points]
Definition 1: Let $\bm{X}$   be a Banach space, and the functional $ f $  is defined within the neighborhood $ D $  of point $ x ^ * \in \bm {X}$  . If there exists a neighborhood $ D_1 \subset D$  of $ x ^ * $ , such that for all $ x \in D_1 $ 
\begin{equation}\label{eq:(dingyi.1.1)}
f(x^*)<=f(x)
\end{equation}
Then $ x ^ * $ is called a local minimum, and $ f (x ^ *) $ is called a local minimum.
\end{theorem}
\begin{theorem} [necessary condition for functional stationarity]
If the functional $ f $ reaches its extremum at $ x ^ * $ and has a first-order variation $ \delta $ at this point, then
\begin{equation}\label{eq:(dingyi.1.2)}
\delta f(x^*)=0
\end{equation}
\end{theorem}
\begin{lemma} [Functional variation Lemma]
Assuming that the function $ f (x) $ is continuous over the interval [a, b], any function $ \eta (x) $ has an n-th order continuous derivative over the interval [a, b], and for a positive number m (m=0,1,..., n), when the condition is satisfied
\begin{equation}\label{eq:(dingyi.1.3)}
\eta_{k}(a)=\eta_{k}(b)=0
\end{equation}
When, if the points are accumulated
\begin{equation}\label{eq:(dingyi.1.4)}
\int_ {a,b}f (x)\eta(x)dx=0
\end{equation}   
Then there are
\begin{equation}\label{eq:(dingyi.1.5)}
f(x)\equiv 0
\end{equation}   
\end{lemma}
Symbolic convention: This article adopts tensor analysis component notation and Einstein summation convention. The subscript $ j$  represents the $ jth$  component, the subscript $ i$ after the comma represent partial derivatives of $ i$  variables. For example, $ y_j, y_{j, i}, y_{j, ip}$  represent the $ jth$  component of $ y$  , the $ jth$  component of $ y$  calculates partial derivatives of the $ ith$   component, the $ jth$  component of $ y$  calculates partial derivatives of the $ ith$   component, and then calculates partial derivatives of the $ pth$ component. $ L_{y_j, i}$  represents $ L$  partial derivative of $ y_{j, i}$  .

\section {Multidimensional Variation Method with Movable Boundaries} \label{sec: bianfen}
The variation method for movable boundaries of one-dimensional variables is relatively mature, and there are also related studies on the variation method for fixed boundaries in multidimensional space. However, there is currently little research on the variation method for movable boundaries in multidimensional space. Therefore, this article first derives the variation method for movable boundaries in multidimensional space, seeking a theoretical basis for the principle of minimum virtual work in mechanics.

\subsection {general theoretical derivation}
\numberwithin{equation}{subsection}
Assuming $ \bm {X}$  is the $ m $ dimensional Euclid space $ E_m $ , the independent variables $ \bm {x}=[x_1, x_2... x_m] ^ T \in \bm {X}$  , $ \bm {Y}$  is the vector space $ C2 (E_m) $ space 
 composed of all n-dimensional vectors with second-order continuous partial derivatives on $ E_m $ , $ \bm {y}=[y_1, y_2... y_n] ^ T \in \bf {Y}$  , control boundary $ \bm {U}$  is a $ k $  dimensional piecewise continuous function, $ \bm {u}=[u_1, u_2... u_p] ^ T \in \bf {U}$  , $ L$  and $ \varphi$  are all continuous functions with respect to their independent variables. $ L$  and $ \varphi$  are continuously differentiable for $ y_j, y_{j, i}, y_{j, ip}$  , and $ L$  satisfies the Lipschitz condition for variables $ y_j, y_{j, i}, y {j,  ip}$  . Consider the following functional
\begin{equation}\label{eq:(bianfen.1.1)}
J = \iiint_D L[x_i,y_j,y_{j,i},y_{j,ip},u_k] dv + \iint_S \varphi[\bar {x}_i ,\bar {y}_j ,\bar {y}_ {j,i}] ds
\end{equation}
Where $ D $ represents the integration domain and $ S $ represents the boundary.
variation analysis of equation \ref{eq:(bianfen.1.1)} yields
\begin{multline} \label{eq:(bianfen.1.2)}    
\delta J = \iiint_{D+\delta D} L[x_i,y_j+\delta y_j,y_{j,i}+\delta y_{j,i},y_{j,ip}+\delta y_{j, ip},u_k+\delta u_k] dv 
\\-\iiint_D L[x_i,y_j,y_{j,i},y_{j,ip},u_k] dv 
\\+ \iint_S \varphi[\bar {x}_i +\delta \bar {x}_i ,\bar {y}_j +\delta \bar {y}_j ,\bar {y}_ {j,i}+\delta \bar {y}_ {j,i}] ds
- \iint_S \varphi[\bar {x}_i ,\bar {y}_j ,\bar {y}_ {j,i}] ds
\\= \iiint_D \{L[x_i,y_j+\delta y_j,y_{j,i}+\delta y_{j,i},y_{j,ip}+\delta y_{j,ip},u_k+\delta u_k]-L[x_i,y_j,y_{j,i},y_{j,ip},u_k] \}dv 
\\+\iiint_{\delta D} L[x_i,y_j+\delta y_j,y_{j,i}+\delta y_{j,i},y_{j,ip}+\delta y_{j, ip},u_k+\delta u_k] dv 
\\+\iint_S \{\varphi[\bar {x}_i +\delta \bar {x}_i ,\bar {y}_j +\delta \bar {y}_j ,\bar {y}_ {j,i}+\delta \bar {y}_ {j,i}] -\varphi[\bar {x}_i ,\bar {y}_j ,\bar {y}_ {j,i}]\} ds
\end{multline} 
Perform Taylor expansion on the first term of equation \ref{eq:(bianfen.1.2)} and omit higher-order terms
\begin{multline}\label{eq:(bianfen.1.3)}    
\iiint_D \{L[x_i,y_j+\delta y_j,y_{j,i}+\delta y_{j,i},y_{j,ip}+\delta y_{j,ip},u_k+\delta u_k]-L[x_i,y_j,y_{j,i},y_{j,ip},u_k] \}dv 
\\=\iiint_D \{L_{y_j} \delta y_j + L_{y_{j,i}} \delta y_{j,i}+L_{y_{j,ip}} \delta y_{j,ip} +L_{u_k} \delta u_k\}dv 
\end{multline}
According to the fractional integral method, there are
\begin{multline}\label{eq:(bianfen.1.4)}
L_{y_{j,i}} \delta y_{j,i}
=\frac{\partial L}{\partial y_{j,i}} \delta {\frac{\partial y_j}{\partial x_i}} 
=\frac{\partial L}{\partial y_{j,i}} \frac{\partial {\delta y_j}}{\partial x_i}
=\frac{\partial} {\partial x_i} \left(L_{y_{j,i}} \delta y_j \right)-\delta y_j \frac{\partial L_{y_{j,i}} } {\partial x_i} 
\\=\left(L_{y_{j,i}} \delta y_j \right)_{,i}-\delta y_j \left(L_{y_{j,i}}\right)_{,i }
\end{multline}
As can be seen from \ref{eq:(bianfen.1.4)}
\begin{multline}\label{eq:(bianfen.1.5)}
L_{y_{j,ip}} \delta y_{j,ip}
=\left(L_{y_{j,ip}} \delta y_{j,i} \right)_{,p}-\delta y_{j,i} \left(L_{y_{j,ip}}\right)_{,p}
\\=\left(L_{y_{j,ip}} \delta y_{j,i} \right)_{,p}-\left[\left(\delta y_{i} \left(L_{y_{j,ip}}\right)_{,p}\right)_{,i}-\delta y_{i} \left(L_{y_{j,ip}}\right)_{,ip}\right]
\\=\left(L_{y_{j,ip}} \delta y_{j,i} \right)_{,p}-\left(\delta y_{i} \left(L_{y_{j,ip}}\right)_{,p}\right)_{,i}+\delta y_{i} \left(L_{y_{j,ip}}\right)_{,ip}
\end{multline}
Substituting \ref{eq:(bianfen.1.4)} \ref{eq:(bianfen.1.5)} into \ref{eq:(bianfen.1.3)} yields
\begin{multline} \label{eq:(bianfen.1.6)}
\iiint_D \{L_{y_j} \delta y_j + L_{y_{j,i}} \delta y_{j,i}+L{y_{j,ip} \delta y_{j,ip} +L_{u_k} \delta u_k}\} dv 
\\=\iiint_D L_{y_j} \delta y_j +\left(L_{y_{j,i}} \delta y_j \right)_{,i}-\delta y_j \left(L_{y_{j,i}}\right)_{,j } +\left(L_{y_{j,ip}} \delta y_{j,i} \right)_{,p}-\left(\delta y_{i} \left(L_{y_{j,ip}}\right)_{,p}\right)_{,i}
\\+\delta y_{i} \left(L_{y_{j,ip}}\right)_{,ip}+L_{u_k} \delta u_k dv 
\\=\iiint_D \{[L_{y_j} - \left(L_{y_{j,i}}\right)_{,i}+\left(L_{y_{j,ip}}\right)_{,ip}]\delta y_j +[\left(L_{y_{j,i}} \delta y_j \right) -\left(\delta y_{i} \left(L_{y_{j,ip}}\right)_{,p}\right)]_{,i}
\\+\left(L_{y_{j,ip}} \delta y_{j,i} \right)_{,p}+L_{u_k} \delta u_k\}dv 
\end{multline}
According to Gauss's formula, there is
\begin{equation} \label{eq:(bianfen.1.7)}
\iiint_D [\left(L_{y_{j,i}} \delta y_j \right) -\left(\delta y_{i} \left(L_{y_{j,ip}}\right)_{,p}\right)]_{,i}dv
=\iint_S [\left(L_{y_{j,i}} \delta y_j \right) -\left(\delta y_{i} \left(L_{y_{j,ip}}\right)_{,p}\right)]n_i da
\end{equation} 
\begin{equation} \label{eq:(bianfen.1.8)}
\iiint_D \left(L_{y_{j,ip}} \delta y_{j,i} \right)_{,p}dv
=\iint_S \left(L_{y_{j,ip}} \delta y_{j,i} \right)n_p da
\end{equation} 
Where $ n_i=cos (\alpha_i) $ represents the direction cosine of the vector with respect to the coordinate axis $ x_i $ .
Substituting \ref{eq:(bianfen.1.7)} \ref{eq:(bianfen.1.8)} into \ref{eq:(bianfen.1.6)} yields
\begin{multline} \label{eq:(bianfen.1.9)}
\iiint_D \{L_{y_j} \delta y_j + L_{y_{j,i}} \delta y_{j,i}+L{y_{j,ip} \delta y_{j,ip}+L_{u_k} \delta u_k
}\} dv 
\\=\iiint_D \{[L_{y_j} - \left(L_{y_{j,i}}\right)_{,i}+\left(L_{y_{j,ip}}\right)_{,ip}]\delta y_j+L_{u_k} \delta u_k\} dv 
\\+\iint_S [\left(L_{y_{j,i}} \delta y_j \right) -\left(\delta y_{i} \left(L_{y_{j,ip}}\right)_{,p}\right)]n_i da
+\iint_S \left(L_{y_{j,ip}} \delta y_{j,i} \right)n_p da
\end{multline} 
For the second term of $ \ref{eq:(bianfen.1.2)}$  , applying the mean value theorem of integrals, we obtain
\begin{multline} \label{eq:(bianfen.1.10)}
\iiint_{\delta D} L[x_i,y_j+\delta y_j,y_{j,i}+\delta y_{j,i},y_{j,ip}+\delta y_{j, ip},u_k+\delta u_k] dv 
\\=\iint_S L[x_i+\theta_i \delta_i x_i,y_j+\delta y_j,y_{j,i}+\delta y_{j,i},y_{j,ip}+\delta y_{j,ip},u_k+\delta u_k]\delta \bar{n} da
\end{multline} 
Among them, $ 0<\theta_i<1 $ . According to the continuity of the $ L$  functional,
\begin{equation} \label{eq:(bianfen.1.11)}    
L[x_i+\theta_i \delta_i x_i,y_j+\delta y_j,y_{j,i}+\delta y_{j,i},y_{j,ip}+\delta y_{j,ip},u_k+\delta u_k]
\\=L[x_i,y_j,y_{j,i},y_{j, ip},u_k]+\varepsilon_1
\end{equation} 
Substituting \ref{eq:(bianfen.1.11)} into \ref{eq:(bianfen.1.10)} yields
\begin{multline} \label{eq:(bianfen.1.12)}    
\iiint_{\delta D} L[x_i,y_j+\delta y_j,y_{j,i}+\delta y_{j,i},y_{j,ip}+\delta y_{j, ip},u_k+\delta u_k] dv 
\\=\iint_{S} L[x_i,y_j,y_{j,i},y_{j,ip},u_k]\delta \bar{n} da+\iint_{S} \varepsilon_1 \delta \bar{n} da
\\=\iint_{S} L[x_i+ x_i,y_j+,y_{j,i},u_k]\delta \bar{n} da+\varepsilon_1 \delta D
\\=\iint_{S} L[x_i+ x_i,y_j+,y_{j,i},u_k]\delta \bar{n} da+o(\delta D)
\end{multline} 
When $ \delta x_i, \delta y_j, \delta y_ {j, i}, \delta y_ {j, ip} \to 0 $ , $ \varepsilon_1 \to 0 $ , then \ref{eq:(bianfen.1.12)} can omit high-order small quantities and become
\begin{multline} \label{eq:(bianfen.1.13)}    
\iiint_{\delta D} L[x_i,y_j+\delta y_j,y_{j,i}+\delta y_{j,i},y_{j,ip}+\delta y_{j, ip},u_k+\delta u_k] dv 
\\=\iint_{S} L[x_i+ x_i,y_j+,y_{j,i},u_k]\delta \bar{n} da
\end{multline} 

By conducting Taylor expansion analysis on the third term of \ref{eq:(bianfen.1.2)} and omit the higher order term, it can be concluded that
\begin{multline} \label{eq:(bianfen.1.14)}
\iint_S \{\varphi[\bar {x}_i +\delta \bar {x}_i ,\bar {y}_j +\delta \bar {y}_j ,\bar {y}_ {j,i}+\delta \bar {y}_ {j,i}] -\varphi[\bar {x}_i ,\bar {y}_j ,\bar {y}_ {j,i}]\} ds
=\\\iint_S [(\varphi_{\bar {x}_i } \delta \bar {x}_i  + \varphi_{\bar {y}_j } \delta \bar{y} _j+\varphi_{\bar {y}_ {j,i}} \delta \bar {y}_ {j,i})] ds
\end{multline} 
By substituting \ref{eq:(bianfen.1.9)}, \ref{eq:(bianfen.1.13)}, and \ref{eq:(bianfen.1.14)} into \ref{eq:(bianfen.1.2)}, we can obtain
\begin{multline} \label{eq:(bianfen.1.15)}    
\delta J = \iiint_D \{[L_{y_j} - \left(L_{y_{j,i}}\right)_{,i}+\left(L_{y_{j,ip}}\right)_{,ip}]\delta y_j \}dv \\+\iint_S [\left(L_{y_{j,i}} \delta y_j \right) -\delta y_{i} \left(L_{y_{j,ip}}\right)_{,p}]n_i da+
\\
\iint_S \left(L_{y_{j,ip}} \delta y_{j,i} \right)n_p da
+\iint_S L[x_i,y_j,y_{j,i},y_{j,ip},u_k]\delta \bar{n} da+ 
\\\iint_S [(\varphi_{\bar {x}_i } \delta \bar {x}_i  + \varphi_{\bar {y}_j } \delta \bar{y} j
+\varphi_{\bar {y}_ {j,i}} \delta \bar {y}_ {j,i}] ds
+\iiint_D [L_{u_k} \delta u_k] dv
\\=\iiint_D \{[L_{y_j} - \left(L_{y_{j,i}}\right)_{,i}+\left(L_{y_{j,ip}}\right)_{, ip}]\delta y_j \}dv 
+\iint_S [\left(L_{y_{j,i}}  \right) - \left(L_{y_{j,ip}}\right)_{, p}]n_i \delta y_j da+
\\\iint_S \left(L_{y_{j,ip}}  \right)n_p \delta y_{j,i} da
+\iint_S L[x_i,y_j,y_{j,i},y_{j,ip},u_k]\delta \bar{n} da
+ 
\\\iint_S [(\varphi_{\bar {x}_i } \delta \bar {x}_i  + \varphi_{\bar {y}_j } \delta \bar{y} _j+\varphi_{\bar {y}_ {j,i}} \delta \bar {y}_ {j,i}] ds
+\iiint_D [L_{u_k} \delta u_k] dv
\end{multline} 
Due to the movable boundary, therefore
\begin{equation} \label{eq:(bianfen.1.16)}    
\delta y_j|_{(x_i=\bar {x}_i )}=\delta \bar{y} _j-y_{j,l}\delta \bar {x}_l
\end{equation} 
\begin{equation} \label{eq:(bianfen.1.17)}    
\delta y_{j,i}|_{(x_i=\bar {x}_i )}=\delta \bar{y} _ {j,i}-y_ {j,ir}\delta \bar {x}_r
\end{equation} 
Substituting \ref{eq:(bianfen.1.16)} \ref{eq:(bianfen.1.17)} into \ref{eq:(bianfen.1.15)} yields
\begin{multline} \label{eq:(bianfen.1.18)}    
\delta J = \iiint_D \{[L_{y_j} - \left(L_{y_{j,i}}\right)_{,i}+\left(L_{y_{j,ip}}\right)_{,ip}]\delta y_j \}dv \\+\iint_S [\left(L_{y_{j,i}}  \right) - \left(L_{y_{j,ip}}\right)_{,p}]n_i (\delta \bar{y} _j-y_{j,l}\delta \bar {x}_l ) da
\\+\iint_S \left(L_{y_{j,ip}}  \right)n_p (\delta \bar{y} _ {j,i}-y_ {j,ir}\delta \bar {x}_r )) da
+\iint_S L[x_i,y_j,y_{j,i},y_{j,ip},u_k]\delta \bar{n} da
\\+ \iint_S [(\varphi_{\bar {x}_i } \delta \bar {x}_i  + \varphi_{\bar {y}_j } \delta \bar{y} _j+\varphi_{\bar {y}_ {j,i}} \delta \bar {y}_ {j,i}] da
+\iiint_D [L_{u_k} \delta u_k] dv
\\=\iiint_D \{[L_{y_j} - \left(L_{y_{j,i}}\right)_{,i}+\left(L_{y_{j,ip}}\right)_{, ip}]\delta y_j \}dv 
+\iint_S L[x_i,y_j,y_{j,i},y_{j,ip},u_k]\delta \bar{n} da
\\+ \iint_S [\varphi_{\bar {x}_i } \delta \bar {x}_i  ] da
+\iint_S \{[\left(L_{y_{j,i}}  \right) - \left(L_{y_{j,ip}}\right)_{,p}]n_i (-y_{j,l})\delta \bar {x}_l \} da
\\+\iint_S \left(L_{y_{j,ip}}  \right)n_p (-y_{j,ir})\delta \bar {x}_r ) da
+\iint_S \{[\left(L_{y_{j,i}}  \right) -\left(L_{y_{j,ip}}\right)_{,p}]n_i + \varphi_{\bar {y}_j } \}\delta \bar{y} _j da
\\+\iint_S [\left(L_{y_{j,ip}}  \right)n_p +\varphi_{\bar {y}_ {j,i}}] \delta \bar {y}_ {j,i} da+\iiint_D [L_{u_k} \delta u_k] dv
\end{multline} 
And because
\begin{equation} \label{eq:(bianfen.1.19)}    
\delta \bar {x}_i = n_i \delta \bar{n}
\end{equation} 
Substituting \ref{eq:(bianfen.1.19)} into \ref{eq:(bianfen.1.18)} yields
\begin{multline} \label{eq:(bianfen.1.20)}    
\delta J = \iiint_D \{[L_{y_j} - \left(L_{y_{j,i}}\right)_{,i}+\left(L_{y_{j,ip}}\right)_{,ip}]\delta y_j \}dv \\+\iint_S L[x_i,y_j,y_{j,i},y_{j,ip},u_k]\delta \bar{n} da
+ \iint_S [\varphi_{\bar {x}_i } n_i \delta \bar{n} ] da
\\+\iint_S \{[\left(L_{y_{j,i}}  \right) -\left(L_{y_{j,ip}}\right)_{,p}]n_i (-y_{j,l})n_l \delta \bar{n}\} da
\\+\iint_S \left(L_{y_{j,ip}}  \right)n_p (-y_{j,ir}) n_r \delta \bar{n}) da
+\iint_S \{[\left(L_{y_{j,i}}  \right) -\left(L_{y_{j,ip}}\right)_{,p}]n_i + \varphi_{\bar {y}_j } \}\delta \bar{y} _j da
\\+\iint_S [\left(L_{y_{j,ip}}  \right)n_p +\varphi_{\bar {y}_ {j,i}}] \delta \bar {y}_ {j,i} da+\iiint_D [L_{u_k} \delta u_k] dv
\\= \iiint_D \{[L_{y_j} - \left(L_{y_{j,i}}\right)_{,i}+\left(L_{y_{j,ip}}\right)_{, ip}]\delta y_j \}dv
\\+\iint_S \{L[x_i,y_j,y_{j,i},y_{j,ip},u_k]+ \varphi_{\bar {x}_i } n_i 
-y_{j,l} L_{y_{j,i}} n_i n_l   + y_{j,l} \left(L_{y_{j,ip}}\right)_{,p} n_i n_l
\\-y_{j,ip} L_{y_{j,ir}} n_i  n_r \}\delta \bar{n} da
+\iint_S \{[\left(L_{y_{j,i}}  \right) -\left(L_{y_{j,ip}}\right)_{,p}]n_i + \varphi_{\bar {y}_j } \}\delta \bar{y} _j da
\\+\iint_S [\left(L_{y_{j,ip}}  \right)n_p +\varphi_{\bar {y}_ {j,i}}] \delta \bar {y}_ {j,i} da+\iiint_D [L_{u_k} \delta u_k] dv
\end{multline} 
According to the condition for taking the extremum of the functional, $ \delta J=0 $ , \ref{eq:(bianfen.1.20)} is changed to
\begin{multline} \label{eq:(bianfen.1.21)}    
\delta J = \iiint_D \{[L_{y_j} - \left(L_{y_{j,i}}\right)_{,i}+\left(L_{y_{j,ip}}\right)_{, ip}]\delta y_j \}dv
\\+\iint_S \{L[x_i,y_j,y_{j,i},y_{j,ip},u_k]+ \varphi_{\bar {x}_i } n_i 
-y_{j,l} L_{y_{j,i}} n_i n_l   + y_{j,l} \left(L_{y_{j,ip}}\right)_{,p} n_i n_l
\\-y_{j,ip} L_{y_{j,ir}} n_i  n_r \}\delta \bar{n} da
\\+\iint_S \{[\left(L_{y_{j,i}}  \right) -\left(L_{y_{j,ip}}\right)_{,p}]n_i + \varphi_{\bar {y}_j } \}\delta \bar{y} _j da
+\iint_S [\left(L_{y_{j,ip}}  \right)n_p 
\\+\varphi_{\bar {y}_ {j,i}}] \delta \bar {y}_ {j,i} da
+\iiint_D [L_{u_k} \delta u_k] dv=0
\end{multline} 
According to the functional variation lemma, since $ \delta y_j $ is an arbitrary value, the Euler equation can be obtained
\begin{equation} \label{eq:(bianfen.1.22)}    
L_{y_j} - \left(L_{y_{j,i}}\right)_{,i}+\left(L_{y_{j,ip}}\right)_{,ip} =0
\end{equation}
Next, let's discuss the boundary conditions. Since $ \delta \bar {n}$  is an arbitrary value, we can obtain
\begin{equation} \label{eq:(bianfen.1.23)}    
L[x_i,y_j,y_{j,i},y_{j,ip},u_k]+ \varphi_{\bar {x}_i } n_i 
-y_{j,l} L_{y_{j,i}} n_i n_l   + y_{j,l} \left(L_{y_{j,ir}}\right)_{,r} n_i n_l 
-y_{j,ip} L_{y_{j,ip}} n_i  n_p =0
\end{equation}
Since $ \delta \bar {y} _j $ is an arbitrary value, we can obtain
\begin{equation} \label{eq:(bianfen.1.24)}    
[\left(L_{y_{j,i}}  \right) -\left(L_{y_{j,ip}}\right)_{,p}]n_i + \varphi_{\bar {y}_j}=0
\end{equation}
Due to $ \delta \bar {y}_ {j, i}$  is an arbitrary value, which can be obtained
\begin{equation} \label{eq:(bianfen.1.25)}    
\left(L_{y_{j,ip}}  \right)n_p +\varphi_{\bar {y}_ {j,i}} =0
\end{equation}
Since $ \delta u_k $ is an arbitrary value, we can obtain
\begin{equation} \label{eq:(bianfen.1.26)}    
L_{u_k} =0
\end{equation}

Next, based on the boundary condition\ref{eq:(bianfen.1.24)} and \ref{eq:(bianfen.1.25)} simplify \ref{eq:(bianfen.1.23)} to obtain
\begin{multline} \label{eq:(bianfen.1.27)}    
L[x_i,y_j,y_{j,i},y_{j,ip},u_k]+ \varphi_{\bar {x}_i } n_i 
-y_{j,l} L_{y_{j,i}} n_i n_l   + y_{j,l} \left(L_{y_{j,ir}}\right)_{,r} n_i n_l 
-y_{j,ir} L_{y_{j,ip}} n_i  n_r
\\=L[x_i,y_j,y_{j,i},y_{j,ip},u_k]+ \varphi_{\bar {x}_i } n_i 
-y_{j,l} [L_{y_{j,i}} - \left(L_{y_{j,ir}}\right)_{,r}]n_i n_l 
-y_{j,ir} (L_{y_{j,ip}} n_p)  n_r 
\\=L[x_i,y_j,y_{j,i},y_{j,ip},u_k]+ \varphi_{\bar {x}_i } n_i 
+\bar{y}_{j,l}\varphi_{\bar {y}_j} n_l 
+\bar{y}_{j,ir}\varphi_{\bar {y}_ {j,i}} n_r =0
\end{multline}

Thus, the basic equations \ref{eq:(bianfen.1.22)} \text{-} \ref{eq:(bianfen.1.26)} of multidimensional spatial variation method was obtained.

\subsection {multidimensional spatial variation method with sharp points}\label{subsec:sharpPoint}
\numberwithin{equation}{subsection}
In the previous section, we derived the multidimensional space variation method. When a multidimensional space is divided into multiple volume domains by different boundaries, that is, when the boundaries have sharp points, the expressions of the Euler equation and boundary conditions are different. We will discuss in detail below.

The extreme value curve $ y$  discussed above belongs to the $ C2 (E_m) $ space, but when the first derivative of $ y$  is continuous but the second derivative is discontinuous, the first derivative on the extreme value curve has a sharp point. Below we derive the minimum principle of functionals with this situation.
Assuming there are $ q-1 $ sharp points, the boundary conditions are divided into $ q+1 $ parts, namely $ S_0, S_1,..., S_q $ , and the volume is divided into $ q $ individual elements, namely $ D_1, D2,..., D2 $ . So equation \ref{eq:(bianfen.1.3)} can be changed to
\begin{equation}\label{eq:(bianfen.2.1)}
J = \sum_{r=1}^{q}\iiint_{D_r} L^{r}[x_i,y_j,y_{j,i},y_{j,ip},u_k] dv + \sum_{r=0}^{q}\iint_{S_r} \varphi^{r}[\bar {x}_i ,\bar {y}_j ,\bar {y}_ {j,i}] ds
\end{equation}
The table $ r $ above $ L$  and $ \varphi $ represents the $ r $ th individual product element. Since there is $ n_r=- n_ {r-1}$  at the boundary $ S_r $ , we can obtain
\begin{multline} \label{eq:(bianfen.2.2)}    
\delta J = 
\sum_{r=1}^{q} \iiint_{D_r} \{[L^r_{y_j} - \left(L^r_{y_{j,i}}\right)_{,i}+\left(L^r_{y_{j,ip}}\right)_{,ip}]\delta y_j \}dv +\iint_{S_0} \{L^0 [x_i,y_j,y_{j,i},y_{j,ip},u_k]
\\+ \varphi^0 _{\bar{x}_i} n_i -y_{j,l} L^0_{y_{j,i}} n_i n_l   + y_{j,l} \left(L^0_{y_{j,ip}}\right)_{,p} n_i n_l 
-y_{j,ip} L^0_{y_{j,ip}} n_i  n_p \}\delta \bar{n} da
\\+\sum_{r=1}^{q-1}\{\iint_{S_r} \{L^{r-1}[x_i,y_j,y_{j,i},y_{j,ip},u_k]+ \varphi^{r-1}_{\bar{x}_i} n_i 
-y_{j,l} L^{r-1}_{y_{j,i}} n_i n_l   + y_{j,l} \left(L^{r-1}_{y_{j,ip}}\right)_{,p} n_i n_l 
\\-y_{j,ip} L^{r-1}_{y_{j,ip}} n_i  n_p \}\delta \bar{n} da
\\-\iint_{S_r} \{L^{r}[x_i,y_j,y_{j,i},y_{j,ip},u_k]+ \varphi^{r}_{\bar{x}_i} n_i 
-y_{j,l} L^{r}_{y_{j,i}} n_i n_l   + y_{j,l} \left(L^{r}_{y_{j,ip}}\right)_{,p} n_i n_l 
\\-y_{j,ip} L^{r}_{y_{j,ip}} n_i  n_p \}\delta \bar{n} da\}
\\+\iint_{S_{q}} \{L^q [x_i,y_j,y_{j,i},y_{j,ip},u_k]+ \varphi^q_{\bar{x}_i} n_i 
-y_{j,l} L^q_{y_{j,i}} n_i n_l   + y_{j,l} \left(L^q_{y_{j,ip}}\right)_{,p} n_i n_l 
\\-y_{j,ip} L^q_{y_{j,ip}} n_i  n_p \}\delta \bar{n} da
\\+\iint_{S_0} \{[\left(L^0_{y_{j,i}}  \right) -\left(L^0_{y_{j,ip}}\right)_{,p}]n_i + \varphi^0_{\bar{y}_j} \}\delta \bar{y} _j da
\\+\sum_{r=1}^{q-1} \{\iint_{S_r} \{[\left(L^{r-1}_{y_{j,i}}  \right) -\left(L{r-1}_{y_{j,ip}}\right)_{,p}]n_i + \varphi^{r-1}_{\bar{y}_j} \}\delta \bar{y} _j da\
\\-\iint_{S_r} \{[\left(L^r_{y_{j,i}}  \right) -\left(L^r_{y_{j,ip}}\right)_{,p}]n_i + \varphi^r_{\bar{y}_j} \}\delta \bar{y} _j da\}
\\+\iint_{S_{q}} \{[\left(L^q_{y_{j,i}}  \right) -\left(L^q_{y_{j,ip}}\right)_{,p}]n_i + \varphi^q_{\bar{y}_j} \}\delta \bar{y} _j da
\\+\iint_{S_0} [\left(L^0_{y_{j,ip}}  \right)n_p +\varphi^0_{\bar{y}_{j,i}}] \delta \bar{y}_{j,i} da
+\sum_{r=1}^{q-1} \{\iint_{S_{r-1}} [\left(L_{y_{j,ip}}  \right)n_p +\varphi_{\bar{y}_{j,i}}] \delta \bar{y}_{j,i} da
\\-\iint_{S_r} [\left(L_{y_{j,ip}}  \right)n_p +\varphi_{\bar{y}_{j,i}}] \delta \bar{y}_{j,i} da\}
\\+\iint_{S_{q}} [\left(L^q_{y_{j,ip}}  \right)n_p +\varphi^q_{\bar{y}_{j,i}}] \delta \bar{y}_{j,i} da+\sum_{r=1}^{q} \iiint_{D_r} [L_{u_k} \delta u_k] dv=0
\end{multline} 
The Euler equation can be obtained from \ref{eq:(bianfen.2.2)}
\begin{equation} \label{eq:(bianfen.2.3)}    
[L^r_{y_j} - \left(L^r_{y_{j,i}}\right)_{,i}+\left(L^r_{y_{j,ip}}\right)_{, ip}]=0,j=1,2...n,r=1,2...q
\end{equation}
And boundary conditions
\begin{subequations}\label{eq:(bianfen.2.4)}  
\begin{multline}  
L^0 [x_i,y_j,y_{j,i},y_{j,ip},u_k]+ \varphi^0 _{\bar {x}_i } n_i 
-y_{j,l} L^0_{y_{j,i}} n_i n_l   + y_{j,l} \left(L^0_{y_{j,ip}}\right)_{,p} n_i n_l 
\\-y_{j,ip} L^0_{y_{j,ip}} n_i  n_p =0
\end{multline}
\begin{multline}  
\{L^{r-1}[x_i,y_j,y_{j,i},y_{j,ip},u_k] 
-y_{j,l} L^ {r-1}_ {y_{j,i}} n_i n_l   + y_{j,l} \left(L^ {r-1}_ {y_{j,ip}}\right)_{,p} n_i n_l 
-y_{j,ip} L^ {r-1}_ {y_{j,ip}} n_i  n_p \}
\\-\{L^{r}[x_i,y_j,y_{j,i},y_{j,ip},u_k]
-y_{j,l} L^ {r}_ {y_{j,i}} n_i n_l   + y_{j,l} \left(L^ {r}_ {y_{j,ip}}\right)_{,p} n_i n_l 
-y_{j,ip} L^ {r}_ {y_{j, ip}} n_i  n_p
\\+ \varphi^ {r-1}_ {\bar {x}_i } n_i
+ \varphi^ {r}_ {\bar {x}_i } n_i \}=0
\end{multline}
\begin{equation}  
L^q [x_i,y_j,y_{j,i},y_{j,ip},u_k]+ \varphi^q_{\bar {x}_i } n_i 
-y_{j,l} L^q_{y_{j,i}} n_i n_l   + y_{j,l} \left(L^q_{y_{j,ip}}\right)_{,p} n_i n_l 
-y_{j,ip} L^q_{y_{j,ip}} n_i  n_p 
\end{equation}
\end{subequations}
\begin{subequations}\label{eq:(bianfen.2.5)}  
\begin{equation} 
[(L^0_{y_{j,i}}) -(L^0_{y_{j,ip}})_{,p}]n_i+\varphi ^0_{\bar{y}_j} =0
\end{equation}
\begin{equation}     
[\left(L^ {r-1}_ {y_{j,i}}  \right) -\left(L^ {r-1}_ {y_{j,ip}}\right)_{,p}]n_i -[\left(L^r_{y_{j,i}}  \right) -\left(L^r_{y_{j,ip}}\right)_{, p}]n_i   + \varphi^ {r-1}_ {\bar {y}_j }+ \varphi^r_{\bar {y}_j}= 0
\end{equation}
\begin{equation}  
[\left(L^q_{y_{j,i}}  \right) -\left(L^q_{y_{j,ip}}\right)_{,p}]n_i + \varphi^q_{\bar {y}_j}= 0
\end{equation}
\end{subequations}
\begin{subequations}\label{eq:(bianfen.2.6)}  
\begin{equation}   
\left(L^0_{y_{j,ip}}  \right)n_p +\varphi^0_{\bar {y}_ {j,i}} =0
\end{equation}
\begin{equation} 
\left(L^ {r-1}_ {y_{j,ip}}  \right)n_p  -\left(L^ {r}_ {y_{j, ip}}  \right)n_p +\varphi^ {r-1}_ {\bar {y}_ {j,i}}+\varphi^ {r}_ {\bar {y}_ {j,i}}=0
\end{equation}
\begin{equation}    
\left(L^q_{y_{j,ip}}  \right)n_p +\varphi^q_{\bar {y}_ {j,i}} =0
\end{equation}
\end{subequations}
governing equation 
\begin{equation} \label{eq:(bianfen.2.7)}    
L_{u_k} =0
\end{equation}
The formulas  \ref{eq:(bianfen.2.5)} and \ref {eq:(bianfen.2.6)} are generalizations of the Weierstrass Edelmann corner condition.

Based on the assumption of continuity of the curve at the inflection point, the continuity equation at inflection point $ r $ is obtained
\begin{equation} \label{eq:(bianfen.2.8)}    
y^ {r-1}_ {j}(x_i)-y^ {r}_ {j}(x_i)=0
\end{equation}
\begin{equation} \label{eq:(bianfen.2.9)}    
y^ {r-1}_ {j,i}(x_i)-y^ {r}_ {j,i}(x_i)=0
\end{equation}
From this, the basic equations of multidimensional variation method with sharp points are obtained, which can be used to directly solve variation problems in multidimensional space, such as directly deriving the mechanical basic equations in three-dimensional space.

\subsection {Variational method when multiple elements intersect on the same boundary}\label{subsec: duotiji}
\numberwithin{equation}{subsection}

In this section, finite element terminology is used. In the common boundary conditions considered in $\ ref {subec: sharpPoint} $, there are only two elements on the same boundary. This does not apply when there are multiple elements at the same boundary. Below, we derive the situation when there are multiple elements at the same boundary.

Assuming there are $ q $units，namely$ D_1, D_2,..., D_q $ , and $ b $ boundaries，namely $ S_1,..., S_b $ . So equation \ref{eq:(bianfen.1.3)} can be changed to
\begin{equation}\label{eq:(bianfen.duotiji.1)}
J = \sum_{r=1}^{q}\iiint_{D_r} L^{r}[x_i,y_j,y_{j,i},y_{j,ip},u_k] dv + \sum_{r=1}^{b}\iint_{S_r} \varphi^{r}[\bar {x}_i ,\bar {y}_j ,\bar {y}_ {j,i}] ds
\end{equation}

Assuming there are $n ^ r $elements at the same node, which means there are $n ^ r $elements sharing this boundary at the same node, considering
\begin{equation} \label{eq:(bianfen.duotiji.2)}    
\delta \bar{n}^r da=\delta \bar{n}^r_m da_m= \delta \bar{n}^r_m n^r_m da
\end{equation}
we can obtain
\begin{multline} \label{eq:(bianfen.duotiji.3)}    
\delta J = 
\sum_{r=1}^{q} \iiint_{D_r} \{[L^r_{y_j} - \left(L^r_{y_{j,i}}\right)_{,i}+\left(L^r_{y_{j,ip}}\right)_{,ip}]\delta y_j \}dv +\sum_{r=1}^{q} \iiint_{D_r} [L^r_{u_k} \delta u_k]dv
\\ +\sum_{r=1}^{b}\iint_{S_r} \sum_{t=1}^{n^t}\{L^t[x_i,y_j,y_{j,i},y_{j,ip},u_k]+ \varphi^{r}_{\bar{x}_i} n^t_i +\bar{y}_{j,l}\varphi^t_{\bar {y}_j} n^t_l
-\bar{y}_{j,ip}\varphi^t_{\bar {y}_ {j,i}} n^t_p \}\delta \bar{n}^r da
\\+\sum_{r=1}^{b} \iint_{S_r} \sum_{t=1}^{n^t}\{\left[L^{t}_{y_{j,i}}-\left(L^{t}_{y_{j,ip}}\right)_{,p}\right]n^t_i + \varphi^{t}_{\bar{y}_j} \}\delta \bar{y} _j da
\\+\sum_{r=1}^{b} \iint_{S_{r}} \sum_{t=1}^{n^t} \{L_{y_{j,ip}}  n^t_p+\varphi^t_{\bar{y}_{j,i}}\}\delta \bar{y}_{j,i} da
\\= \sum_{r=1}^{q} \iiint_{D_r} \{[L^r_{y_j} - \left(L^r_{y_{j,i}}\right)_{,i}+\left(L^r_{y_{j,ip}}\right)_{,ip}]\delta y_j \}dv +\sum_{r=1}^{q} \iiint_{D_r} [L^r_{u_k} \delta u_k]dv
\\ +\sum_{r=1}^{b}\iint_{S_r} \sum_{t=1}^{n^t}\{ \{L^{t}[x_i,y_j,y_{j,i},y_{j,ip},u_k]+ \varphi^{t}_{\bar{x}_i} n^t_i +\bar{y}_{j,l}\varphi^t_{\bar {y}_j} n^t_l
-\bar{y}_{j,ip}\varphi^t_{\bar {y}_ {j,i}} n^t_p \}\delta \bar{n}^t_m n^t_m da
\\+\sum_{r=1}^{b} \iint_{S_r} \sum_{t=1}^{n^t}\{\{\left[L^{t}_{y_{j,i}}   -\left(L{t}_{y_{j,ip}}\right)_{,p}\right]n^t_i + \varphi^{t}_{\bar{y}_j} \}\delta \bar{y} _j da
\\+\sum_{r=1}^{b} \iint_{S_{r}} \sum_{t=1}^{n^t}\{[L_{y_{j,ip}}  n^t_p+\varphi^t_{\bar{y}_{j,i}}\} \delta \bar{y}_{j,i} da
=0
\end{multline}

The Euler equation can be obtained from \ref{eq:(bianfen.2.2)}
\begin{equation} \label{eq:(bianfen.duotiji.4)}    
L^r_{y_j} - \left(L^r_{y_{j,i}}\right)_{,i}+\left(L^r_{y_{j,ip}}\right)_{, ip}=0,j=1,2...n,r=1,2...q
\end{equation}

governing equation 
\begin{equation} \label{eq:(bianfen.duotiji.5)}    
L^r_{u_k} =0,k=1,2...m,r=1,2...q
\end{equation}
And boundary conditions
\begin{equation}  \label{eq:(bianfen.duotiji.6)}     
\sum_{t=1}^{n^r} \left[L^{t}_{y_{j,i}} -\left(L^{t}_{y_{j,ip}}\right)_{,p}] n^t_{i} + \varphi^{t}_{\bar{y}_j} \right ] =0,j=1,2...n,r=1,2...b
\end{equation}
\begin{equation} \label{eq:(bianfen.duotiji.7)}  
\sum_{t=1}^{n^r} \left(L^t_{y_{j,ip}} n^t_{p}+\varphi^t_{\bar{y}_{j,i}}\right)
=0,i=1,2...n,j=1,2...n,r=1,2...b
\end{equation}
\begin{multline} \label{eq:(bianfen.duotiji.8)}   
\sum_{t=1}^{n^r} \{L^{t}[x_i,y_j,y_{j,i},y_{j,ip},u_k]+ \varphi^{t}_{\bar{x}_i} n^t_i 
+y_{j,l} n^t_l \varphi^{t}_{\bar{y}_j}
+y_{j,ip} \varphi^t_{\bar{y}_{j,i}} n^t_p  \}n^t_{m}=0
\\,m=1,2...n,r=1,2...b
\end{multline}

\section {Principle of Minimum in Multidimensional Space} \label{sec: jxz}
In section \ref{sec: bianfen}, the control variable $ u_k $ can take the entire space, but sometimes the control variable is subject to various constraints, and the variation method derived above is no longer applicable. Other methods need to be sought to derive new formulas. The Pontryagin minimum principle can be used to obtain the minimum value when the control variable is constrained, but its integral variable is a one-dimensional time variable $ t $ , which is not applicable to multidimensional spatial variables. This article uses its idea to derive the minimum value of the multidimensional spatial functional when the control variable is constrained through incremental method combined with needle-like variation. This article is called the principle of multidimensional spatial minimum.
\numberwithin{equation}{subsection}
Assuming the control variable $ u_k $ is piecewise continuous and all other conditions are the same as $ \ref{sec: bianfen}$  , classical variation method cannot be used. The following analysis will be conducted using the incremental method, considering the following functional
\begin{equation}\label{eq:(jxz.1.1)}
J = \iiint_D L[x_i,y_j,y_{j,i},y_{j,ip},u_k] dv + \iint_S \varphi[\bar{x}_i,\bar{y}_j,\bar{y}_{j,i}] ds
\end{equation}
The formula is \ref{eq:(bianfen.1.3)}, repeated here for convenience.
Perform incremental analysis on \ref{eq:(jxz.1.1)}, and the incremental expression for functional $ j$  is
\begin{multline} \label{eq:(jxz.1.2)}    
\Delta J = \iiint_{D+\Delta D} L[x_i,y_j+\Delta y_j,y_{j,i}+\Delta y_{j,i},y_{j,ip}+\Delta y_{j, ip},u_k+\Delta u_k] dv 
\\-\iiint_D L[x_i,y_j,y_{j,i},y_{j,ip},u_k] dv 
\\+ \iint_S \varphi[\bar {x}_i +\Delta \bar {x}_i ,\bar {y}_j +\Delta \bar {y}_j ,\bar {y}_ {j,i}+\Delta \bar {y}_ {j,i}] ds
- \iint_S \varphi[\bar {x}_i ,\bar {y}_j ,\bar {y}_ {j,i}] ds
\\= \iiint_D \{L[x_i,y_j+\Delta y_j,y_{j,i}+\Delta y_{j,i},y_{j,ip}+\Delta y_{j,ip},u_k+\Delta u_k]
\\-L[x_i,y_j,y_{j,i},y_{j,ip},u_k] \}dv 
\\+\iiint_{\Delta D} L[x_i,y_j+\Delta y_j,y_{j,i}+\Delta y_{j,i},y_{j,ip}+\Delta y_{j, ip},u_k+\Delta u_k] dv 
\\+\iint_S \{\varphi[\bar {x}_i +\Delta \bar {x}_i ,\bar {y}_j +\Delta \bar {y}_j ,\bar {y}_ {j,i}+\Delta \bar {y}_ {j,i}] -\varphi[\bar {x}_i ,\bar {y}_j ,\bar {y}_ {j,i}]\} ds
\end{multline} 
For the first term of the above equation, Taylor expansion is performed according to the differentiability of L with respect to $ y_j, y_ {j, i}, y_ {j,ip}$, but for $ u_k $ it is only continuous but not differentiable.
\begin{multline} \label{eq:(jxz.1.3)}
\iiint_D \{L[x_i,y_j+\Delta y_j,y_{j,i}+\Delta y_{j,i},y_{j,ip}+\Delta y_{j,ip},u_k+\Delta u_k]-L[x_i,y_j,y_{j,i},y_{j,ip},u_k] \}dv 
\\=\iiint_D \{L[x_i,y_j+\Delta y_j,y_{j,i}+\Delta y_{j,i},y_{j,ip}+\Delta y_{j, ip},u_k+\Delta u_k]-
L[x_i,y_j,y_{j,i},y_{j, ip},u_k+\Delta u_k]+
\\L[x_i,y_j,y_{j,i},y_{j, ip},u_k+\Delta u_k]-
L[x_i,y_j,y_{j,i},y_{j,ip},u_k] \}dv 
\\=\iiint_D L_{y_j} \Delta y_j + L_{y_{j,i}} \Delta y_{j,i}+L_{y_{j,ip}} \Delta y_{j,ip} +o_1(\rho)
\\+[L(x_i,y_j,y_{j,i},y_{j,ip},u_k+\Delta u_k)-L(x_i,y_j,y_{j,i},y_{j,ip},u_k)]dv
\end{multline}
Among them, $ \rho_1=\| \delta y_j, \delta y_ {j, i}, \delta y_ {j, ip} \| $ , $ o (\rho) $ is a high-order infinitesimal of $ \delta y_j, \delta y_ {j, i}, \delta y_ {j, ip}$  .
According to the Lipschitz condition,
\begin{equation} \label{eq:(jxz.1.4)}
\iiint_D \{L_{y_j} \Delta y_j + L_{y_{j,i}} \Delta y_{j,i}+L{y_{j,ip}} \Delta y_{j,ip} \}dv \leq  a \rho_{1}    
\end{equation}
Since $ L$  is continuous with $ u_k $ , there must exist $ b (x_i) $ , such that
\begin{equation} \label{eq:(jxz.1.5)}
\iiint_D [L(x_i,y_j,y_{j,i},y_{j,ip},u_k+\Delta u_k)-L(x_i,y_j,y_{j,i},y_{j,ip},u_k)]dv \leq b(x_i) 
\end{equation}
Among them,
\begin{equation} \label{eq:(jxz.1.6)}    
b(x_i)=
\begin{cases}
0, & \text(\Delta u_k=0)\\
b, & \text(\Delta u_k\neq 0)
\end{cases}
\end{equation}
Therefore, \ref{eq:(jxz.1.3)} can be changed to
\begin{multline} \label{eq:(jxz.1.7)}
\iiint_D \{L[x_i,y_j+\Delta y_j,y_{j,i}+\Delta y_{j,i},y_{j,ip}+\Delta y_{j,ip},u_k+\Delta u_k]
\\-L[x_i,y_j,y_{j,i},y_{j,ip},u_k] \}dv 
\\=\iiint_D \{L_{y_j} \Delta y_j + L_{y_{j,i}} \Delta y_{j,i}+L_{y_{j,ip}} \Delta y_{j,ip}
+[L(x_i,y_j,y_{j,i},y_{j,ip},u_k+\Delta u_k)
\\-L(x_i,y_j,y_{j,i},y_{j,ip},u_k)]\} dv 
+o(\rho_1) \leq a \rho_1+b(x_i)
\end{multline} 
Applying the mean value theorem of integrals to the second term of $ \ref{eq:(jxz.1.2)}$  , we obtain
\begin{multline} \label{eq:(jxz.1.8)}
\iiint_{\Delta D} L[x_i,y_j+\Delta y_j,y_{j,i}+\Delta y_{j,i},y_{j,ip}+\Delta y_{j, ip},u_k+\Delta u_k] dv 
\\=\iint_S L[x_i+\theta_i \Delta_i x_i,y_j+\Delta y_j,y_{j,i}+\Delta y_{j,i},y_{j,ip}+\Delta y_{j,ip},u_k+\Delta u_k]\Delta \bar{n} da
\end{multline} 
Among them, $ 0<\theta_i<1 $ . According to the continuity of the $ L$  functional,
\begin{multline} \label{eq:(jxz.1.9)}    
L[x_i+\theta_i \Delta_i x_i,y_j+\Delta y_j,y_{j,i}+\Delta y_{j,i},y_{j,ip}+\Delta y_{j,ip},u_k+\Delta u_k]
\\=L[x_i,y_j,y_{j,i},y_{j, ip},u_k]+\varepsilon_1
\end{multline} 
Substituting \ref{eq:(jxz.1.12)} into \ref{eq:(jxz.1.11)} yields
\begin{multline} \label{eq:(jxz.1.10)}    
\iiint_{\Delta D} L[x_i,y_j+\Delta y_j,y_{j,i}+\Delta y_{j,i},y_{j,ip}+\Delta y_{j, ip},u_k+\Delta u_k] dv 
\\=\iint_{S} L[x_i,y_j,y_{j,i},y_{j,ip},u_k]\Delta \bar{n} da+\iint_{S} \varepsilon_1 \Delta \bar{n} da
\\=\iint_{S} L[x_i+ x_i,y_j+,y_{j,i},u_k]\Delta \bar{n} da+\varepsilon_1 \Delta D
\\=\iint_{S} L[x_i+ x_i,y_j+,y_{j,i},u_k]\Delta \bar{n} da+o(\rho_2)
\end{multline} 
By conducting Taylor expansion analysis on the third term of \ref{eq:(jxz.1.2)}, it can be concluded that
\begin{multline} \label{eq:(jxz.1.11)}
\iint_S \{\varphi[\bar {x}_i +\Delta \bar {x}_i ,\bar {y}_j +\Delta \bar {y}_j ,\bar {y}_ {j,i}+\Delta \bar {y}_ {j,i}] -\varphi[\bar {x}_i ,\bar {y}_j ,\bar {y}_ {j,i}]\} ds
=\\\iint_S [(\varphi_{\bar {x}_i } \Delta \bar {x}_i  + \varphi_{\bar {y}_j } \Delta \bar{y} _j+\varphi_{\bar {y}_ {j,i}} \Delta \bar {y}_ {j,i})+o(\rho_3)] ds
\end{multline} 
Where $ \rho_3=\| \delta \bar {x}_i ,\Delta \bar{y} _j,\Delta \bar {y}_ {j, i} \| $ is the high-order infinitesimal of $ \delta \bar {x}_i,\delta \bar {y} _j $ .
Substitute \ref{eq:(jxz.1.7)}, \ref{eq:(jxz.1.10)}, and \ref{eq:(jxz.1.11)} into \ref{eq:(jxz.1.2)} yields
\begin{multline} \label{eq:(jxz.1.12)}    
\Delta J = \iiint_D L_{y_j} \Delta y_j + L_{y_{j,i}} \Delta y_{j,i}+L_{y_{j,ip}} \Delta y_{j,ip}
+[L(x_i,y_j,y_{j,i},y_{j,i p},u_k+\Delta u_k)-
\\L(x_i,y_j,y_{j,i},y_{j,i p},u_k)]dv +o(\rho_1) +
\iint_{S} L[x_i+ x_i,y_j+,y_{j,i},u_k]\Delta \bar {n}  da+
\\ o(\rho _{2})+\iint_S [(\varphi_{\bar {x}_i } \Delta \bar {x}_i+ \varphi_{\bar {y}_j } \Delta \bar{y} _j+\varphi_{\bar {y}_ {j,i}} \Delta \bar {y}_ {j,i})+o(\rho _{3})] ds
\end{multline} 
Substituting \ref{eq:(bianfen.1.23)} into \ref{eq:(jxz.1.12)} yields
\begin{multline} \label{eq:(jxz.1.13)}    
\Delta J = \iiint_D \{[L_{y_j} - \left(L_{y_{j,i}}\right)_{,i}+\left(L_{y_{j,ip}}\right)_{, ip}]\Delta y_j \}dv
\\+\iint_S \{L[x_i,y_j,y_{j,i},y_{j,ip},u_k]+ \varphi_{\bar {x}_i } n_i 
-y_{j,l} L_{y_{j,i}} n_i n_l   + y_{j,l} \left(L_{y_{j,ip}}\right)_{,p} n_i n_l 
\\-y_{j,ip} L_{y_{j,ip}} n_i  n_p \}\Delta \bar{n} da
+\iint_S \{[\left(L_{y_{j,i}}  \right) -\left(L_{y_{j,ip}}\right)_{,p}]n_i 
\\+ \varphi_{\bar {y}_j } \}\Delta \bar{y} _j da
+\iint_S [\left(L_{y_{j,ip}}  \right)n_p +\varphi_{\bar {y}_ {j,i}}] \Delta \bar {y}_ {j,i} da
\\+\iiint_D [L(x_i,y_j,y_{j,i},y_{j,ip},u_k+\Delta u_k)-L(x_i,y_j,y_{j,i},y_{j,ip},u_k)] dv+o(\rho_1)+o(\rho_2)+o(\rho_3)
\end{multline} 
Since the function $ L$  is continuously differentiable with respect to $ y_i, y_ {j, i}, y_ {j, ip}$  and continuous with respect to $ u_k $ , when $ \rho_1, \rho_2, \rho_3 \to 0 $ , higher-order terms can be omitted, and \ref{eq:(jxz.1.13)} is changed from increment $ \delta J $ to first-order variation $ \delta J $ , that is
\begin{multline} \label{eq:(jxz.1.14)}    
\delta J = \iiint_D \{[L_{y_j} - \left(L_{y_{j,i}}\right)_{,i}+\left(L_{y_{j,ip}}\right)_{, ip}]\delta y_j \}dv 
\\+\iint_S \{L[x_i,y_j,y_{j,i},y_{j,ip},u_k]+ \varphi_{\bar {x}_i } n_i 
-y_{j,l} L_{y_{j,i}} n_i n_l   + y_{j,l} \left(L_{y_{j,ip}}\right)_{,p} n_i n_l 
\\-y_{j,ip} L_{y_{j,ip}} n_i  n_p \}\delta \bar{n} da
+\iint_S \{[\left(L_{y_{j,i}}  \right) -\left(L_{y_{j,ip}}\right)_{,p}]n_i + \varphi_{\bar {y}_j } \}\delta \bar{y} _j da
\\+\iint_S [\left(L_{y_{j,ip}}  \right)n_p +\varphi_{\bar {y}_ {j,i}}] \delta \bar {y}_ {j,i} da
\\+\iiint_D [L(x_i,y_j,y_{j,i},y_{j,ip},u_k+\delta u_k)-L(x_i,y_j,y_{j,i},y_{j,ip},u_k)] dv
\end{multline} 
According to the condition of taking the extremum of the functional $ \delta J=0 $ , and based on the variation lemma, since $ \delta y_j $ is an arbitrary value, the Euler equation can be obtained
\begin{equation} \label{eq:(jxz.1.15)}    
L_{y_j} - \left(L_{y_{j,i}}\right)_{,i}+\left(L_{y_{j,ip}}\right)_{,ip} =0
\end{equation}
Next, let's discuss the boundary conditions. Since $ \delta \bar {n}$  is an arbitrary value, we can obtain
\begin{equation} \label{eq:(jxz.1.16)}    
L[x_i,y_j,y_{j,i},y_{j,ip},u_k]+ \varphi_{\bar {x}_i } n_i 
-y_{j,l} L_{y_{j,i}} n_i n_l   + y_{j,l} \left(L_{y_{j,ip}}\right)_{,p} n_i n_l 
-y_{j,ip} L_{y_{j,ip}} n_i  n_p =0
\end{equation}
Since $ \delta \bar {y} _j $ is an arbitrary value, we can obtain
\begin{equation} \label{eq:(jxz.1.17)}    
[\left(L_{y_{j,i}}  \right) -\left(L_{y_{j,ip}}\right)_{,p}]n_i + \varphi_{\bar {y}_j}=0
\end{equation}
Due to $ \delta \bar {y}_ {j, i}$  is an arbitrary value, which can be obtained
\begin{equation} \label{eq:(jxz.1.18)}    
\left(L_{y_{j,ip}}  \right)n_p +\varphi_{\bar {y}_ {j,i}} =0
\end{equation}
Let's discuss the last term in \ref{eq:(jxz.1.14)}, assuming $ \delta u_k $ is a needle-like variation, as shown below:
\begin{equation} \label{eq:(jxz.1.19)}    
u_k(x_j)+\delta u_k(x_j)=
\begin{cases}
u_k(x_j), & \text(x_j \notin [\sigma_j,\sigma_j+\varepsilon l])\\
\bar {u}_k (x_j), & \text(x_j \in [\sigma_j,\sigma_j+\varepsilon l])
\end{cases}
\end{equation}
Among them, $ \sigma_j$  is any continuous point on the optimal control, $ l>0 $ is a certain number, and $ \varepsilon>0 $ is a sufficiently small number. When satisfying the Euler equation and boundary conditions mentioned above, $ j$  takes a minimum value. Let the minimum point be $ x ^ * $ 
\begin{multline} \label{eq:(jxz.1.20)}    
\iiint_D [L(x_i,y_j,y_{j,i},y_{j,ip},u_k+\delta u_k)-L(x^*_i,y^*_j,y^*_{j,i},y_{j,ip},u^*_k)] dv
\\=\iiint_{D_{\varepsilon l}}[L(x_i,y_j,y_{j,i},y_{j,ip},u_k+\delta u_k)-L(x^*_i,y^*_j,y^*_{j,i},y^*_{j,ip},u^*_k)] dv
\\={D_{\varepsilon l}} [L(x_i,y_j,y_{j,i},y_{j,ip},u_k(\sigma_j+\theta_j \varepsilon l)+\delta u_k(\sigma_j+\theta_j \varepsilon l))
\\- L(x^*_i,y^*_j,y^*_{j,i},y^*_{j,ip},u^*_k(\sigma_j+\theta_j \varepsilon l))]
\\={D_{\varepsilon l}} [L(y_j,y_{j,i},y_{j,ip},\bar {u}_k (\sigma_j)- L(y^*_j,y^*_{j,i},y^*_{j,ip},u^*_k(\sigma_j))]+o({D_{\varepsilon l}}) \geq 0
\end{multline}
Since $ D_ {\varepsilon l}>0 $ , \ref{eq:(jxz.1.29)} is divided by $ D_ {\varepsilon l}$  at both ends simultaneously
\begin{equation} \label{eq:(jxz.1.21)}    
L(y^*_j,y^*_{j,i},y^*_{j,ip},\bar {u}_k (\sigma_j)) \geq L(y^*_j,y^*_{j,i},y^*_{j,ip},u^*_k(\sigma_j))
\end{equation}
Or write it as
\begin{equation} \label{eq:(jxz.1.22)}    
L(y^*_j,y^*_{j,i},y^*_{j,ip},u^*_k(\sigma_j)) \leq L(y^*_j,y^*_{j,i},y^*_{j,ip},\bar {u}_k (\sigma_j))
\end{equation}
According to the needle-like variation, if equation \ref{eq:(jxz.1.22)} holds for any continuous point in space $ \bm \Omega $ , then \ref{eq:(jxz.1.22)} can be expressed as
\begin{equation} \label{eq:(jxz.1.23)}    
L(y^*_j,y^*_{j,i},y^*_{j,ip},u^*_k(\sigma_j)) = min\underset{\bar u \in \Omega}{}L(y^*_j,y^*_{j,i},y^*_{j,ip},\bar {u}_k )
\end{equation}
In the formula, $ \sigma_j $ takes over all continuous points in the control domain $ \bm \Omega $ . Since $ u $ is assumed to be piecewise continuous, a finite number of discontinuous points do not affect the integration. Therefore, $ \sigma_j $ takes over the entire control domain $ \bm \Omega $ , that is
\begin{equation} \label{eq:(jxz.1.24)}    
L[y^*_j(x_i),y^*_{j,i}(x_i),y^*_{j,ip}(x_i),u^*_k(x_i)] = min\underset{\bar u \in \Omega}{}L[y^*_j(x_i),y^*_{j,i}(x_i),y^*_{j,ip}(x_i),\bar {u}_k (x_i)]
\end{equation}
In summary, when taking the minimum value of $j$  , it satisfies \ref{eq:(jxz.1.25)}, \ref{eq:(jxz.1.26)}, \ref{eq:(jxz.1.27)}, \ref{eq:(jxz.1.28)}, and \ref{eq:(jxz.1.29)}, that is
\begin{equation} \label{eq:(jxz.1.25)}    
L_{y_j} - \left(L_{y_{j,i}}\right)_{,i}+\left(L_{y_{j,ip}}\right)_{,ip} =0,j=1,2...n
\end{equation}
\begin{multline} \label{eq:(jxz.1.26)}    
L[x_i,y_j,y_{j,i},y_{j,ip},u_k]+ \varphi_{\bar {x}_i } n_i 
-y_{j,l} L_{y_{j,i}} n_i n_l   + y_{j,l} \left(L_{y_{j,ip}}\right)_{,p} n_i n_l 
\\-y_{j,ip} L_{y_{j,ip}} n_i  n_p =0
\end{multline}
\begin{equation} \label{eq:(jxz.1.27)}    
[\left(L_{y_{j,i}}  \right) -\left(L_{y_{j,ip}}\right)_{,p}]n_i + \varphi_{\bar {y}_j}=0
\end{equation}
\begin{equation} \label{eq:(jxz.1.28)}    
\left(L_{y_{j,ip}}  \right)n_p +\varphi_{\bar {y}_ {j,i}} =0
\end{equation}
\begin{equation} \label{eq:(jxz.1.29)}    
L[y^*_j(x_i),y^*_{j,i}(x_i),y^*_{j,ip}(x_i),u^*_k(x_i)] = min\underset{\bar u \in \Omega}{}L[y^*_j(x_i),y^*_{j,i}(x_i),y^*_{j,ip}(x_i),\bar {u}_k (x_i)]
\end{equation}
The above formula $\ref{eq:(jxz.1.25)} \text{-} \ref{eq:(jxz.1.29)}$ is the basic equation of the minimum principle in multidimensional space, which can be used for optimal control in multidimensional space when the control domain is a closed set. The only difference from when the control domain is an open set is that the formula \ref{eq:(jxz.1.29)} is inconsistent.

\section {principle minimum virtual work in Mechanics} \label{sec: zxxgyl}
In the first two sections, the multidimensional variation method and the principle of multidimensional minimum for movable boundaries were derived. Next, they were applied to mechanics to obtain the principle of minimum virtual work applicable to movable boundaries in mechanics.
\subsection {principle minimum virtual work for spatial problems}
\numberwithin{equation}{subsection}
According to the relationship between strain and displacement
\begin{equation}\label{eq:(zxxgyl.1.1)}
\varepsilon_{ji}=\frac{1}{2}(y_{j,i}+y_{i,j})
\end{equation}
Then the strain energy function $ U (\varepsilon_ {ji}) $ of the spatial structure can be regarded as $ U (y_ {j, i}) $ . Based on the relationship between stress-strain and strain energy function
\begin{equation}\label{eq:(zxxgyl.1.2)}
\sigma_{ji}=U_{\varepsilon_{ji}} =[U_{y_{j,i}}(\varepsilon_{ji})]_{y_{j,i}}=U_{y_{j,i}}
\end{equation}
In order to give the name of the principle of minimum virtual work derived in this article a clearer physical meaning, the potential energy is given a negative sign and referred to as
The virtual work of the structure, the functional expression of the virtual work is
\begin{equation}\label{eq:(zxxgyl.1.3)}
J = \iiint_D [b_j y_j+u_k y_k-U(y_{j,i})] dv+\iint_{S_{p}} \bar {p}_j  \bar {y}_j ds
\end{equation}
be
\begin{equation}\label{eq:(zxxgyl.1.4)}
L = b_j y_j+u_k y_k-U(y_{j,i}), \varphi=\bar {p}_j  \bar {y}_j
\end{equation}
According to \ref{eq:(bianfen.1.22)}, the Euler equation, also known as the mechanical equilibrium equation, can be obtained
\begin{equation} \label{eq:(zxxgyl.1.5)}    
L_{y_j} - \left(L_{y_{j,i}}\right)_{,i}+\left(L_{y_{j,ip}}\right)_{,ip} =
b_j+u_j-(L_{y_{j,i}})_{,i}=
b_j+u_j+\sigma_{ji,i}
=0,j=1,2...3
\end{equation}
This equation is the equilibrium differential equation.

According to \ref{eq:(bianfen.1.24)}, the boundary conditions that can be obtained are:
\begin{equation} \label{eq:(zxxgyl.1.6)}    
[\left(L_{y_{j,i}}  \right) -\left(L_{y_{j,ip}}\right)_{,p}]n_i + \varphi_{\bar {y}_j } 
=L_{y_{j,i}}n_i+\bar {p}_j  
=\sigma_ {ji}n_i -\bar {p}_j =0
\end{equation}

According to \ref{eq:(bianfen.1.23)}, the active boundary equation can be obtained
\begin{multline} \label{eq:(zxxgyl.1.7)}    
L+ \varphi_{\bar {x}_i } n_i 
-y_{j,l} L_{y_{j,i}} n_i n_l   + y_{j,l} \left(L_{y_{j,ip}}\right)_{,p} n_i n_l 
-y_{j,ip} L_{y_{j,ip}} n_i  n_p 
\\=b_j y_j+u_k y_k-U+\varepsilon_{ji} \sigma_{ji} n_i n_l 
=0
\end{multline}
For linear elastic bodies, due to the independence of energy and direction, the strain energy at the boundary is
\begin{equation} \label{eq:(zxxgyl.1.8)}    
U=\frac{1}{2} \varepsilon_{ji} \sigma_{ji}=\frac{1}{2} \varepsilon_{jl} \sigma_{ji}n_l n_i
\end{equation}
substituting \ref{eq:(zxxgyl.1.8)} into \ref{eq:(zxxgyl.1.7)}yields，
\begin{equation} \label{eq:(zxxgyl.1.9)}    
b_j y_j+u_k y_k+\frac{1}{2} \varepsilon_{ji} \sigma_{ji} n_i n_l =0
\end{equation}
Substituting\ref{eq:(zxxgyl.1.6)}again yields
\begin{equation} \label{eq:(zxxgyl.1.10)}    
b_j y_j+u_k y_k+\frac{1}{2} \bar {p}_j \bar{\varepsilon_j} =0
\end{equation}
Where $b_j y_j+u_k y_k $is the virtual work density done by the volumetric force, $1/2 \bar {p}_j \bar{\varepsilon_j} $ is the surface strain energy density. From this equation, it can be seen that for a movable boundary mechanics system in the integral domain, the sum of the virtual work density of the volume force and the strain energy density of the surface force must be zero on the movable boundary. This equation is temporarily referred to as the virtual work boundary condition.

When $ u_k $ is unrestricted
\begin{equation} \label{eq:(zxxgyl.1.11)}    
L_{u_k} =y_k=0
\end{equation}
This equation is called the control boundary condition.

For a given possible displacement, the displacement boundary is naturally satisfied, that is
\begin{equation} \label{eq:(zxxgyl.1.12)}    
y_j=\bar{y}_j
\end{equation}

According to the above deduction process, it can be seen that a mechanical system with movable boundaries not only satisfies the geometric equation \ref{eq:(zxxgyl.1.1)}, the equilibrium equation (Euler equation) \ref{eq:(zxxgyl.1.5)}, the constitutive equation \ref{eq:(zxxgyl.1.2)}, and the boundary condition \ref{eq:(zxxgyl.1.12)} when the boundary is fixed, but also needs to satisfy the virtual work boundary \ref{eq:(zxxgyl.1.10)} and the control boundary \ref{eq:(zxxgyl.1.11)}. The above equations are called the basic equations of a movable boundary mechanical system. This also proves that the basic equation of the movable boundary mechanics system is an extension of the basic equation of the conventional mechanics system.

Further analysis will be conducted on the control boundary conditions below，the formula \ref{eq:(zxxgyl.1.1)} is called the strong form of the governing equation. From the above equation, it can be seen that when the control load is not restricted, the displacement $ y_k $ is always zero. This situation is applicable to the prestressed tendon configuration of prestressed concrete continuous beams. For example, when a simply supported beam is subjected to a load, if the configured prestressed tendon causes the displacement of the structure to be zero, the bending moment is zero. At this time, the tendon overcomes all the bending moments and is the optimal tendon.
According to \ref{eq:(jxz.1.29)}
\begin{multline} \label{eq:(zxxgyl.1.13)}    
L[y^*_j(x_i),y^*_{j,i}(x_i),(x_i),u^*_k(x_i)] = min\underset{\bar u \in \Omega}{}L[y^*_j(x_i),y^*_{j,i}(x_i),(x_i),\bar {u}_k (x_i)]
\\=min\underset{\bar u \in \Omega}{}[b_j y_j+u_k y_k-U(\varepsilon_{ji})]
\end{multline}
According to the conservation of mechanical energy, the actual work done by external forces is equal to the strain energy of the system, and the actual work of all loads is equal to $ \alpha $ times the imaginary work, with $ 0<\alpha<1 $ . For a linear elastic system, $ \alpha=1/2 $ , that is
\begin{equation} \label{eq:(zxxgyl.1.14)}  
U(\varepsilon_{ji})=\frac{1}{2} (b_j y_j+u_k y_k)
\end{equation}
So change \ref{eq:(zxxgyl.1.9)} to
\begin{equation} \label{eq:(zxxgyl.1.15)}    
L[y^*_j(x_i),y^*_{j,i}(x_i),(x_i),u^*_k(x_i)] =min\underset{\bar u \in \Omega}{}\frac{1}{2}(b_j y_j+u_k y_k)
\end{equation}
This equation is also known as the strong form of the governing equation.
For ease of calculation, the above equation is simplified to the expression of only the control load $ u_k $ . According to the actual stress process of the structure, when a fixed load is applied, positive work is generated on the system, storing equivalent internal energy. When a variable load is applied, negative work needs to be applied to the system to reduce the total work. When negative work is applied to a certain extent, the control load will do positive work, increasing the internal energy of the system. When the fixed load continues to increase, the internal energy of the system also continues to increase until it exceeds the original internal energy of the system. At this time, the control effect has the opposite effect. The specific situation is detailed in the following examples.
Based on the above analysis, in order to make $ \ref{eq:(zxxgyl.1.11)}$  hold, the virtual work density $ w_k $ of the control load can be minimized, that is
\begin{equation} \label{eq:(zxxgyl.1.16)}    
w_k=min(\int_{0}^{u_k} y_k  d u_k)
\end{equation}
Due to the fact that equations \ref{eq:(zxxgyl.1.8)}, \ref{eq:(zxxgyl.1.11)}, and \ref{eq:(zxxgyl.1.16)} hold true everywhere in the domain, they are called strong forms of control conditions. It is worth noting that $ \Omega $ here is the region where the control load acts.

Usually, the control load $ u_k $ is subject to various limitations. The following further discusses the control load $ u_k $ : assuming it is a concentrated load acting on $ x_i $ , it can be expressed as a dirac function
\begin{equation} \label{eq:(zxxgyl.1.17)}    
u_k = \iiint_D u_ky_k h(x-x_i) dv
\end{equation}
There are variations of it
\begin{multline} \label{eq:(zxxgyl.1.18)}    
\delta\iiint_D u_ky_kh(x-x_i)dv = \iiint_D\delta( u_k y_k h(x-x_i) )dv
\\=\iiint_D  [y_k \delta u_k h(x-x_i) +u_k h(x-x_i) \delta y_k] dv
\\=y_k \delta u_k+u_k \delta y_k
\end{multline}
The $ L$  function in equation \ref{eq:(zxxgyl.1.3)} can be expressed as
\begin{equation}\label{eq:(zxxgyl.1.19)}
L = L(y_{j,i})-b_j y_j-u_kh(x-x_i) y_k
\end{equation}

After expressing the concentrated load in the above form, it can be substituted into the integral domain for solution, which is sometimes more convenient.
When $ u_k $ is a concentrated load and is not restricted, it can be seen from \ref{eq:(zxxgyl.1.11)} that the displacement of the concentrated load location $ y_k=0$ 
When $ u_k $ is a distributed load, it usually cannot take any value, that is, it cannot satisfy the functional variation lemma. For example, if the distributed load takes a certain value $ C $ in the integration domain, \ref{eq:(zxxgyl.1.8)} can no longer be used, but \ref{eq:(zxxgyl.1.16)} needs to be used, that is, the entire integration domain should take the minimum value of $ L$  .
For specific mechanical structures, it is difficult to require \ref{eq:(zxxgyl.1.11)} or \ref{eq:(zxxgyl.1.16)} to hold everywhere in the domain, and its weak form, namely the integral form, can be used
\begin{equation}\label{eq:(zxxgyl.1.20)}
W_{min} = min[\iiint_\Omega (\int_{0}^{u_k} y_k  d u_k)dv]
\end{equation}
When the control load can cover the entire space, make $ W_ {min}=0 $ , that is
\begin{equation}\label{eq:(zxxgyl.1.21)}
W_{min} =\iiint_\Omega(\int_{0}^{u_k} y_k  d u_k)dv= 0
\end{equation}
When the control load is constant within the control domain, the control equation \ref{eq:(zxxgyl.1.20)} becomes the minimum integral of displacement, that is, it satisfies
\begin{equation}\label{eq:(zxxgyl.1.22)}
W_{min} = min[\int_{0}^{u_k} (\iiint_\Omega y_k  dv)d u_k]
\end{equation}
The validity of the above equation is obvious, because when the control load is constant, it can be moved beyond the integral sign.
When the control load is a constant within the control domain and this constant is taken over the entire space, $ Y_{min}=0 $ , that is
\begin{equation}\label{eq:(zxxgyl.1.23)}
Y_{min} =\iiint_\Omega y_kdv= 0
\end{equation}

In summary, the principle of minimum virtual work is obtained: 
For a movable boundary mechanical system, the exact solution of the mechanical system minimizes the total virtual work of the system among all possible displacements. In addition to satisfying the equilibrium equations, constitutive equations，geometric relationships, and force and displacement boundary conditions of conventional mechanical systems, the system also needs to meet control conditions and virtual work boundary conditions.

That is to say, in order to minimize the virtual work of the system, the sum of the virtual work density of the volumetric force and the strain energy density of the surface force must be zero, and the displacement at the control load must also be zero.

The above description is based on the assumption that the control load is not subject to any limits, When the control load is limited,  When the system's virtual work is minimized, it no longer needs to satisfy \ref {eq:(zxxgyl.1.11)}, but needs to satisfy one of the equations \ref {eq:(zxxgyl.1.15)}, \ref{eq:(zxxgyl.1.16)}, \ref{eq:(zxxgyl.1.20)} \text{-} \ref{eq:(zxxgyl.1.23)}.

According to \ref{eq:(zxxgyl.1.5)} \text{-} \ref{eq:(zxxgyl.1.12)}, it can be seen that when When the control load $u_k=0 $ and the integration domain is fixed , the principle of minimum virtual work degenerates into the minimum potential energy principle, indicating that the minimum potential energy principle is a special case of the minimum imaginary work principle.

\subsection {principle minimum virtual work of one-dimensional space problems}
\numberwithin{equation}{subsection}
Firstly, derive the minimum principle of one-dimensional space based on the content of the third section, for the form such as
\begin{equation}\label{eq:(zxxgyl.2.1)}
J = \int_{x_{0}}^{x_{1}} L(x,y,y_{,x},y_{,xx},u)dx+\varphi(\bar {x}_ {0},\bar {y}_ {0},\bar {u}_ {0},\bar {x}_ {1},\bar {y}_ {1},\bar {u}_ {1},\bar {y}_ {0,x},\bar {y}_ {1,x})
\end{equation}
According to \ref{eq:(bianfen.1.22)}, substituting \ref{eq:(zxxgyl.2.1)} yields the Euler equation for $ y$  
\begin{equation} \label{eq:(zxxgyl.2.2)}    
L_{y_j} - \left(L_{y_{j,i}}\right)_{,i}+\left(L_{y_{j,ip}}\right)_{,ip} =L_{y} - 
\left(L_{y_{,x}}\right)_{,x}+\left(L_{y_{,xx}}\right)_{,xx}=0
\end{equation}
Similar to \ref{eq:(zxxgyl.2.2)}, obtain the Euler equation about $ u $ 
\begin{equation} \label{eq:(zxxgyl.2.3)}    
L_{u}=0
\end{equation}
When the endpoint $ x_ {1}$  is mutable, according to \ref{eq:(bianfen.1.23)}, and dividing $ y_j $ into $ y$  and $ u $ , we can obtain the virtual work boundary conditions
\begin{multline} \label{eq:(zxxgyl.2.4)}    
L+ \varphi_{\bar {x}_i } n_i 
-y_{j,l} L_{y_{j,i}} n_i n_l   + y_{j,l} \left(L_{y_{j,ip}}\right)_{,p} n_i n_l 
-y_{j,ip} L_{y_{j,ip}} n_i  n_p 
\\=[L+ \varphi_{x_{1}}
-y_{,x} L_{y_{,x}}   + y_{,x} \left(L_{y_{,xx}}\right)_{,x} 
-y_{,xx} L_{y_{,xx}} 
]|_{x=x_1}  
0
\end{multline}
This equation is the requirement for the activity boundary $ x_ {1}$  .
According to \ref{eq:(bianfen.1.24)}, the shear boundary conditions can be obtained
\begin{equation} \label{eq:(zxxgyl.2.5)}    
[L_{y_{j,i}}   -\left(L_{y_{j,ip}}\right)_{,p}]n_i + \varphi_{\bar {y}_j } 
=[L_{y_{,x}}  -\left(L_{y_{,xx}}\right)_{,x} + \varphi_{\bar {y}_ {1} }]|_{x=x_1}  
0
\end{equation}
According to \ref{eq:(bianfen.1.25)}, the bending moment boundary conditions can be obtained
\begin{equation} \label{eq:(zxxgyl.2.6)}    
\left(L_{y_{j,ip}}  \right)n_p +\varphi_{\bar {y}_ {j,i}} 
=[L_{y_{,xx}}  +\varphi_{y_{1,x}}]|_{x=x_1} =  
0
\end{equation}
When $ u $ is unconstrained, the boundary conditions for $ u $ 
\begin{equation} \label{eq:(zxxgyl.2.7)}    
\varphi_{\bar {u}_ {1} }|_{x=x_1} = 
0
\end{equation}
Similarly, when the endpoint $ x_0 $ is mutable, there is
\begin{equation} \label{eq:(zxxgyl.2.8)}    
[L+ \varphi_{x_{1}}
-y_{,x} L_{y_{,x}}   + y_{,x} \left(L_{y_{,xx}}\right)_{,x} 
-y_{,xx} L_{y_{,xx}} 
]|_{x=x_0}  =
0
\end{equation}
\begin{equation} \label{eq:(zxxgyl.2.9)}    
[\left(L_{y_{,x}}  \right) -\left(L_{y_{,xx}}\right)_{,x} + \varphi_{y_{0} }]|_{x=x_0}  =
0
\end{equation}
\begin{equation} \label{eq:(zxxgyl.2.10)}    
[L_{y_{,xx}}  +\varphi_{y_{0,x}}]|_{x=x_0}  = 
0
\end{equation}
When $ u $ is unconstrained, the boundary conditions for $ u $ 
\begin{equation} \label{eq:(zxxgyl.2.11)}    
\varphi_{\bar {u}_ {0} }|_{x=x_0}=  
0
\end{equation}

Based on the minimum principle of one-dimensional space mentioned above, the principle minimum virtual work of beam element is derived in detail, which refers to the virtual work done by the beam under external force
\begin{multline} \label{eq:(zxxgyl.2.12)}    
W= \int_{x_{0}}^{x_{1}} L(x,y,y_{,x},y_{,xx},u)dx+\varphi(\bar {x}_ {0},\bar {y}_ {0},\bar {u}_ {0},\bar {x}_ {1},\bar {y}_ {1},\bar {u}_ {1},\bar {y}_ {0,x},\bar {y}_ {1,x})
\\
=\int_{x_{0}}^{x_{1}} qydx+\bar {F}_0 \bar {y}_ {0}+\bar {M}_0 \bar {y}_ {0,x}+\bar {F}_1 \bar {y}_ {1}+\bar {M}_1 \bar {y}_ {1,x}
\end{multline}
Internal energy is
\begin{equation} \label{eq:(zxxgyl.2.13)}    
U=\int_{x_{0}}^{x_{1}} \frac{1} {2}EIy_ {,xx}^2dx
\end{equation}
Define the total virtual work of the system as the external force work minus the internal energy
\begin{multline} \label{eq:(zxxgyl.2.14)}    
J= W-U=\int_{x_{0}}^{x_{1}} (qy-\frac{1} {2}EIy_ {,xx}^2)dx+\bar {F}_0 \bar {y}_ {0}+\bar {M}_0 \bar {y}_ {0,x}+\bar {F}_1 \bar {y}_ {1}+\bar {M}_1 \bar {y}_ {1,x}
\end{multline}
The essence of this equation is the negative value of potential energy.
From this, it can be inferred that,
\begin{equation} \label{eq:(zxxgyl.2.15)}    
L =(qy-\frac{1} {2}EIy_ {,xx}^2)
\end{equation}
\begin{equation} \label{eq:(zxxgyl.2.16)}    
\varphi=\bar {F}_0 \bar {y}_ {0}+\bar {M}_0 \bar {y}_ {0,x}+\bar {F}_1 \bar {y}_ {1}+\bar {M}_1 \bar {y}_ {1,x}
\end{equation}
According to the formula \ref{eq:(zxxgyl.2.2)} \text{-} \ref{eq:(zxxgyl.2.11)}, the differential equation of the beam is:
\begin{equation} \label{eq:(zxxgyl.2.17)}    
L_{y} - 
\left(L_{y_{,x}}\right)_{,x}+\left(L_{y_{,xx}}\right)_{,xx}
=q-(EIy_{,xx})_{,xx}
0
\end{equation}
For endpoint $ x_ {1}$  , the moving boundary condition is
\begin{multline} \label{eq:(zxxgyl.2.18)}    
[L+ \varphi_{x_{1}}
-y_{,x} L_{y_{,x}}   + y_{,x} \left(L_{y_{,xx}}\right)_{,x} 
-y_{,xx} L_{y_{,xx}} 
]|_{x=x_1}
\\=[(qy-\frac{1} {2}EIy_ {,xx}^2)+y_{,x} \left(EIy_{,xx}\right)_{,x} 
-y_{,xx} EIy_{,xx}]|_{x=x_1}  
0
\end{multline}
Shear boundary conditions
\begin{equation} \label{eq:(zxxgyl.2.19)}    
[L_{y_{,x}}  -(L_{y_{,xx}})_{,x} + \varphi_{\bar {y}_ {1} }]|_{x=x_1}
=[-(EIy_{,xx})_{,x}+ \bar {F}_1 ]|_{x=x_1}
0
\end{equation}
Bending moment boundary condition
\begin{equation} \label{eq:(zxxgyl.2.20)}    
[L_{y_{,xx}}  +\varphi_{\bar {y}_ {1,x}}]|_{x=x_1}  =[EIy_{,xx}  +\bar {M}_1 ]|_{x=x_1}   
0
\end{equation}
When controlling the load at endpoint $ X_1$ , that is, $ \bar {F}_1 , \bar {M}_1 $ When treated as a control variable, the control conditions
\begin{subequations}\label{eq:(zxxgyl.2.21)} 
\begin{equation}    
\bar {y}_ {1}=0
\end{equation}
\begin{equation}    
\bar {y}_ {1,x}=0
\end{equation}    
\end{subequations}
Similarly, when the endpoint $ x0$ is variable, the moving boundary condition is
\begin{multline} \label{eq:(zxxgyl.2.22)}    
[L+ \varphi_{x_{1}}
-y_{,x} L_{y_{,x}}   + y_{,x} \left(L_{y_{,xx}}\right)_{,x} 
-y_{,xx} L_{y_{,xx}} 
]|_{x=x_1}
\\=[(qy-\frac{1} {2}EIy_ {,xx}^2)+y_{,x} \left(EIy_{,xx}\right)_{,x} 
-y_{,xx} EIy_{,xx}]|_{x=x_0}  
0
\end{multline}
Shear boundary conditions
\begin{equation} \label{eq:(zxxgyl.2.23)}    
[L_{y_{,x}}  -(L_{y_{,xx}})_{,x} + \varphi_{\bar {y}_ {0} }]|_{x=x_0}
=[-(EIy_{,xx})_{,x}+ \bar {F}_0 ]|_{x=x_0}
0
\end{equation}
Bending moment boundary condition
\begin{equation} \label{eq:(zxxgyl.2.24)}    
[L_{y_{,xx}}  +\varphi_{\bar {y}_ {0,x}}]|_{x=x_1}  =[EIy_{,xx}  +\bar {M}_0 ]|_{x=x_0}   
0
\end{equation}
When controlling the load at endpoint $ X_0 $ , that is, $ \bar {F}_0 , \bar {M}_0 $ When treated as a control variable, the control conditions
\begin{subequations}\label{eq:(zxxgyl.2.25)} 
\begin{equation}    
\bar {y}_ {0}=0
\end{equation}
\begin{equation}    
\bar {y}_ {0,x}=0
\end{equation}    
\end{subequations}
According to \ref{eq:(zxxgyl.2.21)} and \ref{eq:(zxxgyl.2.25)}, it can be seen that the displacement at the control load $ u $ is zero, and the work done by the external load $ u $ is zero. At this point, the maximum potential energy is taken. When a beam is subjected to a fixed load, the fixed load does work on it, storing internal energy. When a control load is applied, the control load will also do work on it. In order to reduce the total work, the control load must be negative. When the total work is reduced and treated as positive, the total work will increase. Therefore, when the displacement is zero, the total work is minimized and the internal energy reaches its minimum value.

\subsection {principle minimum virtual work for one-dimensional space problems with sharp points}
The extreme value curve Y discussed above belongs to the C2 (E) space, but when the first derivative of Y is continuous but the second derivative is discontinuous, the first derivative on the extreme value curve has a sharp point. The minimum principle of functional in this case is derived based on sections 3.2 and 3.3. Assuming there are q-1 sharp points, the boundary conditions are divided into q+1 parts, namely $S_0, S_1,..., S_{q+1}$, and the volume is divided into q individual elements, namely $D_1, D_2,..., D_{q+1}$. The symbols in this section do not use the Einstein summation convention.
\begin{multline}\label{eq:(zxxgyl.3.1)}
J = \sum_{r=1}^{q}\int_{x_{r-1}}^{x_{r}} L^{r}(x_r,y_r,y_{r,x},y_{r,xx},u_r)dx+\varphi(\bar {x}_ {r},\bar {y}_ {r},\bar {y}_ {r,x},\bar {u}_ {r})
\\=\sum_{r=1}^{q}\int_{x_{r-1}}^{x_{r}} (q_r y_r-\frac{1} {2}EIy_ {r,xx}^2)dx+\sum_{r=1}^{q}\bar {F}_r \bar {y}_ {r}+\bar {M}_r \bar {y}_ {r,x}
\end{multline}
According to \ref{eq:(bianfen.2.3)}, substituting \ref{eq:(jxz.3.1)} yields the Euler equation
\begin{equation} \label{eq:(zxxgyl.3.2)}    
L^ {r}_ {y_{r}} - 
\left(L^ {r}_ {y_{r,x}}\right)_{,x}+\left(L^ {r}_ {y_{r,xx}}\right)_{,xx}=q_r-(EIy_{r,xx})_{,xx}
=0,r=1,2...q
\end{equation}
\begin{equation} \label{eq:(zxxgyl.3.3)}    
L^ {r}_ {u_r}=y_r=0,r=1,2...q
\end{equation}
This equation has little significance because when $ y_r=0 $ , the beam does not deform, meaning there is no load acting on it.
The boundary condition for the intermediate active boundary $ x_ {r}$  is
\begin{multline} \label{eq:(zxxgyl.3.4)}    
[(L^ {r}-y_ {r,x} L_{y_{r,x}}   + y_{r,x} \left(L_{y_{r,xx}}\right)_{r,x}
-y_{r,xx} L_{y_{r,xx}} 
)-\\((L^{r+1}-y_{r+1,x} L_{y_{r+1,x}}   + y_{r+1,x} \left(L_{y_{r+1,xx}}\right)_{r+1,x}
\\-y_{r+1,xx} L_{y_{r+1,xx}} 
)+\varphi_{\bar {x}_ {r}}]|_{x=x_r}  
\\=[(q_r y_r-\frac{1} {2}EIy_ {r,xx}^2)+y_{r,x} \left(EIy_{r,xx}\right)_{r,x} 
-y_{r,xx} EIy_{r,xx}]
\\-[(q_r y_{r+1}-\frac{1} {2}EIy_ {r+1,xx}^2)+y_{r+1,x} \left(EIy_{r+1,xx}\right)_{r+1,x} 
-y_{r+1,xx} EIy_{r+1,xx}]
=0
\\,r=1,2...q
\end{multline}
According to \ref{eq:(bianfen.2.4)}, the shear boundary conditions can be obtained
\begin{multline} \label{eq:(zxxgyl.3.5)}    
[(L^ {r}_ {y_{r,x}} -\left(L^ {r}_ {y_{r,xx}}\right)_{r,x} )-(L^{r+1}_{y_{r+1,x}} -\left(L^{r+1}_{y_{r+1,xx}}\right)_{r+1,x}) + \varphi_{\bar {y}_ {r} }]|_{x=x_r}
\\=[(EIy_{r+1,xx})_{,x}-(EIy_{r,xx})_{,x})+ \bar {F}_r ]|_{x=x_r}
=0,r=1,2...q
\end{multline}
According to \ref{eq:(bianfen.2.5)}, the bending moment boundary conditions can be obtained
\begin{multline} \label{eq:(zxxgyl.3.6)}    
[L^ {r}_ {y_{r,xx}} - L^{r+1}_{y_{r+1,xx}}+\varphi_{\bar {y}_ {r,x}}]|_{x=x_r}
\\=[EIy_{r,xx}  -EIy_{r+1,xx}  +\bar {M}_r ]|_{x=x_r}
=0,r=1,2...q
\end{multline}
When $ u_r $ is unconstrained, the control condition is
\begin{subequations}\label{eq:(zxxgyl.3.7)} 
\begin{equation}    
\bar {y}_ {r}=0
\end{equation}
\begin{equation}    
\bar {y}_ {r,x}=0
\end{equation}    
\end{subequations}
For the boundary conditions of endpoints $ x_0 and x_ {1+1}$  , please refer to the previous section for details.
\subsection {measure of control effect}
To measure the optimal control, the optimal control index is given as follows:
\begin{equation}\label{eq:(zxxgyl.4.1)} 
op=1-W/W_0
\end{equation}  
Among them, $ W $ is the total virtual work of the structure, including the virtual work of fixed loads and control loads, and $ W-0 $ is the virtual work when only fixed loads are present.
According to the definition of virtual work, $ W0 $ is always positive, and $ W $ can be positive or negative. Therefore, when $ op=1 $ , it indicates optimal control; when $ 0<op<1 $ , it indicates that the control has played a certain role; and when $ op<0 $ , it indicates that the control has exceeded the effect of the original load and belongs to over control. Given this definition, it can effectively measure whether the control load is appropriate, such as whether the prestressed tendons of prestressed concrete continuous beams are configured too much, and whether the cables of cable-stayed bridges are over tensioned.
\section {Coordinate transformation algorithm}\label{sec:coordinate}
\subsection {coordinate transformation}
When performing specific calculations, coordinate changes are usually required, so it is necessary to study coordinate transformation formulas. Assuming that the local coordinate system (old coordinate system) is represented by $x_i$, the global coordinate system (new coordinate system) is represented by $x_ {j'}$, and the radial axis $\bm {r}$ can be expressed as
\begin{equation} \label{eq:(algorithm.1.1)}    
    \bm{r}=x_i \bm{e_i}=x_{j'} \bm{e_{j'}}
\end{equation}

Decompose the components in the new coordinates into the old coordinates to obtain
\begin{equation} \label{eq:(algorithm.1.2)}    
    x_{j'}=b_{j'i}x_i
\end{equation}
For a plane coordinate system, the above equation can be expanded to
\begin{equation} \label{eq:(algorithm.1.3)}    
    x_{1'}=b_{1'1}x_1+b_{1'2}x_2
    =cos(\alpha)x_1-sin(\alpha)x_2
\end{equation}
Among them, $\ alpha $is the angle between $e1 $and $e2 {1 '} $.
\begin{equation} \label{eq:(algorithm.1.4)}    
    x_{2'}=b_{2'1}x_1+b_{2'2}x_2
    =sin(\alpha)x_1+cos(\alpha)x_2
\end{equation}
Similarly, decomposing the components in the old coordinates into the new coordinates yields
\begin{equation} \label{eq:(algorithm.1.5)}    
    x_i=a_{ij'}x_{j'}
\end{equation}
For a plane coordinate system, the above equation can be expanded to
\begin{equation} \label{eq:(algorithm.1.6)}    
    x_{1}=b_{11'}x_{1'}+b_{12'}x_{2'}
    =cos(\alpha)x_{1'}+sin(\alpha)x_{2'}
\end{equation}
Among them, $\ alpha $is the angle between $e1 $and $e2 {1 '} $.
\begin{equation} \label{eq:(algorithm.1.7)}    
    x_{2}=b_{21'}x_1+b_{22'}x_2'
    =-sin(\alpha)x_1'+cos(\alpha)x_2'
\end{equation}

Assuming that the component of vector $y $in the local coordinate system is $y_j $and the component in the global coordinate system is $y_ {j '} $, then there is
\begin{equation} \label{eq:(algorithm.1.8)}    
    \bm{y}=y_j \bm{e_j}=y_{i'} \bm{e_{i'}}
\end{equation}
Obtain the expression for the coordinate transformation relationship of vector components based on the coordinate transformation relationship of vectors
\begin{equation} \label{eq:(algorithm.1.9)}    
    y_{j'}=b_{j'i}y_i
\end{equation}
\begin{equation} \label{eq:(algorithm.1.10)}    
    y_i=a_{ij'}y_{j'}
\end{equation}
From the above two equations, it can be seen that the matrix composed of $b_{j'i'},a_{ij}$ is reciprocal, that is
\begin{equation} \label{eq:(algorithm.1.11)}
  b_{j'k}a_{ki'}=\delta_{j'i'}
\end{equation}
In the Cartesian coordinate system, the transformation coefficient matrix is orthogonal
\begin{equation}\label {eq:(algorithm.1.12)}
  b_{j'k}b_{i'k}=\delta_{j'i'}
\end{equation}
From the above two equations, we can also obtain
\begin{equation}\label {eq:(algorithm.1.13)}
  a_{ki'}=b_{i'k}
\end{equation}
When calculating the derivative relationship during coordinate transformation, both the global coordinate system and the local coordinate system are considered as Cartesian coordinate systems. Therefore, $a_{ij'},b_{j'i}$ are constants
\begin{multline} \label{eq:(algorithm.1.14)}    
    y_{j,i}=(a_{jj'}y_{j'})_{,i}=a_{jj'} y_{j',i}=a_{jj'} y_{j',i'} x_{i',i}
    =a_{jj'} y_{j',i'} (b_{i'k}x_k)_{,i}
    \\=a_{jj'} y_{j',i'}b_{i',k}\delta_{ki}
    =a_{jj'} b_{i',i} y_{j',i'}
\end{multline}
Similarly, it can be concluded that
\begin{equation} \label{eq:(algorithm.1.15)}    
    y_{j',i'}=b_{j'j} a_{i,i'}y_{j,i}
\end{equation}
\begin{equation} \label{eq:(algorithm.1.16)}    
    y_{j,ip}=a_{jj'} b_{i',i} b_{p',p} y_{j',i'p'}
\end{equation}
\begin{equation} \label{eq:(algorithm.1.17)}    
    y_{j',i'p'}=b_{j'j} a_{i,i'} a_{p,p'} y_{j,ip}
\end{equation}

To convert \ref{eq:(bianfen.1.1)} to the global coordinate system, there are
\begin{multline}\label{eq:(algorithm.1.18)}
J = \iiint_D L[x_i,y_j,y_{j,i},y_{j,ip},u_k] dv + \iint_S \varphi[\bar {x}_i ,\bar {y}_j ,\bar {y}_ {j,i}] ds
\\= \iiint_D G_0[x_{i'},y_{j'},y_{j',i'},y_{j',i'p'},u_k'] dv + \iint_S \varPhi 0[\bar {x}_{i'} ,\bar {y}_{j'} ,\bar {y}_ {j',i'}] ds
\\=\iiint_{D1} G_0[x_{i'},y_{j'},y_{j',i'},y_{j',i'p'},u_k']Jv dV + \iint_{S1} \varPhi 0[\bar {x}_{i'} ,\bar {y}_{j'} ,\bar {y}_ {j',i'}] |Js|dS
\\=\iiint_{D1} G[x_{i'},y_{j'},y_{j',i'},y_{j',i'p'},u_k'] dV + \iint_{S1} \varPhi [\bar {x}_{i'} ,\bar {y}_{j'} ,\bar {y}_ {j',i'}] dS
\end{multline}
Where 
\\$Jv=|x_{i, j '}|$ is the Jacobian determinant,
\\ $Js=|x_{i,j'}|,
\\ G=Jv G_0 ,
\\\varPhi = Js \varPhi_0 , 
\\G_0[x_{i'},y_{j'},y_{j',i'},y_{j',i'p'},u_k'] \\=L[a_{ii'}x_{i'},a_{jj'}y_{j'},a_{jj'}b_{i'i}y_{j',i'},a_{jj'} b_{i'i} b_{p'p}{y_{j',i'p'}},a_{kk'}u_k'], 
\\\varPhi_0 [\bar {x}_{i'} ,\bar {y}_{j'} ,\bar {y}_ {j',i'}] =\varphi[\bar {a}_{ii'} \bar{x_i'} ,\bar {a}_{jj'} \bar {y}_{j'} ,\bar {a}_{jj'}\bar{a}_{Ii}\bar {Y}_{JI}]$.

According to the derivation in Section \ref{sec: bianfen}, the Euler equation \ref{eq:(bianfen.1.19)} is
\begin{equation} \label{eq:(algorithm.1.19)}    
G_{y_{j'}} - \left(G_{y_{j',i'}}\right)_{,i'}+\left(G_{y_{j',i'p'}}\right)_{,i'p'} =0
\end{equation}

The following proves that the Euler equation has invariance.
\begin{equation} \label{eq:(algorithm.1.20)}    
G_{y_{j'}}=\frac{\partial (Jv L)}{\partial y_{j'}}
=Jv \frac{\partial L}{\partial y_{j'}}
=Jv \frac{\partial L}{\partial y_j} \frac{\partial y_j}{\partial y_{j'}}
=Jv a_{jj'} L_{y_j} 
\end{equation}
Similarly
\begin{multline} \label{eq:(algorithm.1.21)}    
\left(G_{y_{j',i'}}\right)_{,i'}=\frac{\partial (G_{y_{j',i'})}}{\partial x_{i'}}
=\frac{\partial (G_{y_{j',i'})}}{\partial x_i}  \frac{\partial x_i}{\partial x_{i'}}=\frac{\partial G_{y_{j',i'}}}{\partial x_i}  a_{ii'}
=a_{ii'}\frac{\partial }{\partial x_i} (\frac{\partial G}{\partial y_{j',i'}}) 
\\=a_{ii'}\frac{\partial }{\partial x_i} \left[(\frac{\partial G}{\partial y_{m,n}}) (\frac{\partial y_{m,n}}{\partial y_{j',i'}})\right]
=a_{ii'}(\frac{\partial y_{m,n}}{\partial y_{j',i'}})\frac{\partial }{\partial x_i} (\frac{\partial G}{\partial y_{m,n}})
\end{multline}

Substituting \ref{eq:(algorithm.1.7)} into \ref{eq:(algorithm.1.14)} and replacing the dummy label yields
\begin{multline} \label{eq:(algorithm.1.22)}    
\left(G_{y_{j',i'}}\right)_{,i'}
=a_{ii'}(\frac{\partial (a_{mj'}b_{i'n} y_{j',i'})}{\partial y_{j',i'}})\frac{\partial }{\partial x_i} (\frac{\partial G}{\partial y_{m,n}})
=a_{ii'}(a_{mj'}b_{i'n})\frac{\partial }{\partial x_i} (\frac{\partial G}{\partial y_{m,n}})
\\=Jv a_{mj'} \delta_{in} (L_{y_{m,n}})_{,i}
=Jv a_{mj'} (L_{y_{m,n}})_{,n}
=Jv a_{jj'} (L_{y_{j,i}})_{,i}
\end{multline}
Similarly
\begin{multline}\label{eq:(algorithm.1.23)}    
\left(G_{y_{j',i'p'}}\right)_{,i'p'}
=a_{ii'} a_{pp'}\left(G_{y_{j',i'p'}}\right)_{,ip}
=a_{ii'} a_{pp'} a_{jj'}\left(G_{y_{j,i'p'}}\right)_{,ip}
\\=a_{ii'} a_{pp'} a_{jj'} b_{i'm} b_{p'n}\left(G_{y_{j,mn}}\right)_{,ip}
=a_{jj'} \delta_{im} \delta_{pn} \left(G_{y_{j,mn}}\right)_{,ip}
=a_{jj'} \left(G_{y_{j,mn}}\right)_{,mn}
\\=Jv a_{jj'} \left(L_{y_{j,mn}}\right)_{,mn}
=Jv a_{jj'} \left(L_{y_{j,ip}}\right)_{,ip}
\end{multline}
From the above proof, it can be found that the subscript of $a $follows the following pattern: when the subscript is not differentiated, the subscript is shifted before, the subscript is differentiated once (including composite differentiation), the subscript is located after, the subscript is differentiated twice, and the subscript is shifted before.

substituting \ref{eq:(algorithm.1.10)} yields
\begin{equation} \label{eq:(algorithm.1.24)}    
Jv  a_{jj'} L_{y_j}- Jv a_{jj'}(L_{y_{j,i}})_{,i}+Jv a_{jj'}\left(L_{y_{j,ip}}\right)_{,ip} =0
\end{equation}
Divide both ends simultaneously by $Jv a_ {jj '} $to obtain
\begin{equation} \label{eq:(algorithm.1.25)}    
L_{y_j}- (L_{y_{j,i}})_{,i}+\left(L_{y_{j,ip}}\right)_{,ip} =0
\end{equation}

This equation is the Euler equation before coordinate transformation, consistent with \ref{eq:(algorithm.1.12)}, which also proves the invariance of the Euler equation.
Similarly, it can be proven that the boundary condition is when $\delta \ bar {Y}_J $When free, according to \ref{eq:(bianfen.1.24)}, it can be obtained
\begin{equation} \label{eq:(algorithm.1.26)}    
\left[\left(G_{y_{j',i'}}  \right) -\left(G_{y_{j',i'p'}}\right)_{,p'}\right]n_{i'} + \varPhi_{\bar{y_{j'}}}=0
\end{equation}

In the Cartesian coordinate system, the conversion coefficient $x_ {i'n}=a_ {ni '},\\x_ {i'n} a_ {ii'}=\ delta {in}, Jv=1, Js=1 $, prove that the boundary conditions have invariance in the Cartesian coordinate system.
because
\begin{equation} \label{eq:(algorithm.1.27)}    
b_{i'n}=a_{ni'}
\end{equation}
\begin{multline} \label{eq:(algorithm.1.28)}    
[G_{y_{j',i'}}-\left(G_{y_{j',i'p'}}\right)_{,p'}]n_{i'} + \varPhi_{\bar {y}_{j'}}
\\=[Jv a_{jj'} b_{i'i} L_{y_{j,i}} -Jv a_{jj'} b_{i'i}b_{p'p}  a_{mp'} \left(L_{y_{j,ip}}\right)_{,m}]b_{i'k}n_k + \varPhi_{\bar {y}_{j'}}
\\=[Jv a_{jj'} b_{i'i}b_{i'k} L_{y_{j,i}} -Jv a_{jj'} b_{i'i} b_{p'p}  a_{mp'} b_{i'k}\left(L_{y_{j,ip}}\right)_{,m}]n_k + |Js| a_{jj'} \varphi_{\bar {y}_j}
\\=[Jv a_{jj'} \delta_{ik} L_{y_{j,i}} -Jv a_{jj'} \delta_{ik}\delta_{mp}\left(L_{y_{j,ip}}\right)_{,m}]n_k + |Js| a_{jj'} \varphi_{\bar {y}_j}
\\=[Jv a_{jj'} L_{y_{j,k}} -Jv a_{jj'} \left(L_{y_{j,km}}\right)_{,m}]n_k + |Js| a_{jj'} \varphi_{\bar {y}_j}
\\=a_{jj'} [ L_{y_{j,i}}   -  \left(L_{y_{j,ip}}\right)_{,p}]n_i + a_{jj'} \varphi_{\bar {y}_j}=0
\end{multline}
The above equation can be divided by $a_ {jj '} $to obtain,
\begin{equation} \label{eq:(algorithm.1.29)}    
 [ L_{y_{j,i}}   -  \left(L_{y_{j,ip}}\right)_{,p}]n_i + \varphi_{\bar {y}_j}=0
\end{equation}

This equation is the boundary condition before coordinate transformation, which is consistent with \ref{eq:(algorithm.1.12)}, and also proves the invariance of boundary conditions in Cartesian coordinate system.

Similarly, other boundary conditions can be obtained, according to \ref{eq:(bianfen.1.25)}
\begin{equation} \label{eq:(algorithm.1.31)}    
\left(G_{y_{j',i'p'}}  \right)n_p' +\varPhi_{\bar {y}_ {j',i'}} =0
\end{equation}
according to
\ref{eq:(bianfen.1.26)}
\begin{equation} \label{eq:(algorithm.1.32)}    
G_{u_k'} =0
\end{equation}
according to
\ref{eq:(bianfen.1.27)}
\begin{equation} \label{eq:(algorithm.1.33)}    
G[x_i',y_j',y_{j',i'},y_{j',i'p'},u_k']+ \varPhi_{\bar {x}_i' } n_i' 
+\bar{y}_{j',l'}\varPhi_{\bar {y}_j'} n_l' 
+\bar{y}_{j',i'r'}\varPhi_{\bar {y}_ {j',i'}} n_r' =0
\end{equation}

\subsection {Convert the boundary function to the global coordinate system}
\numberwithin{equation}{subsection}
According to the derivation of $\ref {subsec:duotiji} $, the case where the same node has multiple elements is obtained. In this case, multiple elements generally have their own coordinate systems and require coordinate transformation.

Due to the presence of $n ^ r $elements at the boundary $S_r$, the transformation relationship from the local coordinate system to the entire system is as follows
\begin{equation} \label{eq:(algorithm.2.3)}    
n^r_i=a_{ik'}n^r_{k'}
\end{equation}

Project the boundary conditions of \ref{eq:(bianfen.duotiji.3)} onto the global coordinate system $x'_i $, and obtain
\begin{multline} \label{eq:(algorithm.2.6)}    
\delta J = \sum_{r=1}^{q} \iiint_{D_r} \{[L^r_{y_j} - \left(L^r_{y_{j,i}}\right)_{,i}+\left(L^r_{y_{j,ip}}\right)_{,ip}]\delta y_j \}dv +\sum_{r=1}^{q} \iiint_{D_r} [L_{u_k} \delta u_k]dv
\\ +\sum_{r=1}^{b}\iint_{S_r} \{\sum_{t=1}^{n^r}\{L^{t}[x_i,y_j,y_{j,i},y_{j,ip},u_k]+ \varphi^{t}_{\bar{x}_i} n^t_i +\bar{y}_{j,l}\varphi^t_{\bar {y}_j} n^t_l
-\bar{y}_{j,ip}\varphi^t_{\bar {y}_ {j,i}} n^t_p \}n^r_{m'}\}\delta \bar{n}^r_{m'}  \}da
\\+\sum_{r=1}^{b}\iint_{S_r} \{\sum_{t=1}^{n^t} \{[\left(L^{t}_{y_{j,i}}  \right) -\left(L{t}_{y_{j,ip}}\right)_{,p}] a_{im'} a_{jn'} n^t_{m'} + a_{jn'} \varphi^{t}_{\bar{y}_j} \}\delta \bar{y} _{n'} \}da
\\+\sum_{r=1}^{b}\iint_{S_{r}}\{\sum_{t=1}^{n^r} [\left(L_{y_{j,ip}}  \right)a_{jm'} b_{n'i}a_{pk'}n^t_{k'}+a_{jm'} b_{n'i}\varphi_{\bar{y}_{j,i}}] \delta \bar{y}_{m',n'} \} da
=0
\end{multline}

The Euler equation can be obtained from \ref{eq:(bianfen.2.2)}
\begin{equation} \label{eq:(algorithm.2.7)}    
L^r_{y_j} - \left(L^r_{y_{j,i}}\right)_{,i}+\left(L^r_{y_{j,ip}}\right)_{, ip}=0,j=1,2...n,r=1,2...q
\end{equation}
governing equation 
\begin{equation} \label{eq:(algorithm.2.8)}    
L^r_{u_k} =0,k=1,2...m,r=1,2...q
\end{equation}
And boundary conditions
\begin{equation}  \label{eq:(algorithm.2.9)}     
\sum_{t=1}^{n^r} \left[L^{t}_{y_{j,i}} -\left(L^{t}_{y_{j,ip}}\right)_{,p}] a_{im'} a_{jn'} n^t_{m'} + a_{jn'}\varphi^{t}_{\bar{y}_j} \right ] =0,n'=1,2...n,r=1,2...b
\end{equation}
\begin{equation} \label{eq:(algorithm.2.10)}  
\sum_{t=1}^{n^r} \left(L^t_{y_{j,ip}}  a_{jm'} b_{n'i}a_{pk'}n^t_{k'}+a_{jm'} b_{n'i}\varphi_{\bar{y}_{j,i}}\right)
=0,n'=1,2...n,r=1,2...b
\end{equation}
\begin{multline} \label{eq:(algorithm.2.11)}   
\sum_{t=1}^{n^r} \{L^{t}[x_i,y_j,y_{j,i},y_{j,ip},u_k]+ \varphi^{t}_{\bar{x}_i} n^t_i 
+y_{j,l} n^t_l \varphi^{t}_{\bar{y}_j}
+y_{j,ip} \varphi^t_{\bar{y}_{j,i}} n^t_p  \}n^t_{m'}=0
\\,m'=1,2...n,r=1,2...b
\end{multline}

\section {Example} \label{sec: example}
Next, this article will analyze multiple examples, firstly to verify the correctness of the theory proposed in this article, and secondly to illustrate the wide application value of this theory in the field of bridge engineering.
\subsection {cantilever beam}
In order to verify the correctness of the theory proposed in this article, a cantilever beam was used as an example for verification. As the cantilever beam is a statically determinate structure, it is easy to use other methods for verification.

\subsubsection {when concentrated load takes any value}
\numberwithin{equation}{subsection}
Firstly, verify the principle of minimum virtual work under the action of concentrated loads with arbitrary variations. As shown in the figure \ref{fig:jizhongli}, the length of the cantilever beam is $ L$  , the interface moment of inertia is $ i$  , the elastic modulus is $ E $ , the uniformly distributed force is $ q $ , and the concentrated force is $ F $ . Find the value of F so that the internal energy (strain energy) of the structure is minimized (the sum of squares of bending moments is minimized, or the material is saved the most).
\begin{figure}[h!] 
    \centering
    \includegraphics[width=0.75\textwidth,keepaspectratio]{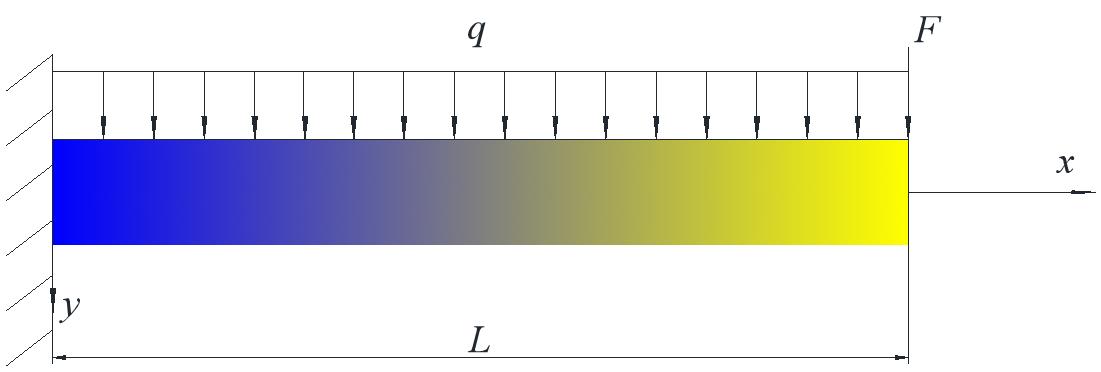} 
    \caption{Schematic diagram of force on cantilever beam}
    \label{fig:Schematic diagram of force on cantilever beam}
\end{figure}
\begin{multline}\label{eq:(example.xbl.1)}
J = \int_{x_{0}}^{x_{1}} L(x,y,y_{,x},y_{,xx},u,u_{,x})dx+\varphi(x_{0},y_{0},u_{0},x_{1},y_{1},u_{1},y_{0,x},u_{0,x},y_{1,x},u_{1,x})\\
=\int_{0}^{L} -\frac{1} {2}EIy ''^2dx+Fy_{L}+\int_{0}^ {L}qy dx
\\=\int_{0}^{L} (-\frac{1} {2}EIy ''^2+qy)dx+Fy_{L}
\end{multline}
Then there are
\begin{equation} \label{eq:(example.xbl.2)}    
L=-\frac{1} {2}EIy ''^2+qy
\end{equation}
\begin{equation} \label{eq:(example.xbl.3)}    
\varphi=Fy_{L}\end{equation}
According to equation \ref{eq:(zxxgyl.2.2)}, the Euler equation for the cantilever beam element is obtained as follows:
\begin{equation}\label{eq:(example.xbl.4)}
L_{y} - \left(L_{y_{,x}}\right)_{,x}+\left(L_{y_{,xx}}\right)_{,xx}
=-EI\frac{d^2y''}{dx^2}+q=0
\end{equation}
According to \ref{eq:(zxxgyl.2.5)} \ref{eq:(zxxgyl.2.6)}, obtain the boundary conditions
\begin{equation}\label{eq:(example.xbl.5)}
EIy''=0 (x=L)
\end{equation}
\begin{equation}\label{eq:(example.xbl.6)}
EI\frac{dy''}{dx}+F=0 (x=L)
\end{equation}
According to the fixed end constraint conditions
\begin{equation}\label{eq:(example.xbl.7)}
y(0)=0 (x=0)
\end{equation}
\begin{equation}\label{eq:(example.xbl.8)}
y'(0)=0 (x=0)
\end{equation}
According to the control conditions obtained from \ref{eq:(zxxgyl.2.21)}
\begin{equation}\label{eq:(example.xbl.9)}
y_L=y(L)=0
\end{equation}
This formula indicates that when the displacement at the control load F is zero, that is, when the work done by F is zero, the minimum value of potential energy is taken at this time. According to the principle of energy conservation, the work done by external force is equal to the internal energy stored inside the beam. Therefore, when the displacement is zero, the work done is zero, and the energy converted into the interior is also zero. At this time, the internal energy must be minimized (the converted internal energy is always positive).

From this \ref{eq:(example.xbl.4)} \text{-} \ref{eq:(example.xbl.8)}, the solution to the differential equation can be obtained as
\begin{equation}\label{eq:(example.xbl.10)}
y(x)=\frac{6qL^2x^2-4qLx^3+12FLx^2+qx^4-4Fx^3}{24EI}
\end{equation}
At this time, the virtual power of the system is
\begin{equation}\label{eq:(example.xbl.11)}
W=\frac{L^3(20F^2 + 15FLq + 3L^2q^2)}{120EI}
\end{equation}
According to the control condition \ref{eq:(example.xbl.9)},
\begin{equation}\label{eq:(example.xbl.12)}
y_L=\frac{8FL^3-3L^4q}{24EI}
\end{equation}
Solved to obtain
\begin{equation}\label{eq:(example.xbl.13)}
F=-\frac{3Lq}{8}
\end{equation}
At this point, the potential energy reaches its minimum value
\begin{equation}\label{eq:(example.xbl.14)}
W=\frac{L^5q^2}{640EI}
\end{equation}
Next, using the extremum condition of the function to verify the above conclusion, the derivative of \ref{eq:(example.xbl.10)} is obtained
\begin{equation}\label{eq:(example.xbl.15)}
W'(F)= (L^3*(40*F + 15*L*q))/(120*E*I1)
\end{equation}
\begin{equation}\label{eq:(example.xbl.16)}
W''(F)=L^3/(3*E*I1)
\end{equation}
From \ref{eq:(example.xbl.16)}, it can be obtained that
\begin{equation}\label{eq:(example.xbl.17)}
F=-\frac{3Lq}{8}
\end{equation}
Consistent with \ref{eq:(example.xbl.13)}.
Next, the conclusion above is verified by minimizing the area of the sum of squares of bending moments, and the derivative of \ref{eq:(example.xbl.10)} is obtained
\begin{equation}\label{eq:(example.xbl.18)}
M=(q*x^2)/2 + F*x
\end{equation}
Integrating the square of the bending moment to obtain
\begin{equation}\label{eq:(example.xbl.19)}
IM= (F^2*L^3)/3 - (F*L^4*q)/4 + (L^5*q^2)/20
\end{equation}
Derive it to obtain
From \ref{eq:(example.xbl.16)}, it can be obtained that
\begin{equation}\label{eq:(example.xbl.20)}
F=-\frac{3Lq}{8}
\end{equation}
Consistent with \ref{eq:(example.xbl.13)} through the above verification, it has been demonstrated that the theory proposed in this article is correct and feasible.
In order to display the calculation results more intuitively, the following are the curves of displacement, controllable load work, and fixed load virtual work, i.e. strain energy, as a function of $ F $ . Assuming $ E=1, I=1, q=1, L=40 $ , the results are shown in the following figure
\begin{figure}[h!] 
    \centering
    \includegraphics[width=0.75\textwidth,keepaspectratio]{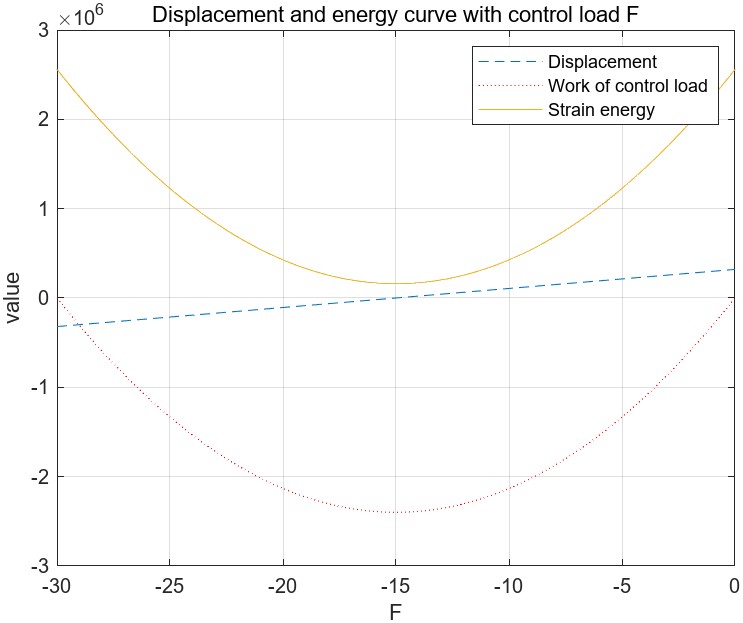} 
    \caption{Displacement and energy change curves}
    \label{fig:Displacement and energy change curves}
\end{figure}
As shown in the figure, with the decrease of $ F $ , the negative work done by the controllable load increases, and the internal energy of the system decreases. When $ F=- {3Lq}/{8}=-15 $ , the displacement $ y_L=0 $ , the work done by the controllable load takes the minimum value, and the strain energy (internal energy) of the system is also minimized. When $ F $ continues to decrease, $ y_L $ changes from positive to negative (i.e. from downward deformation to upward deformation), and the controllable load begins to do positive work, with the total work done increasing (the curve changes upward), and the internal energy of the system begins to increase. When $ F=-30 $ , the effect of the controlled load on the system is greater than that of the fixed load, and side effects begin to occur.
In summary, the principle of minimum virtual work proposed in this article is correct, which can effectively calculate the minimum internal energy, that is, the minimum sum of squared bending moments, thus obtaining the optimal control load and the most material saving.
\subsubsection {when concentrated load is limited}
Next, we will analyze the control variable $ F $ when it is restricted, assuming
\begin{equation}\label{eq:(example.xbl.21)}
\frac{-2Lq}{8}\leq F\leq \frac{-Lq}{8}
\end{equation}
Since the system satisfies \ref{eq:(example.xbl.4)} \text{-} \ref{eq:(example.xbl.8)}, the work done by an external force containing $ F $ can be calculated as $ W$ 
\begin{equation}\label{eq:(example.xbl.22)}
W=Fy_{L}+\int_{0}^ {L}qy dx
=\frac{L^3 {\left(20F^2 +15FLq+3L^2 q^2 \right)}}{60\textrm{E}I }
\end{equation}
By taking the first derivative of $ W $ , we can obtain
\begin{equation}\label{eq:(example.xbl.23)}
W=Fy_{L}+\int_{0}^ {L}qy dx
=\frac{L^3 {\left(20F^2 +15FLq+3L^2 q^2 \right)}}{60\textrm{E}I }
\end{equation}
\begin{equation}\label{eq:(example.xbl.24)}
W'=\frac{L^3 {\left(40F+15Lq\right)}}{60\textrm{E}I }
\end{equation}
Obtain the extremum point as
\begin{equation}\label{eq:(example.xbl.25)}
F=-\frac{3Lq}{8}
\end{equation}
Not within the range of \ref{eq:(example.xbl.21)}, since $ W $ is a quadratic parabola with an upward opening about $ F $ , when $ F=\frac {-2Lq} {8}$  , $ W $ takes the minimum value, that is
\begin{equation}\label{eq:(example.xbl.26)}
W_{min}=W(\frac{-2Lq}{8})=\frac{L^5 q^2 }{120\textrm{E}I }
\end{equation}
This example verifies that the principle minimum virtual work proposed in this article is correct for restrained concentrated loads and can effectively obtain the minimum internal energy of the structure.
\subsubsection {when the distributed load takes any constant}
\begin{figure}[h!] 
    \centering
    \includegraphics[width=0.75\textwidth,keepaspectratio]{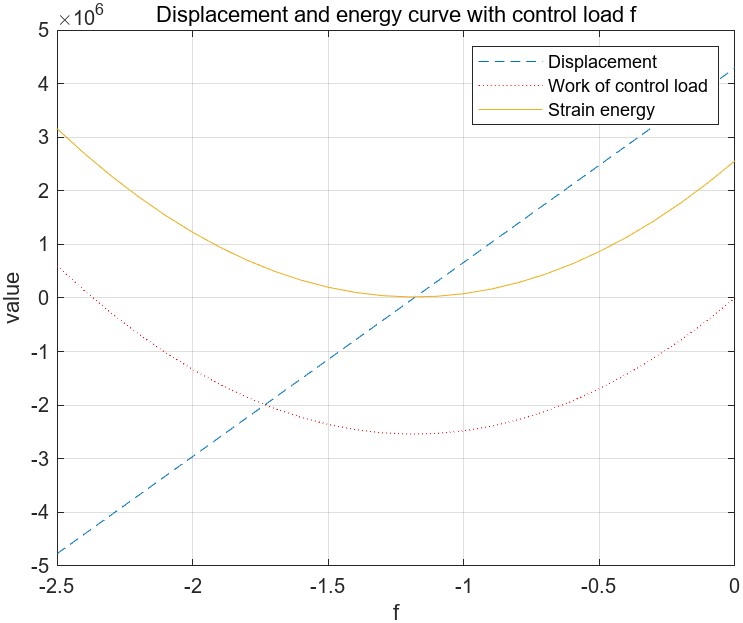} 
    \caption{The controlled load on the cantilever beam is a distributed load.}
    \label{fig:fenbuli}
\end{figure}

When controlling the load to take the distributed force, the control load is the distributed force $ f $ . Find the value of $ f $ to minimize the strain energy, as shown in Figure \ref{fig:fenbuli}
According to equation \ref{eq:(zxxgyl.2.2)}, the Euler equation for a cantilever beam element is
\begin{equation}\label{eq:(example.xbl.27)}
EI\frac{d^2y_1''}{dx^2}-q=0,0\leq x<L1
\end{equation}
\begin{equation}\label{eq:(example.xbl.28)}
EI\frac{d^2y_1''}{dx^2}-q-f=0,L1< x<L
\end{equation}
Obtain boundary conditions based on \ref{eq:(zxxgyl.2.4)} \text{-} \ref{eq:(zxxgyl.2.5)}
\begin{equation}\label{eq:(example.xbl.29)}
y(0)=0 (x=0)
\end{equation}
\begin{equation}\label{eq:(example.xbl.30)}
\frac{d y_1}{dx}=0 (x=0)
\end{equation}
\begin{equation}\label{eq:(example.xbl.31)}
EI\frac{d^2 y_2}{dx^2}=0 (x=L)
\end{equation}
\begin{equation}\label{eq:(example.xbl.32)}
EI\frac{d^3 y_2}{d^3 x}=0 (x=L)
\end{equation}
\begin{equation}\label{eq:(example.xbl.33)}
y_1(x_1)=y_2(x_1)
\end{equation}
\begin{equation}\label{eq:(example.xbl.34)}
\frac{d y_1}{dx}=\frac{d y_2}{dx}(x=x_1)
\end{equation}
\begin{equation}\label{eq:(example.xbl.35)}
\frac{d^2 y_1}{dx^2}=\frac{d^2 y_2}{dx^2}(x=x_1)
\end{equation}
\begin{equation}\label{eq:(example.xbl.36)}
\frac{d^3 y_1}{dx^3}=\frac{d^3 y_2}{dx^3}(x=x_1)
\end{equation}
According to the control conditions obtained from \ref{eq:(zxxgyl.1.23)}
\begin{equation}\label{eq:(example.xbl.37)}
\int_{x_1}^ {L}y_2 dx=0
\end{equation}
This equation indicates that when the displacement integral at the control load F is zero, that is, when the virtual work done by f is zero, the internal energy is at its minimum. This conclusion verifies that the principle minimum virtual work proposed in this paper is correct when the cantilever beam is subjected to distributed loads.
From here \ref{eq:(example.xbl.27)}\text{-}\ref{eq:(example.xbl.36)}, the solution of the differential equation can be obtained as
\begin{equation}\label{eq:(example.xbl.38)}
y_1(x)=\frac{x^2 {\left(6 L^2 f+6L^2 q-6f {x_1 }^2 +qx^2 -4Lfx-4Lqx+4fxx_1 \right)}}{24E_1 I_1 }
\end{equation}
\begin{equation}\label{eq:(example.xbl.39)}
y_2(x)=\frac{fx^4 +f{x_1 }^4 +qx^4 -4fx{x_1 }^3 +6L^2 fx^2 +6L^2 qx^2 -4Lfx^3 -4Lqx^3 }{24E_1 I_1 }
\end{equation}
At this time, the total virtual work is

\begin{multline}\label{eq:(example.xbl.40)}
    W=\frac{1}{120E_1 I_1} (3 L^5 f^2 +6L^5 fq+3L^5 q^2 -10L^2 f^2  {x_1 }^3 -10L^2 fq{x_1 }^3 +5Lf^2 {x_1 }^4 
    \\+5Lfq{x_1 }^4 +2f^2 {x_1 }^5 -fq{x_1 }^5)
\end{multline}

According to the control condition \ref{eq:(example.xbl.9)},
\begin{multline}\label{eq:(example.xbl.41)}
\frac{1}{120E_1 I_1 }(6 L^5 f^2 +6qL^5 f-20L^2 f^2  {x_1 }^3 -10qL^2 f{x_1 }^3 +10Lf^2 {x_1 }^4 
\\+5qLf{x_1 }^4 +4f^2 {x_1 }^5 -qf{x_1 }^5)=0
\end{multline}
Solved to obtain
\begin{equation}\label{eq:(example.xbl.42)}
f=-\frac{q{\left(6L^5 -10L^2 {x_1 }^3 +5L{x_1 }^4 -{x_1 }^5 \right)}}{6L^5 -20L^2 {x_1 }^3 +10L{x_1 }^4 +4{x_1 }^5 }
\end{equation}
At this point, the potential energy reaches its minimum value
\begin{equation}\label{eq:(example.xbl.43)}
W=\frac{q^2 {x_1 }^5 {\left(36 L^3 -28L^2 x_1 +8L {x_1 }^2 -{x_1 }^3 \right)}}{480E_1 I_1 {\left(3 L^3 +6L^2 x_1 +9L {x_1 }^2 +2{x_1 }^3 \right)}}
\end{equation}
Next, using the extremum condition of the function to verify the above conclusion, the derivative of \ref{eq:(example.xbl.11)} is obtained
\begin{multline}\label{eq:(example.xbl.44)}
W'(F)=\frac{1}{120E_1 I_1 }(6 L^5 f+6L^5 q+4f {x_1 }^5 -q{x_1 }^5 -20L^2 f{x_1 }^3 
\\-10L^2 q{x_1 }^3 +10Lf{x_1 }^4 +5Lq{x_1 }^4 )
\end{multline}
From \ref{eq:(example.xbl.16)}, it can be obtained that
\begin{equation}\label{eq:(example.xbl.45)}
f=-\frac{q{\left(6L^5 -10L^2 {x_1 }^3 +5L{x_1 }^4 -{x_1 }^5 \right)}}{6L^5 -20L^2 {x_1 }^3 +10L{x_1 }^4 +4{x_1 }^5 }
\end{equation}
consistent with \ref{eq:(example.xbl.13)}.
In order to display the calculation results more intuitively, the following are the curves of displacement, controllable load work, fixed load virtual work, or strain energy as a function of $ F $ , assuming $ E=1, I=1,q=1,L=40,x_1=20$ , The result is shown in the following figure
\begin{figure}[h!] 
    \centering
    \includegraphics[width=0.75\textwidth,keepaspectratio]{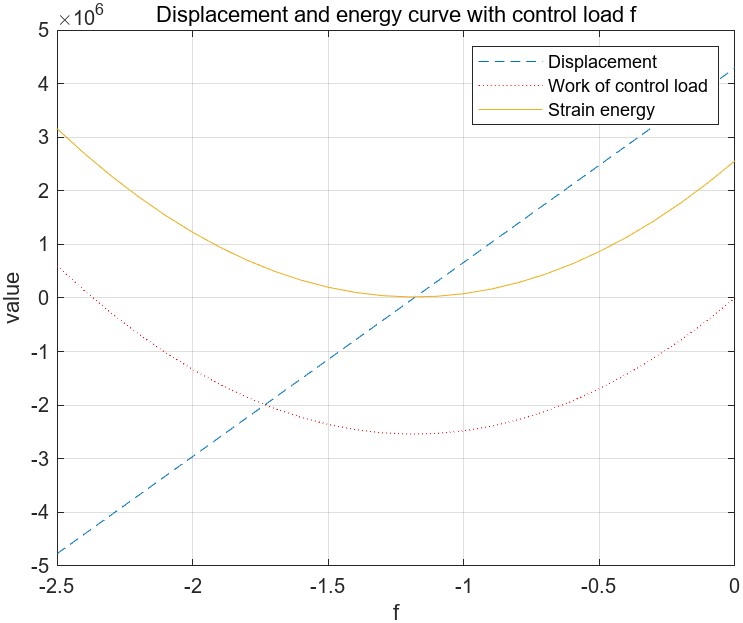} 
    \caption{Displacement and energy change curve - distributed force}
    \label{fig:Displacement and energy change curve - distributed force}
\end{figure}

As shown in the figure, as $ f$  decreases, the negative work done by the controllable load increases, and the internal energy of the system decreases. When $ F=-1.18$ , the displacement $ \int_ {x_1}^ {L}y_2 dx=0 $ , the minimum value of controllable load work is taken, and the strain energy (internal energy) of the system is also minimized. When $ f$  continues to decrease, $ y_L $  changes from positive to negative (i.e. from downward deformation to upward deformation), and the controllable load begins to do positive work, with the total work done increasing (the curve changes upward), and the internal energy of the system begins to increase. When $ f=-2.36 $ , the effect of the controlled load on the system is greater than that of the fixed load, and side effects begin to occur.

In summary, the principle of minimum virtual work proposed in this article is correct, which can effectively calculate the minimum potential energy, that is, the minimum sum of squares of bending moments, thus obtaining the optimal control load and the most material saving.

\subsection {Method for configuring prestressed tendons in prestressed concrete beams}
This example is used to verify the principle minimum virtual work in the strong form of control conditions when the structure is subjected to arbitrary distributed loads, and to illustrate the usage and practical significance of the optimal control index $ op $ . In addition, it is also used to demonstrate the theoretical guidance of this article for the prestressed tendon configuration of prestressed concrete continuous beams.

In bridge engineering, the most commonly used structure is prestressed concrete beams, because prestressed concrete beams have a large span capacity, are fast and convenient to construct, and have good economy. The configuration of prestressed tendon directly affects the stress and material consumption of the structure. However, the current design of prestressed tendon configuration lacks theoretical guidance for optimal prestressed tendon configuration. This section will take simply supported prestressed concrete beams as an example to illustrate the effect of using the principle of minimum virtual work to configure optimal prestressed tendons.
\begin{figure}[h!] 
    \centering
    \includegraphics[width=0.75\textwidth,keepaspectratio]{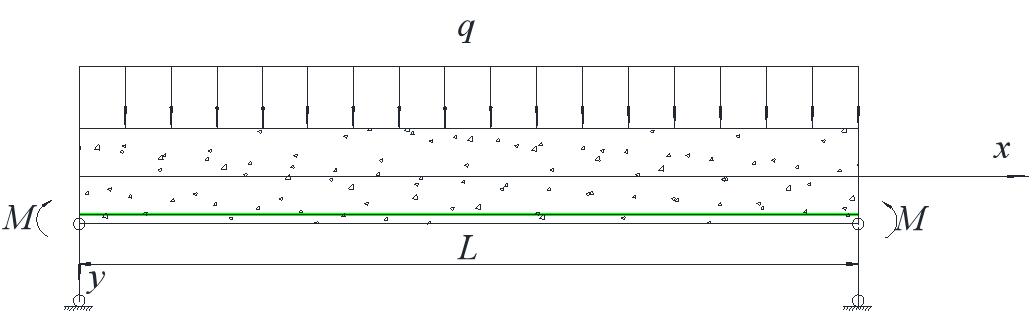} 
    \caption{Simply supported beam - straight tendon}
    \label{fig:Simply supported beam - straight tendon}
\end{figure}

As shown in the figure, it is a simply supported beam with a span of L and a stiffness of EI, subjected to a uniformly distributed force q. Assuming a straight steel beam is installed at its lower edge, it is equivalent to applying a concentrated bending moment M at the end of the beam
\begin{equation}\label{eq:(example.yyl.23)}
M=F h
\end{equation}
According to the principle of minimum virtual work, when the virtual work $ W=M \theta=0 $ , i.e. $ \theta=0 $ , the bending moment $ M=\frac {L ^ 2  q} {12  \textrm {E}  I}$  of the simply supported beam is minimized, and the tendon configured at this time is optimal.
At this point, the total virtual work is equal to the virtual work done by $ q $ , which is $ w=\frac {L ^ 5  q ^ 2} {720  \textrm {E}  I}$  . When no tendon is applied, the virtual work $ w_0=\frac {L ^ 5  q ^ 2} {120  \textrm {E}  I}$  , and the optimal control index for virtual work is achieved
\begin{equation}\label{eq:(example.yyl.24)}
op=1-w_z/w_0=5/6
\end{equation}

From this, it can be seen that when determining the form of the tendon, $ op=5/6 $ is the optimal control tendon. In the design process, the arrangement of the tendon may not necessarily reach the optimal level, but it should be continuously optimized with $ op $ as the goal. This example presents the overall goal of configuring prestressed tendon for prestressed concrete beams, which can guide the design of prestressed tendons, maximize their effectiveness, reduce the amount of prestressed tendons used, and save a significant amount of engineering investment.

Next, we will discuss the principle of minimum virtual work when the tendon of a simply supported beam is in a curved form and passes through the centroid axis of the endpoint. At this point, the action of the tendon on the simply supported beam is equivalent to applying $ M (x) $ , and satisfies $ M (0)=M (L)=0 $ , which satisfies the functional variation lemma. Therefore, the control equation can adopt the strong form of the principle minimum virtual work control equation \ref{eq:(zxxgyl.1.8)}
\begin{figure}[h!] 
    \centering
    \includegraphics[width=0.75\textwidth,keepaspectratio]{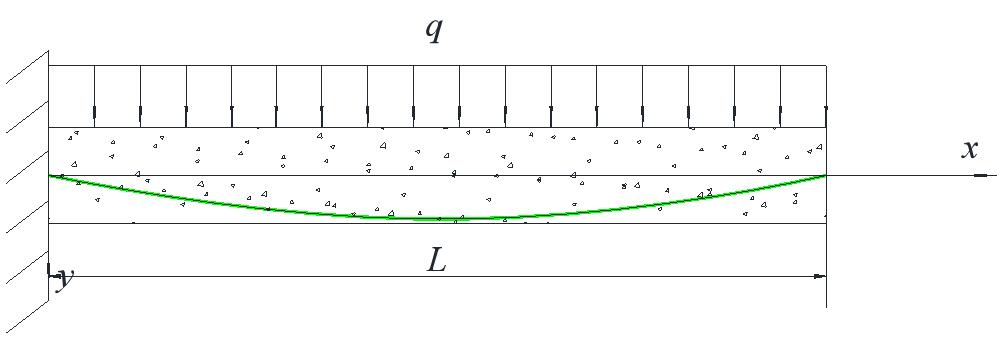} 
    \caption{Simply supported beam - bent tendon}
    \label{fig:Simply supported beam - bent tendon}
\end{figure}
If $ f=M '' $ , then $ V=M '$ . According to \ref{eq:(zxxgyl.2.2)}, the Euler equation for the beam is obtained as follows:
\begin{equation}\label{eq:(example.yyl.25)}
L_{y} - \left(L_{y_{,x}}\right)_{,x}+\left(L_{y_{,xx}}\right)_{,xx}
=EI\frac{d^2y''}{dx^2}-q+M ''=0
\end{equation}
By substituting the boundary conditions, the solution of the differential equation can be obtained
\begin{equation}\label{eq:(example.yyl.26)}
y=-\frac{x{\left(\mathrm {f}-q \right)}{\left(L^3 -2Lx^2 +x^3 \right)}}{24\textrm{E}I }
\end{equation}
According to the strong form of the governing equation \ref{eq:(zxxgyl.1.18)}, we can obtain $ y \equiv0 $ , which is
\begin{equation}\label{eq:(example.yyl.27)}
\frac{Lx^3 {\left(\mathrm {f}-q \right)}}{12\textrm{E}I }-\frac{x^4 {\left(\mathrm {f}-q \right)}}{24\textrm{E}I }-\frac{L^3 x{\left(\mathrm {f}-q \right)}}{24\textrm{E}I }=0
\end{equation}
Solved to obtain
\begin{equation}\label{eq:(example.yyl.28)}
M''=f=q
\end{equation}
Integrate once to obtain shear force, integrate twice to obtain bending moment, substitute into boundary conditions to solve as follows
\begin{equation}\label{eq:(example.yyl.29)}
M'=V=\int_{0,x} q dx=qx-\frac{1} {2}qL
\end{equation}
\begin{equation}\label{eq:(example.yyl.30)}
M=\int_{0,x} V dx =\frac{1} {2}qx (x-L)
\end{equation}
From the above equation, it can be seen that $ M $ is the negative value of the bending moment generated by the fixed load $ q $ , and the two exactly cancel each other out. At this point, the optimal control index $ op=1 $ . This example verifies the correctness of the strong form of the control conditions, which is different from the verification method in the previous section of the cantilever beam example. This example cannot use extreme value verification of the function, which further illustrates the correctness of the principle minimum virtual work and its wide application range.
\subsection {continuous beam with variable support position and reaction force}
\begin{figure}[h!] 
    \centering
    \includegraphics[width=0.75\textwidth,keepaspectratio]{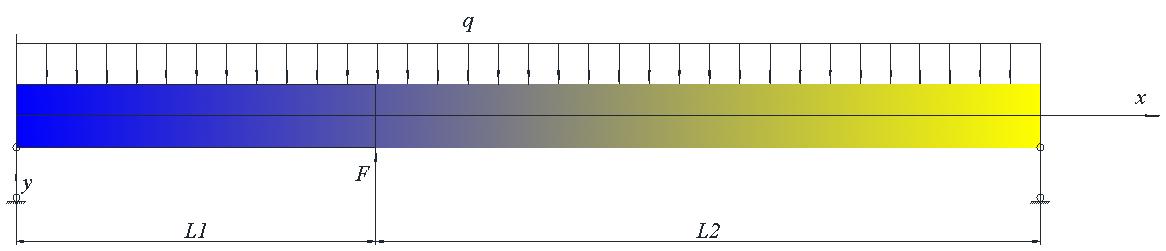} 
    \caption{Diagram of Continuous Beam with Variable Support Position and Reaction Force}
    \label{fig:lianxuliang}
\end{figure}

\numberwithin{equation}{subsection}
In order to verify the principle of minimum virtual work when the support position is variable, the reasonable position of the fulcrum and the optimal reaction force of a continuous beam with two spans are discussed below, as shown in Figure \ref{fig:lianxuliang}

According to the principle of minimum virtual work for continuous beams with sharp points, there are
\begin{multline}\label{eq:(example.zcwz.1)}
J = \sum_{r=1}^{q}\int_{x_{r-1}}^{x_{r}} L^{r}(x_r,y_r,y_{r,x},y_{r,xx},u_r,u_{r,x})dx
\\+\varphi^{r}(\bar {x}_ {r-1},\bar {y}_ {r-1},\bar {u}_ {r-1},\bar {y}_ {r-1,x},\bar {u}_ {r-1,x},\bar {x}_ {r},\bar {y}_ {r},\bar {u}_ {r},\bar {y}_ {r,x},\bar {u}_ {r,x})\\
=\int_{x_0}^{x_1} \frac{1} {2}EIy_1 ''^2dx-\int_{x_0}^ {x_1}qy_1 dx+\int_{x_1}^{x_2} \frac{1} {2}EIy_2 ''^2dx-\int_{x_1}^ {x_2}qy_2 dx-F\bar {y}_ {1}
\\=\int_{x_0}^{x_1}( \frac{1} {2}EIy_1 ''^2-qy_1)dx+\int_{x_1}^{x_2}( \frac{1} {2}EIy_2 ''^2-qy_2)dx-F\bar {y}_ {1}
\end{multline}
Then there are
\begin{equation} \label{eq:(example.zcwz.2)}    
L^1=\frac{1} {2}EIy_ {1,xx}^2-qy_1
\end{equation}
\begin{equation} \label{eq:(example.zcwz.3)}    
L^2=\frac{1} {2}EIy_ {2,xx}^2-qy_2
\end{equation}
\begin{equation} \label{eq:(example.zcwz.4)}    
\varphi=-Fy_{1}
\end{equation}
According to equation \ref{eq:(zxxgyl.3.2)}, the Euler equation is
\begin{equation}\label{eq:example.zcwz.5)}
EI\frac{d^2y_{1,xx}}{dx^2}-q=0
\end{equation}
\begin{equation}\label{eq:(example.zcwz.6)}
EI\frac{d^2y_{2,xx}}{dx^2}-q=0
\end{equation}
According to the \ref{eq:(zxxgyl.3.5)} boundary conditions
\begin{multline} \label{eq:(example.zcwz.7)}    
[(L^ {1}_ {y_{1,x}} -\left(L^ {1}_ {y_{1,xx}}\right)_{,x} )-(L^ {2}_ {y_{2,x}} -\left(L^ {2}_ {y_{2,xx}}\right)_{,x}) + \varphi_{y_{1} }]|_{x=x_1}  
\\=[0-(EIy_{1,xx})_{,x}]-[(]0-(EIy_{1,xx})]-F|_{x=x_1}  \\=-(EIy_{1,xx})_{,x}+(EIy_{2,xx})_{,x}-F|_{x=x_1}
0
\end{multline}
This equation is the boundary condition for shear force.
\\
According to \ref{eq:(zxxgyl.3.6)}
\begin{equation} \label{eq:(example.zcwz.8)}    
[L^ {1}_ {y_{1,xx}} - L^ {2}_ {y_{2,xx}}+\varphi_{y_{1,x}}]|_{x=x_1}   
=[EIy_ {1,xx}-EIy_ {2,xx}]|_{x=x_1}  
0
\end{equation}
This equation is the boundary condition for bending moment.
\\
When $ u_r $ is unconstrained
\begin{equation} \label{eq:(example.zcwz.9)}    
[(L^ {r}_ {u_r} - \left(L^ {r}_ {u_{r,x}}\right)_{,x})-(L^{r+1}_{u_{r+1}} - \left(L^{r+1}_{u_{r+1,x}}\right)_{,x})+\varphi_{u_1}]|_{x=x_r} 
\\=-\bar {y}_1=
0
\end{equation}

This equation is the condition for solving $ F $ .
According to $ y$  being a second-order continuous differentiable function, there is
\begin{equation} \label{eq:(example.zcwz.10)}    
y_1=y_2
=0,(x=x_1)
\end{equation}
\begin{equation} \label{eq:(example.zcwz.11)}    
y_{1,x}=y_{2,x}
=0,(x=x_1)
\end{equation}
\begin{equation} \label{eq:(example.zcwz.12)}    
y_{1,xx}=y_{2,xx}
=0,(x=x_1)
\end{equation}
When $ x=x_1 $ , according to \ref{eq:(zxxgyl.3.3)}, the boundary conditions are
\begin{multline} \label{eq:(example.zcwz.13)}    
[(L^ {1}-y_ {1,x} L_{y_{1,x}}   + y_{1,x} \left(L_{y_{1,xx}}\right)_{,x}
-y_{1,xx} L_{y_{1,xx}} 
-u_{1,x} L_{u_{1,x}})-
\\((L^ {2}-y_ {2,x} L_{y_{2,x}}   + y_{2,x} \left(L_{y_{,xx}}\right)_{2,x}
-y_{2,xx} L_{y_{2,xx}} 
-u_{2,x} L_{u_{2,x}})+\\
\varphi_{x_{1}}]|_{x=x_1}  
\\=\frac{1} {2}EIy_ {1,xx}^2-qy_1-0+y_{1,x} \left(EI{y_{1,xx}}\right)_{,x}-y_{1,xx} \left(EI{y_{1,xx}}\right)
\\-(\frac{1} {2}EIy_ {2,xx}^2-qy_2-0+y_{2,x} \left(EI{y_{2,xx}}\right)_{,x}-y_{2,xx} \left(EI{y_{2,xx}}\right))+0
\\=\frac{1} {2}EIy_ {1,xx}^2-qy_1+y_{1,x} \left(EI{y_{1,xx}}\right)_{,x}-y_{1,xx} \left(EI{y_{1,xx}}\right)
\\-\frac{1} {2}EIy_ {2,xx}^2+qy_2-y_{2,x} \left(EI{y_{2,xx}}\right)_{,x}+y_{2,xx} \left(EI{y_{2,xx}}\right)
\\=\frac{1} {2}EIy_ {1,xx}^2+y_{1,x} \left(EI{y_{1,xx}}\right)_{,x}-y_{1,xx} \left(EI{y_{1,xx}}\right)
\\-\frac{1} {2}EIy_ {2,xx}^2-y_{2,x} \left(EI{y_{2,xx}}\right)_{,x}+y_{2,xx} \left(EI{y_{2,xx}}\right)
=\\y_{1,x} \left(EI{y_{1,xx}}\right)_{,x}
-y_{2,x} \left(EI{y_{2,xx}}\right)_{,x}=0
\\=y_ {1,x}F=
0
\end{multline}
Since $ F $ is arbitrary, therefore
\begin{equation}\label{eq:(example.zcwz.14)}
y_{1,x}=0 
\end{equation}
This equation is for solving the boundary conditions of $ x_1 $ . According to the shear deformation formula $ \gamma=y_ {1,x}- \theta $ , where $ \gamma $ represents the shear deformation of the beam and $ \theta $ represents the angle of rotation of the beam section. Since shear deformation is ignored here, $ y_ {1, x}=\theta $ , i.e. the angle of rotation is equal to the first derivative of displacement. From this, it can be seen that when the turning angle is zero, that is, when the virtual work done by the shear force is zero, the internal energy is minimized. This formula further illustrates that when the angle of the middle support of the continuous beam is zero, the edge to middle span ratio is the most reasonable, which has important value for designing the span ratio of the continuous beam and can achieve the optimal span design of the continuous beam.
According to the hinge conditions at both ends
\begin{equation}\label{eq:(example.zcwz.15)}
y(0)=0 (x=0)
\end{equation}
\begin{equation}\label{eq:(example.zcwz.16)}
y''(0)=0 (x=0)
\end{equation}
\begin{equation}\label{eq:(example.zcwz.17)}
y(L)=0 (x=L)
\end{equation}
\begin{equation}\label{eq:(example.zcwz.18)}
y''(L)=0 (x=L)
\end{equation}
From this \ref{eq:(example.zcwz.7)}\text{-}\ref{eq:(example.zcwz.16)}, it can be seen that there are 10 boundary conditions that can solve two systems of differential equations.
Obtain the solutions of two differential equations as
\begin{equation}\label{eq:(example.zcwz.19)}
x1=\frac{L}{2}
\end{equation}
\begin{equation}\label{eq:(example.zcwz.20)}
y_1(x)=\frac{qx{{\left(L-2x\right)}}^2 {\left(L+4x\right)}}{384E_1 I_1 }
\end{equation}
\begin{equation}\label{eq:(example.zcwz.21)}
y_2(x)=\frac{q{{\left(L-2x\right)}}^2 {\left(5L^2 -9Lx+4x^2 \right)}}{384E_1 I_1 }
\end{equation}
\begin{equation}\label{eq:(example.zcwz.22)}
F=\frac{5Lq}{8}
\end{equation}
This example validates the correctness of the principle minimum virtual work derived in this article when the integral domain is variable, and provides a method for determining the reasonable span of continuous beam bridges.

Next, we will continue to discuss the issue of prestressed tendon configuration for continuous beams. As continuous beams are statically indeterminate structures, secondary reaction forces will be generated when configuring tendons. As shown in the figure, a secondary reaction force R will be generated at the midpoint. Since the vertical displacement of the support is zero, the virtual work generated by the vertical displacement is zero, and it will not increase the internal energy of the system, but will generate a angular displacement. When the virtual work F $ \theta $ of the angular displacement is negative, the internal energy will decrease, and when the virtual work of the angular displacement is positive, the internal energy will increase. Therefore, the arrangement of tendons should be optimized so that the angular displacement is zero. The discovery of this conclusion will directly guide the configuration of prestressed tendon in continuous beams, and macroscopically judge whether the configuration of prestressing is reasonable. The judgment method is simple and feasible, and can be directly judged by the size of the angle $ \theta $ . When the angle $ \theta =0 $ , the prestressed tendons are optimal.
\begin{figure}[h!] 
    \centering
    \includegraphics[width=0.75\textwidth,keepaspectratio]{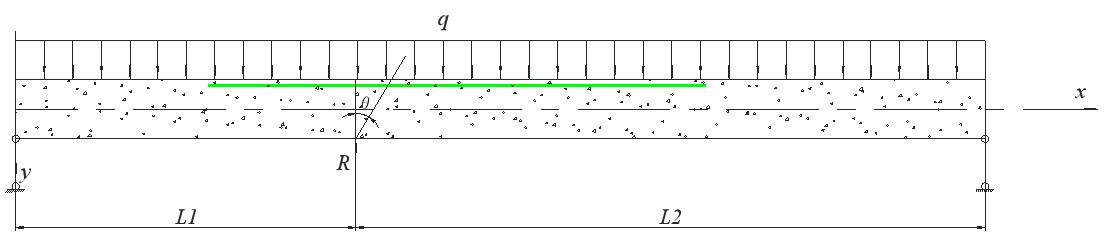} 
    \caption{Continuous beam tendon}
    \label{fig:Continuous beam tendon}
\end{figure}

\subsection {optimal span of continuous rigid frame beam bridge}
\numberwithin{equation}{subsection}
Next, we will discuss the optimal span position for continuous rigid frame bridges to ensure the most reasonable structural stress.
Treating the pier and beam as isolated bodies for stress analysis, it can be seen from the figure that the pier exerts axial force N, bending moment M, and shear force F on the beam, and vice versa. If the axial force $ N $ , bending moment $ M $ , and shear force $ F $ are considered as control loads, then according to the principle of minimum virtual work, it can be known that the vertical displacement, turning angle, and horizontal displacement at the junction of the pier and beam are all equal to zero, and the virtual work is zero. Therefore, the continuous rigid frame is subjected to the most reasonable force. Using this theory, when designing a continuous rigid frame bridge, finite element calculations can be directly performed to adjust the position of the bridge piers, so that the joint between the piers and beams can be moved to zero. This can be closely integrated with finite elements, making the design convenient and fast.
\begin{figure}[h!] 
    \centering
    \includegraphics[width=0.75\textwidth,keepaspectratio]{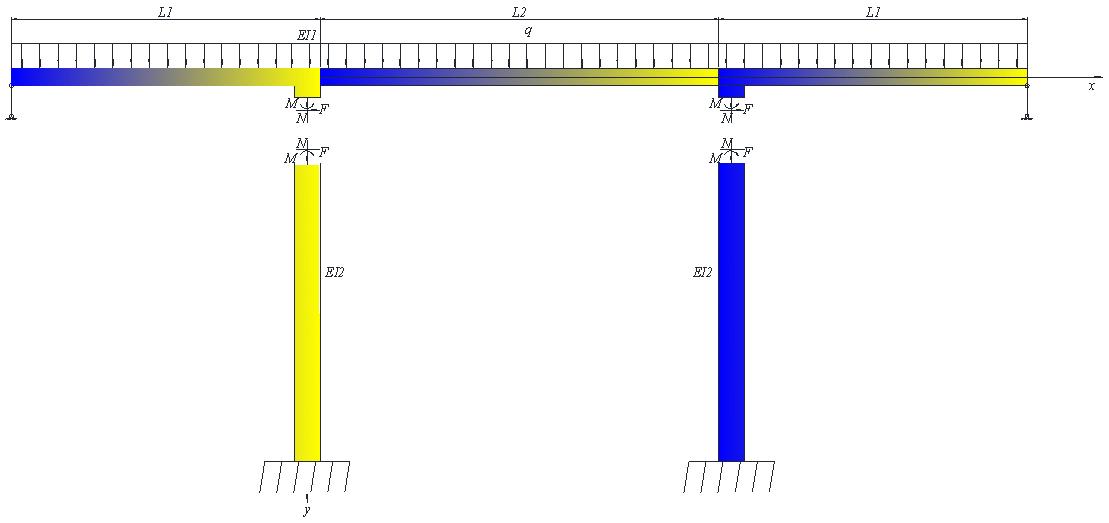} 
    \caption{Stress analysis of continuous rigid frame bridge}
    \label{fig:Stress analysis of continuous rigid frame bridge}
\end{figure}

In this example, the force analysis of the isolation body between the bridge pier and the beam is carried out, and the internal force of the bridge pier on the beam is regarded as the control load, which proves the correctness of the theory in this paper. Moreover, it further illustrates that internal forces can be regarded as control loads, demonstrating the wide applicability of the method proposed in this paper.

\subsection {Simplified Cable Tension Optimization of Cable stayed Bridges}
\numberwithin{equation}{subsection}
For simplicity, in order to explain the optimization of cable forces in cable-stayed bridges, the following simplified model is discussed. As shown in the figure, the structure consists of a vertical bridge tower and a main beam, which are connected by a cable. The main tower is fixed to the beam, and the length of the main beam is $ L_1 $ , the interface moment of inertia is $ I_1 $ , the elastic modulus is $ E_1 $ , the height of the main tower is $ L_2 $ , the interface moment of inertia is $ I2 $ , and the elastic modulus is $ E2 $ . The main beam is subjected to a uniformly distributed force $ q $ . The value of the cable force $ X_1 $ is calculated to minimize the internal energy of the structure (the sum of squares of bending moments is minimized, or the material is saved)
\begin{figure}[h!] 
    \centering
    \includegraphics[width=0.75\textwidth,keepaspectratio]{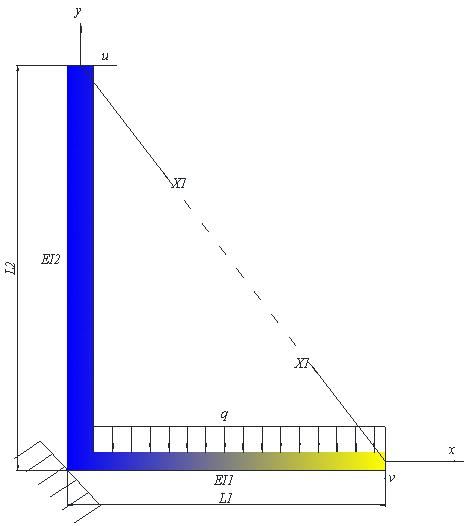} 
    \caption{Simplified diagram of cable-stayed bridge}
    \label{fig:Simplified diagram of cable-stayed bridge}
\end{figure} 
\begin{multline}\label{eq:(example.xlq.1)}
J = \int_{x_{0}}^{x_{1}} L(x,y,y_{,x},y_{,xx},u,u_{,x})dx+\varphi(x_{0},y_{0},u_{0},x_{1},y_{1},u_{1},y_{0,x},u_{0,x},y_{1,x},u_{1,x})\\
=\int_{0}^{L_1} (qv-\frac{1} {2}EIv ''^2)dx+\int_{0}^{L_2} -\frac{1} {2}EIu ''^2dy-X_1 sin\theta v_{L_1}+X_1 cos\theta u_{L_2}
\end{multline}
Then there are
\begin{equation} \label{eq:(example.xlq.2)}    
L^1=-\frac{1} {2}EIv ''^2+qv
\end{equation}
\begin{equation} \label{eq:(example.xlq.3)}    
L^2=-\frac{1} {2}EIu ''^2
\end{equation}
\begin{equation} \label{eq:(example.xlq.4)}    
\varphi=-X_1 sin\theta v_{L_1}+X_1 cos\theta u_{L_2}
\end{equation}
For the main beam, the Euler equation is
\begin{equation}\label{eq:(example.xlq.5)}
E_1 I_1\frac{d^2 v''}{dx^2}-q=0
\end{equation}
The boundary condition is
\begin{equation}\label{eq:(example.xlq.6)}
E_1 I_1 v''^2=0 (x=L_1)
\end{equation}
\begin{equation}\label{eq:(example.xlq.7)}
E_1 I_1\frac{dv''} {dx}-X_1  sin\theta=0 (x=L_1)
\end{equation}
According to the fixed end constraint conditions
\begin{equation}\label{eq:(example.xlq.8)}
v(0)=0 (x=0)
\end{equation}
\begin{equation}\label{eq:(example.xlq.9)}
v'(0)=0 (x=0)
\end{equation}
For the main tower, the Euler equation is
\begin{equation}\label{eq:(5.jxz.10)}
E_2 I_2\frac{d^2 u''}{dx^2}-q=0
\end{equation}
The boundary condition is
\begin{equation}\label{eq:(5.jxz.11)}
E_2 I_2 u''^2=0 (y=L_2)
\end{equation}
\begin{equation}\label{eq:(5.jxz.12)}
E_2 I_2\frac{du''}{dx}+X_1 cos\theta=0 (y=L_2)
\end{equation}
According to the fixed end constraint conditions
\begin{equation}\label{eq:(5.jxz.13)}
u(0)=0 (y=0)
\end{equation}
\begin{equation}\label{eq:(5.jxz.14)}
u'(0)=0 (y=0)
\end{equation}
According to the control conditions obtained from \ref{eq:(zxxgyl.2.7)}
\begin{equation}\label{eq:(5.jxz.15)}
-X_1 (sin\theta V_{L_1} +cos\theta u_{L_2})=0 
\end{equation}
This formula indicates that when the sum of displacements along the direction of $ X_1 $ at the control load $ X_1 $ is zero, i.e. the virtual work done by $ X_1 $ is zero, the internal energy is also minimized, and the cable force of the cable-stayed bridge is most reasonable.
From here \ref{eq:(example.xlq.4)}~\ref{eq:(example.xlq.8)}, the solution of the differential equation can be obtained as
\begin{equation}\label{eq:(5.jxz.16)}
v(x)=\frac{6q{L_1 }^2 x^2 -4qL_1 x^3 -12F\sin \left(a\right)L_1 x^2 +qx^4 +4F\sin \left(a\right)x^3 }{24E_1 I_1 }
\end{equation}
\begin{equation}\label{eq:(5.jxz.17)}
u(y)=-\frac{Fy^3 \cos \left(a\right)-3FL_2 y^2 \cos \left(a\right)}{6E_2 I_2 }
\end{equation}
At this moment, the virtual work is
\begin{multline}\label{eq:(5.jxz.18)}
W=\frac{F^2 {L_2 }^3 {\cos \left(a\right)}^2 }{6E_2 I_2 }-\frac{{L_1 }^3 {\left(20F^2 {\sin \left(a\right)}^2 -15FL_1 q\sin \left(a\right)+3{L_1 }^2 q^2 \right)}}{120E_1 I_1 }+
\\\frac{{L_1 }^4 q{\left(2L_1 q-5F\sin \left(a\right)\right)}}{40E_1 I_1 }-\frac{F\sin \left(a\right){\left(3{L_1 }^4 q-8F{L_1 }^3 \sin \left(a\right)\right)}}{24E_1 I_1 }
\end{multline}
According to the control condition \ref{eq:(zxxgyl.2.7)}
Solved to obtain
\begin{equation}\label{eq:(jxz.5.19)}
F=\frac{3E_2 I_2 {L_1 }^4 q\sin \left(a\right)}{8E_2 I_2 {L_1 }^3 {\sin \left(a\right)}^2 +8E_1 I_1 {L_2 }^3 {\cos \left(a\right)}^2 }
\end{equation}
At this point, the potential energy reaches its minimum value
\begin{equation}\label{eq:(example.xlq.20)}
W=\frac{{L_1 }^5 q^2 {\left(E_2 I_2 {L_1 }^3 {\sin \left(a\right)}^2 -16E_1 I_1 {L_2 }^3 {\sin \left(a\right)}^2 +16E_1 I_1 {L_2 }^3 \right)}}{640E_1 I_1 {\left(E_2 I_2 {L_1 }^3 {\sin \left(a\right)}^2 -E_1 I_1 {L_2 }^3 {\sin \left(a\right)}^2 +E_1 I_1 {L_2 }^3 \right)}}
\end{equation}
Next, using the extremum condition of the function to verify the above conclusion, the derivative of \ref{eq:(example.xlq.20)} is obtained
\begin{multline}\label{eq:(example.xlq.21)}
W'(F)= \frac{F{L_2 }^3 {\cos \left(a\right)}^2 }{3E_2 I_2 }-\frac{{L_1 }^3 {\left(40F{\sin \left(a\right)}^2 -15L_1 q\sin \left(a\right)\right)}}{120E_1 I_1 }-\frac{{L_1 }^4 q\sin \left(a\right)}{8E_1 I_1 }-
\\\frac{\sin \left(a\right){\left(3{L_1 }^4 q-8F{L_1 }^3 \sin \left(a\right)\right)}}{24E_1 I_1 }+\frac{F{L_1 }^3 {\sin \left(a\right)}^2 }{3E_1 I_1 }
\end{multline}
\begin{equation}\label{eq:(example.xlq.22)}
W''(F)=\frac{{L_2 }^3 {\cos \left(a\right)}^2 }{3E_2 I_2 }+\frac{{L_1 }^3 {\sin \left(a\right)}^2 }{3E_1 I_1 }
\end{equation}
From \ref{eq:(example.xlq.22)}, it can be obtained that
\begin{equation}\label{eq:(example.xlq.23)}
F=\frac{3E_2 I_2 {L_1 }^4 q\sin \left(a\right)}{8E_2 I_2 {L_1 }^3 {\sin \left(a\right)}^2 +8E_1 I_1 {L_2 }^3 {\cos \left(a\right)}^2 }
\end{equation}
Consistent with \ref{eq:(jxz.5.19)}.
This example further illustrates that the theory of using the principle minimum virtual work proposed in this article is correct and can theoretically solve the cable force optimization problem of cable-stayed bridges.

Below is a further discussion on the cable force optimization methods for cable-stayed bridges. The commonly used cable force optimization methods in cable-stayed bridge design include zero displacement method, rigid support continuous beam method, zero support reaction method, minimum bending energy method, mathematical optimization algorithm, etc. Among them, the zero displacement method, the rigid support continuous beam method, and the zero support reaction force method all suggest the influence matrix of the displacement or reaction force of the fulcrum through cable force, and then iteratively calculate the cable force at zero displacement. The minimum bending energy method aims to minimize the bending moment of the bridge tower and main beam. Currently, the method of reducing the bending stiffness of the structure is commonly used to achieve the minimum bending energy of cable-stayed bridges. Mathematical programming algorithms mainly set objective functions, such as minimum bending energy, minimum displacement, etc., and suggest the corresponding influence matrix between the cable force and it, continuously optimizing and solving to obtain the optimal cable force method.

The zero displacement method, rigid support continuous beam method, and zero support reaction force method are actually all aimed at achieving zero displacement at the junction of the cable and beam. In fact, these three methods are special cases of this method, which essentially only consider the bending strain energy of the main beam, zero displacement at the support constraint, and zero virtual work done on the beam body. The bending energy method and mathematical programming approach aim to minimize the bending energy or other physical quantities of the bridge tower and main beam. By influencing the matrix and iterating, the optimal cable force is obtained, which has been widely applied in the optimization of cable forces in cable-stayed bridges. However, it has not been explained from the essence of mechanics. The method used in this article explains from the principle of mechanics, that is, the essence is that when the virtual work of the cable force on the tower and beam is zero, the internal energy of the main tower and main beam is minimized. In specific calculations, it can be well combined with finite elements, and the cable force of the cable-stayed bridge can be intuitively viewed through displacement to determine whether it is reasonable.
In this example, for simplicity, only bending deformation was considered, but in reality, axial deformation and shear deformation are also feasible.

\subsection {Optimization of Suspension Rod Force for Arch Bridges}
\numberwithin{equation}{subsection}
The optimal suspension force for a tied arch bridge is discussed below, which ensures the most reasonable stress on the tie beams and arch ribs. The tied arch bridge is connected to the main beam by 7 cables, and a reasonable suspension force design can make the stress on the arch ribs and main beam more uniform, saving more engineering materials.
\begin{figure}[h!] 
    \centering
    \includegraphics[width=0.75\textwidth,keepaspectratio]{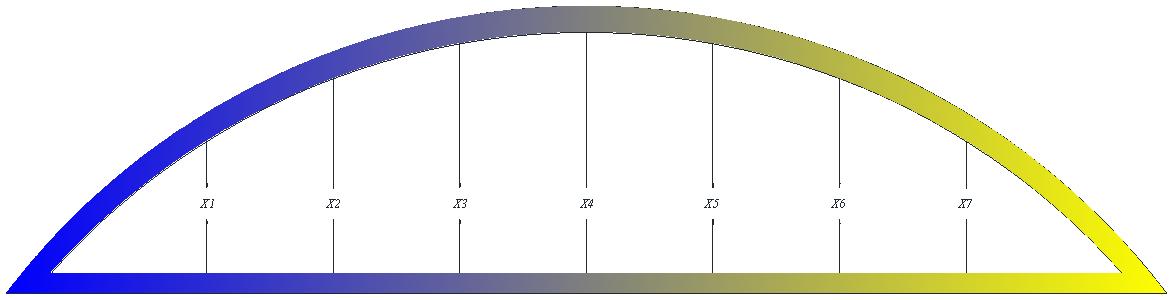} 
    \caption{Schematic diagram of arch bridge structure}
    \label{fig:Schematic diagram of arch bridge structure}
\end{figure}
According to the principle of minimum virtual work, it can be concluded that adjusting the tension of the suspension rod $ X_1 \text{-} X_7 $ so that $ v_i=v'_i $ results in the most reasonable stress on the arch ribs and tie beams. Using this theory, when designing a tied arch bridge, finite element calculations can be directly performed to adjust the position of the cable force, so that the deformation $ v'_i $ of the beam body is equal to the deformation $ v_i $ of the arch ribs. This is equivalent to attaching deformation conditions to the finite element equation.

\begin{figure}[h!] 
    \centering
    \includegraphics[width=0.75\textwidth,keepaspectratio]{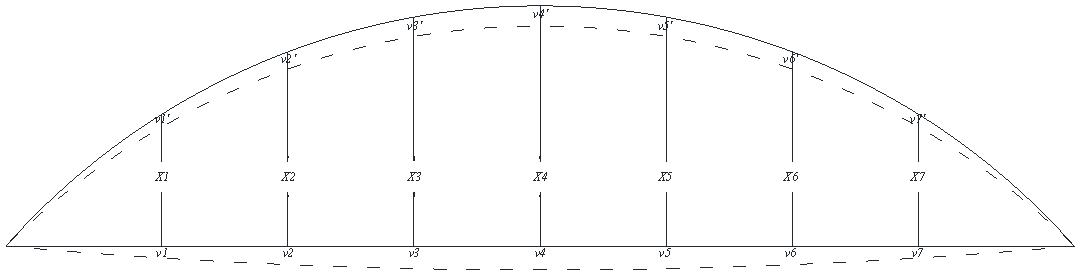} 
    \caption{Schematic diagram of arch bridge displacement}
    \label{fig:Schematic diagram of arch bridge displacement}
\end{figure}
This example is mainly used to illustrate the use of the method proposed in this article for optimizing the suspension force of arch bridges. The traditional methods for optimizing the suspension force of arch bridges include rigid support continuous beam method, zero displacement method, minimum bending energy method, etc. These methods are basically consistent with the cable force optimization method of cable-stayed bridges. The method proposed in this paper can simultaneously minimize the internal energy of the arch ribs and main beams, which is more reasonable than the commonly used arch bridge suspension force optimization method. Moreover, it is explained in principle and can be well combined with finite element method to visually check whether the suspension force of the arch bridge is reasonable through displacement.
\subsection {One dimensional rod element (with only stationary values and no extremum)}
The next few sections mainly discuss the problem of rod elements, with the aim of proving that the method provided in this paper can be used for topology optimization. Assuming there is a one-dimensional rod element, as shown in the figure
\begin{figure}[h!] 
    \centering
    \includegraphics[width=0.75\textwidth,keepaspectratio]{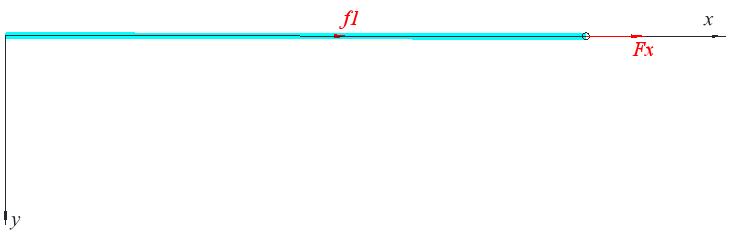} 
    \caption{Simplified truss}
    \label{fig:Simplified truss}
\end{figure}

Its virtual work functional is
\begin{equation}\label{eq:(example.oneRod.1)}
J=\int_{0}^{x_1} (f_1 u_1 -\frac{1} {2}EA_1(u_{1,1})^2)dt+F_1 \bar {u}_ {1}
\end{equation}

Variational analysis of the above equation yields
\begin{multline}\label{eq:(example.oneRod.2)}
\delta J=\int_{0}^{x_1}( -EA_1 u_{1,1} \delta u_{1,1}+f_1 \delta u_1 )dt+F_{1} \delta \bar {u}_ {1}+( -\frac{1} {2}EA_1(u_{1,1})^2+f_1 u_1 )\delta x
\\=\int_{0}^{x_1}[ -EA_1 ((u_{1,1} \delta u_{1})_{,1}-u_{1,1} \delta u_{1,11})+f_1 \delta u_1 ]dt+F_{1} \delta \bar {u}_ {1}
+( -\frac{1} {2}EA_1(u_{1,1})^2+f_1 u_1 )\delta x
\\=\int_{0}^{x_1}[ EA_1 u_{1,11} \delta u_{1}+f_1 \delta u_1 ]dt
-EA_1 u_{1,1} \delta u_{1}|_{L_1}+F_{1} \delta \bar {u}_ {1}+( -\frac{1} {2}EA_1(u_{1,1})^2+f_1 u_1 )\delta x
\end{multline}

Euler equation can be obtained from the above formula.
\begin{equation} \label{eq:(example.oneRod.3)}    
EA_1 u_{1,11} +f_1  =0
\end{equation}

Since $\delta u_ {1}$ is related to the endpoint $x_{1}$, that is, $ \ delta u _ {i} = \ delta u _ {1} (x _ {1}) $,there is.
\begin{equation} \label{eq:(example.oneRod.6)}    
\delta u_ {1}|_{x_{1}}=\delta \bar{u}_ {1}-u_ {1,1}\delta x
\end{equation}

From the condition $\delta J=0$, there is
\begin{multline}\label{eq:(example.oneRod.7)}
0=-EA_1 u_{1,1} \delta u_{1}|_{x_1}+F_{1} \delta \bar {u}_ {1}+( -\frac{1} {2}EA_1(u_{1,1})^2+f_1 u_1 )\delta x
\\=-EA_1 u_{1,1} (\delta \bar{u}_ {1}-u_ {1,1}\delta x_{1'}) 
+( -\frac{1} {2}EA_1(u_{1,1})^2+f_1 u_1 ) \delta x_{1'} +F_{1} \delta \bar {u}_ {1}
\\=(F_{1} -EA_1 u_{1,1}  )\delta \bar {u}_ {1}+[EA_1 u_{1,1}  u_ {1,1}+(-\frac{1}{2}EA_1(u_{1,1})^2+f_1 u_1 )
]\delta x_{1}
\\=(F_{1} -EA_1 u_{1,1}  )\delta \bar {u}_ {1'}+[\frac{1}{2}EA_1(u_{1,1})^2+f_1 u_1 )
]\delta x_{1}
\end{multline}

When the endpoints are fixed, that is, $\delta x_{1}=0$, $\delta u_{1}$ is arbitrary, which is an effective equilibrium condition from the above formula.
\begin{equation} \label{eq:(example.oneRod.8)}    
F_{1} -EA_1 u_{1,1}=0
\end{equation}

When the end point is movable, in addition to the boundary conditions of force, the following virtual work boundary conditions need to be satisfied because $\delta x_{1}$ is arbitrary.
\begin{equation}\label{eq:(example.oneRod.10)}
\frac{1}{2}EA_1(u_{1,1})^2+f_1 \bar{u}_1
=0
\end{equation}

According to the above formula, $f_1 \bar{u}_1<0$, and when $f_1>0$, there is $\bar{u}_1<0$.
\subsection {Two bar elements (variational method in mixed coordinates)
}\label{subsec:twoRod}
Suppose there are two one-dimensional bar elements that intersect at one point, and the coordinates of the intersection point between bar 1 and bar 2 are $(x_{1'},y_{1'})$. Suppose that the axial stiffness of bars 1 and 2 are $ EA _ 1 and EA _ 2 $ respectively, and they are subjected to distributed loads $ F _ 1 and F _ 2 $ respectively, and concentrated loads are applied at the intersection points.
\begin{figure}[h!] 
    \centering
    \includegraphics[width=0.75\textwidth,keepaspectratio]{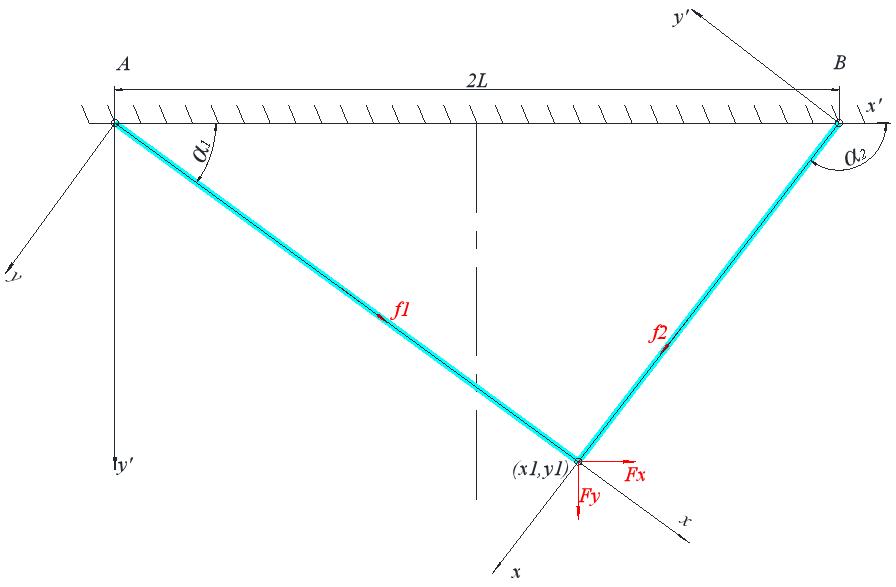} 
    \caption{two truss}
    \label{fig:two truss}
\end{figure}

Its virtual work equation is
\begin{multline}\label{eq:(example.twoRod.1)}
J=-\int_{0}^{L_1} \frac{1} {2}EA_1(u^{\tilde{1}}_{1,1})^2dt+\int_{0}^ {L_1}f^{\tilde{1}}_1 u^{\tilde{1}}_1 dt-\int_{0}^{L_2} \frac{1} {2}EA_2 (u^{\tilde{1}}_{2 ,1})^2dt
\\+\int_{0}^ {L_2}f^{\tilde{2}}_1 u^{\tilde{2}}_1 dt+F_{1'}\bar {u_{1'}}+F_{2'}\bar {u'}_ {1}
\\=\int_{0}^{L_1}( -\frac{1} {2}EA_1(u^{\tilde{1}}_{1,1})^2+f^{\tilde{1}}_1 u^{\tilde{1}}_1 )dt+\int_{0}^{L_2}( -\frac{1} {2}EA_2 (u^{\tilde{1}}_{2 ,1})^2+f^{\tilde{2}}_1 u^{\tilde{2}}_1 )dt+F_1 \bar {u'}_ {1}+F_{2'} \bar {u'}_{2}
\end{multline}

By using the mixed variation in local coordinate system and global coordinate system, substituting \ref{eq:(example.twoRod.1)}, we get
\begin{multline}\label{eq:(example.twoRod.2)}
\delta J=\delta [\int_{0}^{L_1}( -\frac{1} {2}EA_1(u^{\tilde{1}}_{1,1})^2+f^{\tilde{1}}_1 u^{\tilde{1}}_1 )dt+\int_{0}^{L_2}( -\frac{1} {2}EA_2 (u^{\tilde{1}}_{2 ,1})^2+f^{\tilde{2}}_1 u^{\tilde{2}}_1 )dt+F_{1'} \bar {u'}_ {1}+F_{2'} \bar {u'}_{2}]
\\=\int_{0}^{L_1}( -EA_1 u^{\tilde{1}}_{1,1} \delta u^{\tilde{1}}_{1,1}+f^{\tilde{1}}_1 \delta u^{\tilde{1}}_1 )dt+\int_{0}^{L_2} (-EA_2 u^{\tilde{1}}_{2 ,1} \delta u^{\tilde{1}}_{2 ,1} +f^{\tilde{2}}_1 \delta u^{\tilde{2}}_1 )dt+F_{1'} \delta \bar {u}_ {1'}+F_{2'} \delta \bar {u}_{2'}
\\+( -\frac{1} {2}EA_1(u^{\tilde{1}}_{1,1})^2+f^{\tilde{1}}_1 u^{\tilde{1}}_1 )\delta t_1+( -\frac{1} {2}EA_2 (u^{\tilde{1}}_{2 ,1})^2+f^{\tilde{2}}_1 u^{\tilde{2}}_1 ) \delta t_2
\\=\int_{0}^{L_1}[ -EA_1 ((u^{\tilde{1}}_{1,1} \delta u^{\tilde{1}}_{1})_{,1}-u^{\tilde{1}}_{1,1} \delta u^{\tilde{1}}_{1,11})+f^{\tilde{1}}_1 \delta u^{\tilde{1}}_1 ]dt
\\+\int_{0}^{L_2} [-EA_2 ((u^{\tilde{2}}_{1,1} \delta u^{\tilde{2}}_{1})_{,1}-u^{\tilde{2}}_{1,1} \delta u^{\tilde{2}}_{1,11}) +f^{\tilde{2}}_1 \delta u^{\tilde{2}}_1 ]dt
\\+F_{1'} \delta \bar {u}_ {1'}+F_{2'} \delta \bar {u}_{2'}
+( -\frac{1} {2}EA_1(u^{\tilde{1}}_{1,1})^2+f^{\tilde{1}}_1 u^{\tilde{1}}_1 )\delta t_1+( -\frac{1} {2}EA_2 (u^{\tilde{1}}_{2 ,1})^2+f^{\tilde{2}}_1 u^{\tilde{2}}_1 ) \delta t_2
\\=\int_{0}^{L_1}[ EA_1 u^{\tilde{1}}_{1,11} \delta u^{\tilde{1}}_{1}+f^{\tilde{1}}_1 \delta u^{\tilde{1}}_1 ]dt+\int_{0}^{L_2} [EA_2u^{\tilde{2}}_{1,11} \delta u^{\tilde{2}}_{1} +f^{\tilde{2}}_1 \delta u^{\tilde{2}}_1 ]dt
\\-EA_1 u^{\tilde{1}}_{1,1} \delta u^{\tilde{1}}_{1}|_{L_1}-EA_2 u^{\tilde{2}}_{1,1} \delta u^{\tilde{2}}_{1}|_{L_2}
\\+F_{1'} \delta \bar {u}_ {1'}+F_{2'} \delta \bar {u}_{2'}
+( -\frac{1} {2}EA_1(u^{\tilde{1}}_{1,1})^2+f^{\tilde{1}}_1 u^{\tilde{1}}_1 )\delta t_1+( -\frac{1} {2}EA_2 (u^{\tilde{1}}_{2 ,1})^2+f^{\tilde{2}}_1 u^{\tilde{2}}_1 ) \delta t_2
\end{multline}

Euler equation can be obtained from the above formula.
\begin{equation} \label{eq:(example.twoRod.3)}    
EA_1 u^{\tilde{1}}_{1,11} +f^{\tilde{1}}_1  =0
\end{equation}
\begin{equation} \label{eq:(example.twoRod.4)}    
EA_2 u^{\tilde{2}}_{1,11} +f^{\tilde{2}}_1  =0
\end{equation}

Projecting $\delta u^{\tilde{i}}_{1}|_{L_1}$ in the global coordinate system
\begin{equation} \label{eq:(example.twoRod.5)}    
\delta u^{\tilde{i}}_{1}|_{L_1}=cos \alpha_i \delta u_ {1'}|_{(x_{1'},y_{1'})} +sin \alpha_i \delta u_ {2'}|_{(x_{1'},y_{1'})} ,i=1,2
\end{equation}

Because $\delta u_ {i'}$ is related to the intersection $(x_{1'},y_{1'})$, that is, \\$\delta u_ {i'}=\delta u_{i'}(x_{1'},y_{1'})$,then there is
\begin{equation} \label{eq:(example.twoRod.6)}    
\delta u_ {i'}|_{(x_{1'},y_{1'})}=\delta \bar{u}_ {i'}-u_ {i',x_{1'}}\delta x_{1'}-u_ {i',y_{1'}}\delta y_{1'},i=1,2
\end{equation}

Then, assuming the condition $\ delta J=0 $, project all components onto the global coordinate system
\begin{multline}\label{eq:(example.twoRod.7)}
0=-EA_1 u^{\tilde{1}}_{1,1} \delta u^{\tilde{1}}_{1}|_{L_1}-EA_2 u^{\tilde{2}}_{1,1} \delta u^{\tilde{2}}_{1}|_{L_2}
\\+F_{1'} \delta \bar {u}_ {1'}+F_{2'} \delta \bar {u}_{2'}
+( -\frac{1} {2}EA_1(u^{\tilde{1}}_{1,1})^2+f^{\tilde{1}}_1 u^{\tilde{1}}_1 )\delta t_1+( -\frac{1} {2}EA_2 (u^{\tilde{1}}_{2 ,1})^2+f^{\tilde{2}}_1 u^{\tilde{2}}_1 ) \delta t_2
\\=-EA_1 u^{\tilde{1}}_{1,1} (\delta \bar{u}^{\tilde{1}}_{1}-{u}^{\tilde{1}}_{1,1}|_{L_1} \delta t_1)-EA_2 u^{\tilde{2}}_{1,1} (\delta \bar{u}^{\tilde{2}}_{1}-{u}^{\tilde{1}}_{2,1}|_{L_1} \delta t_2)
+F_{1'} \delta \bar {u}_ {1'}+F_{2'} \delta \bar {u}_{2'}
\\+( -\frac{1} {2}EA_1(u^{\tilde{1}}_{1,1})^2+f^{\tilde{1}}_1 u^{\tilde{1}}_1 )\delta t_1+( -\frac{1} {2}EA_2 (u^{\tilde{1}}_{2 ,1})^2+f^{\tilde{2}}_1 u^{\tilde{2}}_1 ) \delta t_2
\\=-EA_1 u^{\tilde{1}}_{1,1} (\delta \bar{u}^{\tilde{1}}_{1})-EA_2 u^{\tilde{2}}_{1,1} (\delta \bar{u}^{\tilde{2}}_{1})
+F_{1'} \delta \bar {u}_ {1'}+F_{2'} \delta \bar {u}_{2'}
\\+( \frac{1} {2}EA_1(u^{\tilde{1}}_{1,1})^2+f^{\tilde{1}}_1 u^{\tilde{1}}_1 )\delta t_1+( \frac{1} {2}EA_2 (u^{\tilde{1}}_{2 ,1})^2+f^{\tilde{2}}_1 u^{\tilde{2}}_1 ) \delta t_2
\\=-EA_1 u^{\tilde{1}}_{1,1} (cos \alpha_1 (\delta \bar{u}_ {1'}) +sin \alpha_1 (\delta \bar{u}_ {2'})-EA_2 u^{\tilde{2}}_{1,1} (cos \alpha_2 (\delta \bar{u}_ {1'})+sin \alpha_2 (\delta \bar{u}_ {2'}))
\\+( \frac{1} {2}EA_1(u^{\tilde{1}}_{1,1})^2+f^{\tilde{1}}_1 u^{\tilde{1}}_1 )\delta t_1+( \frac{1} {2}EA_2 (u^{\tilde{1}}_{2 ,1})^2+f^{\tilde{2}}_1 u^{\tilde{2}}_1 ) \delta t_2
\\+F_{1'} \delta \bar {u}_ {1'}+F_{2'} \delta \bar {u}_{2'}
\\=(F_{1'} -EA_1 u^{\tilde{1}}_{1,1} cos \alpha_1 -EA_2 u^{\tilde{2}}_{1,1} cos \alpha_2 )\delta \bar {u}_ {1'}
+(F_{2'} -EA_1 u^{\tilde{1}}_{1,1} sin \alpha_1 
\\-EA_2 u^{\tilde{2}}_{1,1} sin \alpha_2 )\delta \bar {u}_ {2'}
\\+( \frac{1} {2}EA_1(u^{\tilde{1}}_{1,1})^2+f^{\tilde{1}}_1 u^{\tilde{1}}_1 )\delta t_1+( \frac{1} {2}EA_2 (u^{\tilde{1}}_{2 ,1})^2+f^{\tilde{2}}_1 u^{\tilde{2}}_1 ) \delta t_2
\end{multline}

When the intersection point of two rods is fixed, i.e. $\delta x_ {1 '}=\delta y_ {1'}=0 $, $\delta u_ {1 '}=\delta u_ {2'}=0$ is arbitrary, and the equilibrium condition for the force can be obtained from the above equation
\begin{equation} \label{eq:(example.twoRod.8)}    
F_{1'} -EA_1 u^{\tilde{1}}_{1,1} cos \alpha_1 -EA_2 u^{\tilde{2}}_{1,1} cos \alpha_2=0
\end{equation}
\begin{equation} \label{eq:(example.twoRod.9)}    
F_{2'} -EA_1 u^{\tilde{1}}_{1,1} sin \alpha_1 -EA_2 u^{\tilde{2}}_{1,1} sin \alpha_2=0
\end{equation}

Substitute $\ref{eq:(example.twoRod.7)} $to obtain
\begin{equation}\label{eq:(example.twoRod.10)}
0=( \frac{1} {2}EA_1(u^{\tilde{1}}_{1,1})^2+f^{\tilde{1}}_1 u^{\tilde{1}}_1 )\delta t_1+( \frac{1} {2}EA_2 (u^{\tilde{1}}_{2 ,1})^2+f^{\tilde{2}}_1 u^{\tilde{2}}_1 ) \delta t_2
\end{equation}

Because $t_i=t_i(u_{1'},u_{2'})$,so
\begin{multline} \label{eq:(example.twoRod.11)}    
\delta t_i=t_{i,u_{1'}}\delta u_{1'}+t_{i,u_{2'}}\delta u_{2'}
\\=t_{i,u_{1'}}(u_{1',x_1}\delta x_{1'}+u_{1',y_1}\delta y_{1'})+t_{i,u_{2'}}(u_{2',x_1}\delta x_{1'}+u_{2',y_1}\delta y_{1'})
\\=(t_{i,u_{1'}}u_{1',x_1}+t_{i,u_{2'}}u_{2',x_1})\delta x_{1'}+(t_{i,u_{1'}}u_{1',y_1}+t_{i,u_{2'}}u_{2',y_1}) \delta y_{1'}
\end{multline}
$\ delta t_i $represents the $x$coordinate in the actual configuration of the unit.
Substitute $\ref{eq:(example.twoRod.10)}$to obtain
\begin{multline}\label{eq:(example.twoRod.12)}
(\frac{1} {2}EA_1(u^{\tilde{1}}_{1,1})^2+f^{\tilde{1}}_1 u^{\tilde{1}}_1)[(t_{1,u_{1'}}u_{1',x_1}+t_{1,u_{2'}}u_{2',x_1})\delta x_{1'}+(t_{1,u_{1'}}u_{1',y_1}+t_{1,u_{2'}}u_{2',y_1}) \delta y_{1'}]
\\+( \frac{1} {2}EA_2 (u^{\tilde{1}}_{2 ,1})^2+f^{\tilde{2}}_1 u^{\tilde{2}}_1 )[(t_{2,u_{1'}}u_{1',x_1}+t_{2,u_{2'}}u_{2',x_1})\delta x_{1'}+(t_{2,u_{1'}}u_{1',y_1}+t_{2,u_{2'}}u_{2',y_1}) \delta y_{1'}]
\\=[(\frac{1} {2}EA_1(u^{\tilde{1}}_{1,1})^2+f^{\tilde{1}}_1 u^{\tilde{1}}_1 )(t_{1,u_{1'}}u_{1',x_1}+t_{1,u_{2'}}u_{2',x_1})+(\frac{1} {2}EA_2 (u^{\tilde{1}}_{2 ,1})^2+f^{\tilde{2}}_1 u^{\tilde{2}}_1)
\\(t_{2,u_{1'}}u_{1',x_1}+t_{2,u_{2'}}u_{2',x_1})]\delta x_{1'}
+[(\frac{1} {2}EA_1(u^{\tilde{1}}_{1,1})^2+f^{\tilde{1}}_1 u^{\tilde{1}}_1)(t_{1,u_{1'}}u_{1',y_1}+t_{2,u_{2'}}u_{2',y_1})
\\+( \frac{1} {2}EA_1(u^{\tilde{1}}_{1,1})^2+f^{\tilde{1}}_1 u^{\tilde{1}}_1 )t_{2,u_{2'}}u_{2',y_1})]\delta y_{1'}=0
\end{multline}

When two intersection points are movable, in addition to satisfying the boundary conditions of force, the following virtual work boundary conditions need to be satisfied because $\delta x_{1'},\delta y_{1'}$are arbitrary
\begin{multline}\label{eq:(example.twoRod.13)}
( \frac{1} {2}EA_1(u^{\tilde{1}}_{1,1})^2+f^{\tilde{1}}_1 u^{\tilde{1}}_1 )(t_{1,u_{1'}}u_{1',x_{1'}}+t_{1,u_{2'}}u_{2',x_{1'}})
\\+( \frac{1} {2}EA_2 (u^{\tilde{2}}_{1 ,1})^2+f^{\tilde{2}}_1 u^{\tilde{2}}_1 )(t_{2,u_{1'}}u_{1',x_{1'}}+t_{2,u_{2'}}u_{2',x_{1'}})
\\=[( \frac{1} {2}EA_1(u^{\tilde{1}}_{1,1})^2+f^{\tilde{1}}_1 u^{\tilde{1}}_1 )t_{1,u_{1'}}+( \frac{1} {2}EA_2 (u^{\tilde{2}}_{1 ,1})^2+f^{\tilde{2}}_1 u^{\tilde{2}}_1 )t_{2,u_{1'}}]u_{1',x_{1'}}
\\+[( \frac{1} {2}EA_1(u^{\tilde{1}}_{1,1})^2+f^{\tilde{1}}_1 u^{\tilde{1}}_1 )t_{1,u_{2'}}+(\frac{1} {2}EA_2 (u^{\tilde{2}}_{1 ,1})^2+f^{\tilde{2}}_1 u^{\tilde{2}}_1 )t_{2,u_{2'}}]u_{2',x_{1'}}
=0
\end{multline}
\begin{multline}\label{eq:(example.twoRod.14)}
(\frac{1} {2}EA_1(u^{\tilde{1}}_{1,1})^2+f^{\tilde{1}}_1 u^{\tilde{1}}_1 )(t_{1,u_{1'}}u_{1',y_{1'}}+t_{1,u_{2'}}u_{2',y_{1'}})
 \\+( \frac{1} {2}EA_2(u^{\tilde{2}}_{1,1})^2+f^{\tilde{2}}_1 u^{\tilde{2}}_1 )(t_{2,u_{1'}}u_{1',y_{1'}}+t_{2,u_{2'}}u_{2',y_{1'}})
 \\=[( \frac{1} {2}EA_1(u^{\tilde{1}}_{1,1})^2+f^{\tilde{1}}_1 u^{\tilde{1}}_1 )t_{1,u_{1'}}+( \frac{1} {2}EA_2 (u^{\tilde{2}}_{1 ,1})^2+f^{\tilde{2}}_1 u^{\tilde{2}}_1 )t_{2,u_{1'}}]u_{1',y_{1'}}
\\+[( \frac{1} {2}EA_1(u^{\tilde{1}}_{1,1})^2+f^{\tilde{1}}_1 u^{\tilde{1}}_1 )t_{1,u_{2'}}+(\frac{1} {2}EA_2 (u^{\tilde{2}}_{1 ,1})^2+f^{\tilde{2}}_1 u^{\tilde{2}}_1 )t_{2,u_{2'}}]u_{2',y_{1'}}
=0
\end{multline}
From the above two equations, it can be seen that the displacement rate of the virtual work boundary, weighted by the virtual work density of the boundary, is equal to zero.

According to the displacement boundary, for element 1, there is (in the local coordinate system)
\begin{equation} \label{eq:(example.twoRod.15)}    
u^{\tilde{1}}_{1}(x=0)=0
\end{equation}
For Unit 2, there is (in local coordinate system)
\begin{equation} \label{eq:(example.twoRod.16)}    
u^{\tilde{2}}_{1}(x=0)=0
\end{equation}


For the Euler equation \ref{eq:(example.twoRod.3)},\ref{eq:(example.twoRod.4)} has 4 unknowns, plus $x_{1'},x_{2'}$ unknown, for a total of 6 unknowns. According to \ref{eq:(example.twoRod.10)} \text{-} \ref{eq:(example.twoRod.15)} having 6 boundary conditions, the solutions to all equations can be obtained.

Considering a special case, when two rods are symmetric about $A_1=A_2,F_{1'}=f^{\tilde{2}}_1=f^{\tilde{2}}_1=0$, and $delta x_ {1 '}=0 $, then $cos \alpha_2=-cos \alpha_1,sin \alpha_2=sin \alpha_1$, So from \ref{eq:(example.twoRod.7)}, it can be concluded that the boundary condition of the force becomes
\begin{equation} \label{eq:(example.twoRod.17)}    
-EA_1 u^{\tilde{1}}_{1,1} cos \alpha_1 -EA_1 u^{\tilde{2}}_{1,1} cos \alpha_2=
-EA_1 u^{\tilde{1}}_{1,1} cos \alpha_1 +EA_1 u^{\tilde{2}}_{1,1} cos \alpha_1
=0
\end{equation}
So we can obtain the boundary conditions for forces
\begin{equation} \label{eq:(example.twoRod.18)}    
u^{\tilde{2}}_{1,1} =
u^{\tilde{1}}_{1,1}
\end{equation}
substitute\ref{eq:(example.twoRod.9)}to obtain
\begin{equation} \label{eq:(example.twoRod.19)}    
F_{2'} -EA_1 u^{\tilde{1}}_{1,1} sin \alpha_1 -EA_2 u^{\tilde{2}}_{1,1} sin \alpha_2
=F_{2'} -2EA_1 u^{\tilde{1}}_{1,1} sin \alpha_1
=0
\end{equation}

The boundary condition for virtual work only has equation \ref{eq:(example.twoRod.16)}, but not equation \ref{eq:(example.twoRod.10)}. Equation \ref{eq:(example.twoRod.11)} becomes
\begin{multline}\label{eq:(example.twoRod.20)}
(\frac{1} {2}EA_1(u^{\tilde{1}}_{1,1})^2 )(t_{1,u_{1'}}u_{1',y_{1'}}+t_{1,u_{2'}}u_{2',y_{1'}})
\\ +( \frac{1} {2}EA_2(u^{\tilde{2}}_{1,1})^2)(t_{2,u_{1'}}u_{1',y_{1'}}+t_{2,u_{2'}}u_{2',y_{1'}})
 =\\(\frac{1} {2}EA_1(u^{\tilde{1}}_{1,1})^2 )(t_{1,u_{2'}}u_{2',y_{1'}})
\\ +( \frac{1} {2}EA_2(u^{\tilde{2}}_{1,1})^2)(t_{2,u_{2'}}u_{2',y_{1'}})
=(EA_1(u^{\tilde{1}}_{1,1})^2 t_{1,u_{2'}}u_{2',y_{1'}}
=0
\end{multline}
because$u^{\tilde{2}}_{1,1} \neq 0$,and$t_{1,u_{2'}} \neq 0$,then
\begin{equation} \label{eq:(example.twoRod.21)}    
  u_{2',y_{1'}}=0 
\end{equation}


From the above equation, it can be seen that the virtual work boundary condition is the extremum condition of the vertical displacement $u_ {2 '} $, that is, the first derivative of the vertical displacement with respect to the movable boundary variable (intersection vertical coordinate $y_ {1'} $) is equal to zero. This conclusion is consistent with the conclusion obtained in equation \ref{eq:(example.zcwz.14)}.

To further illustrate the calculation process, a detailed explanation of the solution process will be provided below. During the solution process, it is assumed that$f_1=f_2=0,A_1=A_2=A，F_x=0,x_1=L,x_2=2L$。
according to
\ref{eq:(example.twoRod.3)},\ref{eq:(example.twoRod.4)}, \ref{eq:(example.twoRod.8)} ,\ref{eq:(example.twoRod.9)},\ref{eq:(example.twoRod.15)},\ref{eq:(example.twoRod.16)}，then
\begin{equation} \label{eq:(example.twoRod.22)}    
u^{\tilde{1}}_{1}= \frac{F_2 t\sqrt{L^2 +{y_1 }^2 }}{2A\textrm{E}y_1 }
\end{equation}
\begin{equation} \label{eq:(example.twoRod.23)}    
 u^{\tilde{2}}_{1}= \frac{F_2 t\sqrt{L^2 +{y_1 }^2 }}{2A\textrm{E}y_1 }
\end{equation}

For the intersection point, there are
\begin{equation} \label{eq:(example.twoRod.24)}    
u^{\tilde{1}}_{1}= \frac{F_2(L^2 +{y_1 }^2 )}{2A\textrm{E}y_1 }
\end{equation}
\begin{equation} \label{eq:(example.twoRod.25)}    
 u^{\tilde{2}}_{1}= \frac{F_2(L^2 +{y_1 }^2 )}{2A\textrm{E}y_1 }
\end{equation}

According to\ref{eq:(algorithm.1.9)} ~ \ref{eq:(algorithm.1.10)}，transformation matrix is
\begin{equation} \label{eq:(example.twoRod.26)} 
    Ta_i=
    \begin{bmatrix}
      cos\alpha_i & sin\alpha_i \\
      -sin\alpha_i & cos\alpha_i
    \end{bmatrix}
    i=1,2
\end{equation}
\begin{equation} \label{eq:(example.twoRod.27)} 
    Tb_i=(Ta_i)'=
    \begin{bmatrix}
      cos\alpha_i & -sin\alpha_i \\
      sin\alpha_i & cos\alpha_i
    \end{bmatrix}
    i=1,2
\end{equation}
According to the coordinate transformation
\begin{equation} \label{eq:(example.twoRod.28)} 
     \begin{bmatrix}
      u_{1'}\\
      u_{2'}
    \end{bmatrix}
    =Tb_1 
    \begin{bmatrix}
      u^{\tilde{1}}_{1}\\
      u^{\tilde{1}}_{2}
    \end{bmatrix}
    =Tb_2 
    \begin{bmatrix}
      u^{\tilde{2}}_{1}\\
      u^{\tilde{2}}_{2}
    \end{bmatrix}
\end{equation}
Obtained by solving the second equal sign in the above equation
\begin{equation} \label{eq:(example.twoRod.29)} 
     \begin{bmatrix}
      u^{\tilde{1}}_{2}\\
      u^{\tilde{2}}_{2}
    \end{bmatrix}
    =
    \begin{bmatrix}
      \frac{F_2 L(L^2 +{y_1 }^2 )}{2A\textrm{E}{y_1 }^2 }\\
      -\frac{F_2 L(L^2 +{y_1 }^2 )}{2A\textrm{E}{y_1 }^2 }
    \end{bmatrix}   
\end{equation}
Substitute \ref{eq:(example.twoRod.29)}into\ref{eq:(example.twoRod.28)}to obtain
\begin{equation} \label{eq:(example.twoRod.30)} 
     \begin{bmatrix}
      u_{1'}\\
      u_{2'}
    \end{bmatrix}
    =
    \begin{bmatrix}
      0\\
      \frac{F_2 {{\left(L^2 +{y_1 }^2 \right)}}^{3/2} }{2A\textrm{E}{y_1 }^2 }
    \end{bmatrix}   
\end{equation}
According to\ref{eq:(example.twoRod.21)}we can obtain
\begin{equation} \label{eq:(example.twoRod.31)}    
  -\frac{F_2 {\left(2L^2 -{y_1 }^2 \right)}\sqrt{L^2 +{y_1 }^2 }}{2A\textrm{E}{y_1 }^3 }=0
\end{equation}
then
\begin{equation} \label{eq:(example.twoRod.32)}    
 y_1=\sqrt{2}L
\end{equation}
At this situation, the angle between the two rods is $70.5 $°


\section {Conclusion}
This article proposes the principle of minimum virtual work applicable to mechanical structures and verifies it with multiple specific examples. The main conclusions obtained are as follows:

1. Creatively proposed the principle of minimum virtual work in mechanics. For a movable boundary mechanical system, the exact solution of the mechanical system minimizes the total virtual work of the system among all possible displacements. When the control load  is zero, the principle of minimum virtual work degenerates into the principle of minimum potential energy.

2. The basic equations of the movable boundary mechanics system have been derived, including geometric equations, equilibrium equations, constitutive equations, force and displacement boundary conditions, as well as virtual work boundary conditions and control boundary conditions. The virtual work boundary condition and control boundary condition are additional equations that need to be satisfied by the movable boundary mechanics system. When these two equations are not considered, they degenerate into the basic equations of the mechanics system with fixed boundaries.

3. The general formula for the multidimensional spatial variation method of movable boundaries has been derived through the variation method. Based on this formula, the control equations and boundary conditions of mechanical structures can be quickly obtained.

4. By using the incremental method, the minimum principle of multidimensional space with limited movable boundaries is derived, which extends the Pontryagin's minimum principle in control theory and expands its integration domain from one-dimensional space to multidimensional space.

5. Through multiple simplified bridge cases, the correctness of the theory proposed in this article has been proven, and various optimization design methods and guiding ideas for bridge structures have been proposed. It also demonstrates that the theory proposed in this article has extremely wide applicability and practical value for bridge optimization design.

This article systematically proposes the principle of minimum virtual work, promotes the principle of energy, and promotes the fundamental equations of mechanics，and theoretically solves the structural optimization problem of movable boundaries, providing a feasible path for active control of mechanical structures. It has high theoretical and practical value and can be widely applied in the optimization design of structural engineering such as bridges, promoting the development of structural optimization.
\bibliographystyle{plain}
\bibliography{refs.bib}

\begin{thebibliography}{10}

\bibitem{wuchengwei2024}
Wu~Chengwei.
\newblock Proposal, development, and application of the variational principle
  of parameter variables.
\newblock {\em Computational Mechanics}, 041(001), 2024.

\bibitem{liuzhongkan1986}
Liu Chongkan.
\newblock {\em Applied Functional Analysis}, volume~1.
\newblock Defense Industry Press, 1986.

\bibitem{daigonglian2015}
Wang~Yu Dai~Gonglian, Tang~Lixin.
\newblock Study on reasonable side to middle span ratio of large span railway
  continuous beam arch composite bridge.
\newblock {\em Journal of Railway Science and Engineering}, 12(8), 2015.

\bibitem{fujinlong2014}
Huang~Tianli Fu~Jinlong.
\newblock Applicability study of optimization method for cable force of
  suspension rod of rigid tied arch bridge.
\newblock {\em Journal of Dao Science and Engineering}, 11(4), 2014.

\bibitem{huhaichang1954}
Hu~Haichang.
\newblock On general variational principles in elasticity and normative body
  mechanics.
\newblock {\em Acta Physica Sinica}, 10(3), 1954.

\bibitem{penghaijun2011}
Peng Haijun.
\newblock Mixed variable variational method for solving optimal control
  problems and its application in aerospace control.
\newblock {\em Journal of Automation}, 37(10), 2011.

\bibitem{zhanghongwu2006}
ZHANG Hong-wu.
\newblock Parametric variational principle based elastic-plastic analysis of
  heterogeneous materials with voronoi finite element method.
\newblock {\em Applied Mathematics and Mechanics}, 27(8), 2006.

\bibitem{wayajun2016}
Wang Yajun;~Di Jin.
\newblock Reasonable edge to mid span ratio of equal section continuous beams.
\newblock {\em Journal of Lanzhou University (Natural Science Edition)}, 52(3),
  2016.

\bibitem{sunjiong2018}
Sun Jiong.
\newblock {\em Functional Analysis}, volume~1.
\newblock Higher Education Press, 2018.

\bibitem{huangkezhi2020}
Huang Kezhi.
\newblock {\em Tensor Analysis}, volume~1.
\newblock Tsinghua University Press, 2020.

\bibitem{reza2007}
Reza Memarbashi.
\newblock Variational problems with moving oundaries using decomposition
  method.
\newblock {\em Mathematical Problems in Engineering}, 2007(11), 2007.

\bibitem{zhongwanxie1997}
Zhong Wanxie.
\newblock {\em Variational Principle of Parameter Variables and Its Application
  in Engineering}, volume~1.
\newblock Science Press, 1997.

\bibitem{niuxiangjun1981}
Niu Xiangjun.
\newblock Discrete variational principles of solids - variational principles of
  finite element discrete analysis.
\newblock {\em Applied Mathematics and Mechanics}, 2(5), 1981.

\bibitem{xiaorucheng1998}
Gong~Haifan Xiao~Rucheng.
\newblock Cable force optimization and engineering application of cable stayed
  bridges.
\newblock {\em Journal of Computational Mechanics}, 15(1), 1998.

\bibitem{qianfeng2022}
Qian Feng; Peng Wei; Junbin Lou; Jinbiao Cai;~Rongqiao Xu.
\newblock Analytical solution for quick decision of tied–arch bridge
  parameters at early-design stage based on hellinger–reissner variational
  method, 2022.

\bibitem{daijie2019}
Dai Jie; Qin Fengjiang; Di Jin;~Chen Yongrui.
\newblock Overview of research on cable force optimization methods for cable
  stayed bridges.
\newblock {\em Chinese Journal of Highways}, 1(1), 2019.

\bibitem{zhouyungang2017}
Zhou Yungang.
\newblock Optimization method for constant load cable forces of long span multi
  tower cable stayed bridges.
\newblock {\em Journal of Chongqing University (Natural Science Edition)},
  36(2), 2017.

\bibitem{longyuqiu1983}
Long Yuqiu.
\newblock Generalized variational principle of partition for elastic thick
  plates.
\newblock {\em Applied Mathematics and Mechanics}, 4(2), 1983.

\bibitem{longyuqiu1987}
Long Yuqiu.
\newblock Generalized variational principle and permutation multiplier method
  with multiple arbitrary parameters.
\newblock {\em Applied Mathematics and Mechanics}, 8(7), 1987.

\bibitem{liuzhao2009}
Liu Zhao.
\newblock Determination of optimal suspension rod internal forces for tied arch
  bridges based on energy method.
\newblock {\em Engineering Mechanics}, 1(8), 2009.

\bibitem{laodazhong2011}
Lao~Da Zhong.
\newblock {\em Fundamentals of Variational Method}, volume~1.
\newblock Defense Industry Press, 2011.

\end{thebibliography}
\end{CJK}
\end{document}